\documentclass[aps,prx,twocolumn,showpacs,psfig,superscriptaddress,
longbibliography]{revtex4-1}
\usepackage{amssymb}
\usepackage{algpseudocode}
\usepackage{algorithm}
\usepackage{amsmath}
\usepackage{braket}
\usepackage{booktabs}
\usepackage{chngcntr}
\usepackage{color}
\usepackage{comment}
\usepackage{dsfont}
\usepackage{etoolbox}
\usepackage{epstopdf}
\usepackage[T1]{fontenc}
\usepackage{float}
\usepackage{graphicx}
\usepackage[colorlinks=true,linkcolor=blue,citecolor=blue,urlcolor=blue]{hyperref}
\usepackage[latin9]{inputenc}
\usepackage{indentfirst}
\usepackage{latexsym,bm,euscript}
\usepackage{mathrsfs}
\usepackage{multirow}
\usepackage{natbib}
\usepackage{soul}
\usepackage{subfigure}
\usepackage{times}
\usepackage{tikz}
\usepackage{txfonts}
\usepackage{ulem}
\usepackage{xspace}
\usepackage{xfrac}

\begin{document}
\title{High-field quantum spin liquid transitions and angle-field phase diagram 
of Kitaev magnet $\alpha$-RuCl$_3$}

\author{Han Li}
\affiliation{Kavli Institute for Theoretical Sciences, University of Chinese 
Academy of Sciences, Beijing 100190, China}
\affiliation{CAS Center for Excellence in Topological Quantum Computation, 
University of Chinese Academy of Sciences, Beijing 100190, China}

\author{Wei Li}
\email{w.li@itp.ac.cn}
\affiliation{CAS Key Laboratory of Theoretical Physics, Institute of Theoretical 
Physics, Chinese Academy of Sciences, Beijing 100190, China}
\affiliation{CAS Center for Excellence in Topological Quantum Computation, 
University of Chinese Academy of Sciences, Beijing 100190, China}

\author{Gang Su}
\email{gsu@ucas.ac.cn}
\affiliation{Kavli Institute for Theoretical Sciences, University of Chinese 
Academy of Sciences, Beijing 100190, China}
\affiliation{CAS Center for Excellence in Topological Quantum Computation, 
University of Chinese Academy of Sciences, Beijing 100190, China}

\begin{abstract}
The pursuit of quantum spin liquid (QSL) in the Kitaev honeycomb 
magnets has drawn intensive attention recently. In particular, $\alpha$-RuCl$_3$ 
has been widely recognized as a promising candidate for the Kitaev QSL. 
Although the compound exhibits an antiferromagnetic order under zero field, 
it is believed to be endowed with fractionalized excitations, and can be driven 
to the QSL phase by magnetic fields. Here, based on a realistic 
$K$-$J$-$\Gamma$-$\Gamma'$ model for $\alpha$-RuCl$_3$~\cite{Han2021}, 
we exploit the exponential tensor renormalization group approach to explore 
the phase diagram of the compound under magnetic fields. We calculate the 
thermodynamic quantities, including the specific heat,  Gr\"uneisen parameter, 
magnetic torque, and the magnetotropic susceptibility, etc, under a magnetic field 
with a tilting angle $\theta$ to the $c^*$-axis perpendicular to the honeycomb plane. 
We find an extended QSL in the angle-field phase diagram determined with 
thermodynamic responses. The gapless nature of such field-induced QSL is 
identified from the specific heat and entropy data computed down to very low 
temperatures. The present study provides guidance for future high-field experiments 
for the QSL in $\alpha$-RuCl$_3$ and other candidate Kitaev magnets.
\end{abstract}

\date{\today}
\maketitle

% ====================== Introduction ===================== %
\section{Introduction}
Quantum spin liquids (QSL) constitute an exotic many-body state without 
symmetry-breaking spin order, where a number of unconventional properties 
such as fractionalized excitations and long-range entanglement emerge
\cite{Anderson1973,Balents2010,Zhou2017,JW2019QMats,Broholm2020Science}. 
The celebrated, exactly solvable Kitaev model has attracted enormous attention 
due to the QSL ground state with localized and itinerant Majorana fermions useful 
for fault-tolerant quantum computing~\cite{Kitaev2003,Kitaev2006}. Such remarkable 
properties have incited a flurry of works on the materialization of the Kitaev model in, 
e.g., certain 4$d$- and 5$d$-electron compounds including cations with the low-spin 
$d^5$ electron configuration and the edge-shared ligand octahedra. It yields the 
Kitaev interaction by the synergy of large spin-orbit coupling and Coulomb repulsion 
on a honeycomb lattice~\cite{Jackeli2009}. Moreover, some high-spin $d$- and 
$f$-electron systems beyond the Jackeli-Khaliullin mechanism come forth recently 
which may also realize the Kitaev interaction in the compounds
\cite{Trebst2017arXiv,Winter2017,Janssen2019,Motome2020b}. 

The ruthenium halide $\alpha$-RuCl$_3$ is arguably the most studied 
Kitaev material~\cite{Sears2015,Banerjee2017,Do2017,Do2017,
Banerjee2018,Balz2021,Sears2017, Zheng2017,Baek2017,Jansa2018,
Wulferding2020,Ponomaryov2020,Leahy2017,Modic2021,
Kasahara2018Unusual,Kasahara2018,Yokoi2021Science,
Yamashita2020sample}. 
Although it has a long-range zigzag antiferromagnetic ordered state 
below 7~K ~\cite{Sears2015,Banerjee2017,
Do2017}, proximate Kitaev QSL behaviors at elevated 
temperatures were observed~\cite{Do2017,Banerjee2018}. 
The zigzag spin order is suppressed under an 
in-plane field of around 7~T~\cite{Sears2017,Zheng2017,Banerjee2018,Balz2021}, 
where the possible field-induced QSL phase has been intensively 
studied via multiple experimental probes including the Raman scattering
\cite{Wulferding2020}, terahertz absorption measurements
\cite{Ponomaryov2020}, nuclear magnetic resonance~\cite{Zheng2017,
Baek2017,Jansa2018}, magnetic torque \cite{Leahy2017,Modic2021} 
and thermal Hall conductivity measurements~\cite{Kasahara2018Unusual,
Kasahara2018,Yokoi2021Science,Yamashita2020sample}. 

% ============= Fig1: Phase diagram and Lattice ============= %
\begin{figure}[h!]
\includegraphics[angle=0,width=1\linewidth]{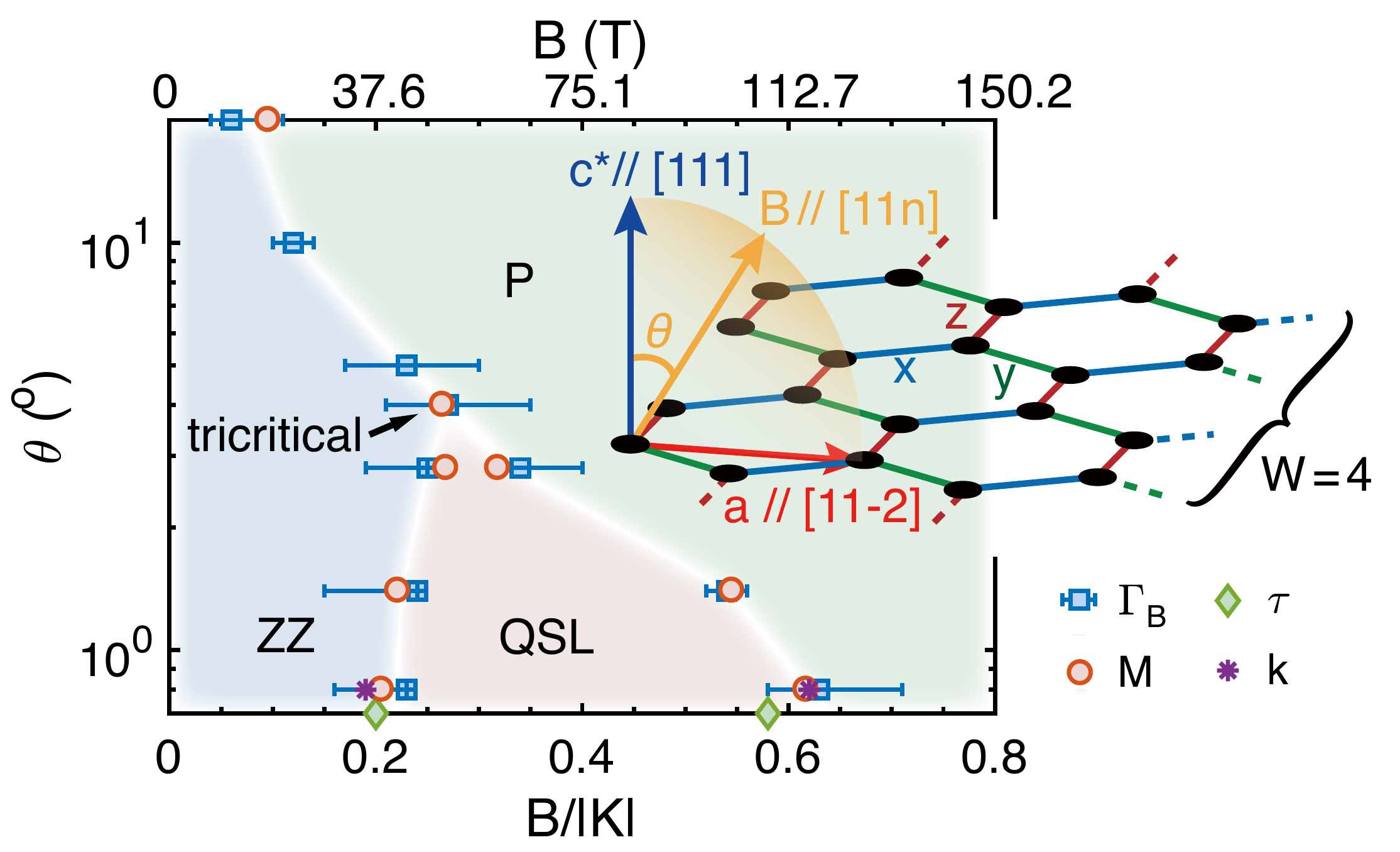}
\renewcommand{\figurename}{\textbf{Fig. }}
\caption{The angle-field phase diagram of the realistic 
$K$-$J$-$\Gamma$-$\Gamma'$ model for $\alpha$-RuCl$_3$. There are 
three phases including zigzag (ZZ), quantum spin liquid (QSL), and the 
polarized (P) states as indicated in the figure. The phase boundaries are 
determined from the responses in Gr\"uneisen parameters $\Gamma_B$, 
magnetic torque $\tau$, and the magnetotropic susceptibility $k$ at $T/|K| 
\simeq 0.01$, in consistent with that from the ground-state magnetization 
curves~\cite{Zhou2022arXiv}. We reveal that the phase transitions between 
ZZ (a ``solid'' order), QSL (liquid-like phase), and the P (a weakly interacting 
``gas''-like system) phases meet at a tricritical point. The inset illustrates the 
honeycomb lattice defined on a cylinder of width $W=4$, where the $x$, $y$, 
and $z$ bonds with bond-directional Kitaev interactions are marked in blue, 
green, and red colors, respectively. The in-plane $a$-axis, out-of-plane $c^*$-axis, 
and the angle $\theta$ of the applied field within the ${ac}^*$-plane are indicated 
by the arrows.
}
\label{Fig:Illus}
\end{figure}
% ================================================= %

On the other hand, in the theoretical studies, 
the accurate microscopic model description 
of $\alpha$-RuCl$_3$ is important for understanding the compound,
which, however, has been unsettled for a long period~\cite{Laurell2020}. 
Recently, some of the authors proposed a Kitaev-Heisenberg-Gamma-Gamma' 
($K$-$J$-$\Gamma$-$\Gamma'$) model with dominant Kitaev interaction 
$K=-25$~meV, nearest-neighbor Heisenberg coupling $J=-0.1|K|$, 
off-diagonal terms $\Gamma=0.3|K|$ and $\Gamma'=-0.02|K|$, 
which puts the major experimental observations in a coherent picture, 
and makes a relevant prediction of QSL states induced by high 
out-of-plane fields~\cite{Han2021}. Such a high-field QSL phase 
is separated from zigzag antiferromagnetic and the polarized phases, 
through two quantum phase transitions (QPTs) at 35~T and 130~T, 
respectively. This theoretical prediction is recently confirmed in high 
pulsed field experiments~\cite{Zhou2022arXiv}.

In this work, we extend the previous theoretical studies to the angle-field 
phase diagram of the realistic $K$-$J$-$\Gamma$-$\Gamma'$ model 
with the thermal tensor network approach~\cite{Chen.b+:2017:SETTN, 
Chen2018,Lih2019}. Through the finite-temperature simulations of the 
specific heat $C_{\rm m}$, Gr\"uneisen parameters $\Gamma_B$, 
magnetic torque $\tau$, and magnetotropic susceptibility $k$, etc., 
we find a high-field QSL phase residing between the zigzag 
antiferromagnetic and the field-polarized phases. We determine 
the transition fields with prominent thermodynamic responses and 
offer concrete theoretical proposal for experimental probes of such 
spin liquid transitions in $\alpha$-RuCl$_3$ and potentially also other 
Kitaev candidate magnets.

% ============ Model and methods ============= %
\section{Model and methods}
The effective spin Hamiltonian of $\alpha$-RuCl$_3$~\cite{Han2021} 
considered in this work reads
\begin{equation}
\begin{split}
H=& \sum_{\langle i,j\rangle_{\gamma}} [K S_i^{\gamma}S_j^{\gamma} 
+ J\,\textbf{S}_i\cdot \textbf{S}_j 
+ \Gamma(S_i^{\alpha}S_j^{\beta}+S_i^{\beta}S_j^{\alpha}) \\
& +\Gamma'(S_i^{\gamma}S_j^{\alpha}+S_i^{\gamma}S_j^{\beta}
+ S_i^{\alpha}S_j^{\gamma}+S_i^{\beta}S_j^{\gamma})],
\end{split}
\label{Eq:HamRuCl3}
\end{equation}
where the summation is over the nearest-neighbor (NN) bond $\langle i, 
j \rangle_\gamma$ with $\gamma = \{ x,y,z\}$ (see inset in 
Fig.~\ref{Fig:Illus}). $K$ denotes the bond-dependent Kitaev interactions, 
$J$ is the Heisenberg term, and $\Gamma$, $\Gamma'$ are the off-diagonal 
symmetric couplings with $\{ \alpha, \beta, \gamma\}$ being the three spin 
components under a cyclic permutation.

The magnetic field $B$ is applied along the direction $[l\ m\ n]$ 
in the spin space $(S^x, S^y, S^z)$, i.e., the Zeeman term is 
$H_{\rm Zeeman} = \frac{B}{\sqrt{l^2+m^2+n^2}}[S^x, S^y, S^z]\cdot 
[l,m,n]^{T}$. Therefore, $H_{[11\bar2]}$ and $H_{[111]}$ correspond 
to the fields applied along the $a$- and $c^*$-axis, respectively. 
The angle between the 
applied field $H_{[11n]}$ and $c^*$-axis within the $ac^*$-plane can 
be represented by $\theta = \arccos(\frac{2+n}{\sqrt{6+3n^2}})\times 
\frac{180^{\circ}}{\pi}$, as depicted in the inset of Fig.~\ref{Fig:Illus}.

Simulations based on the $K$-$J$-$\Gamma$-$\Gamma'$ model can 
well reproduce the low-temperature zigzag order~\cite{Sears2015,
Banerjee2017,Do2017}, double-peaked specific heat~\cite{Kubota2015,
Do2017,Widmann2019}, magnetic anisotropy~\cite{Sears2015,Kubota2015,
Johnson2015,Weber2016,Banerjee2017,Lampen-Kelley2018,Sears2020}, 
magnetization curves~\cite{Kubota2015,Johnson2015,Zheng2017,
Banerjee2018}, and the prominent M-star dynamical spin structure 
factors~\cite{Banerjee2017,Do2017} in $\alpha$-RuCl$_3$ (see a brief 
recapitulation in Appendix \ref{Sec:KJGGP}). Besides, one remarkable 
prediction based on this realistic model is the presence of high-field 
QSL driven by out-of-plane fields~\cite{Han2021}, whose nature is still 
under intensive investigation~\cite{Yu2022}.

Below we employ the exponential tensor renormalization group 
(XTRG)~\cite{Chen2018,Lih2019} method and perform 
finite-temperature calculations on a honeycomb-lattice cylinder 
with total sites $N=W\times L\times 2$, where the width is fixed 
as $W=4$ and length $L$ ranges from 6 to 12, as illustrated in the 
inset of Fig.~\ref{Fig:Illus}. We retained up to $D = 400$ bond states 
with truncation errors $\epsilon \simeq 10^{-4}$ down to the lowest 
temperature $T/|K| \simeq 0.0085$, which guarantees well converged 
results till the lowest temperature (c.f., Appendix \ref{Sec:Benchmark}).

% =========== Fig2: Cm, S, and Gamma_B ================== %
\begin{figure*}[t!]
\includegraphics[angle=0,width=0.99\linewidth]{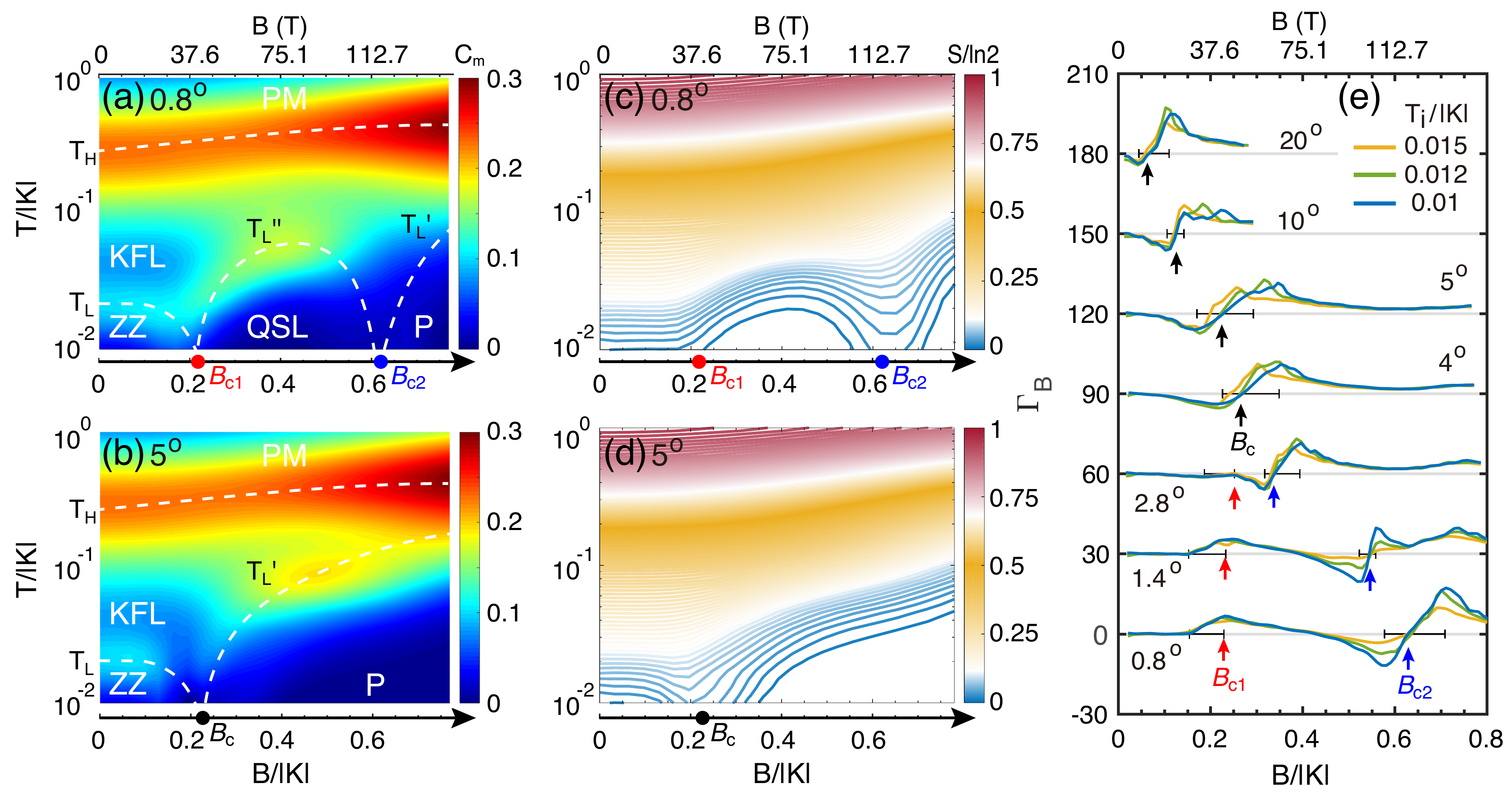}
\renewcommand{\figurename}{\textbf{Fig. }}
\caption{(a,b) Contour plots of the specific heat results in fields applied along 
$\theta=0.8^{\circ}$ and $5^{\circ}$, respectively. The solid dots marked on 
the $T=0$ axis denote the QPTs obtained through the DMRG calculations
\cite{Zhou2022arXiv}. The white dashed lines separating the ZZ, QSL, P, 
Kitaev fractional liquid (KFL), and paramagnetic (PM) phases are guides 
for the eye. (c,d) show the isentropes $S/\ln{2}$ for two different $\theta$ 
angles, where the critical fields ($B_{c_1,c_2}$ and $B_c$) are indicated 
by the dots. These calculations with scanned fields are performed on the 
YC$4\times6\times2$ lattice. (e) The field-dependent Gr\"uneisen parameters 
$\Gamma_B$ at various $\theta$ angles with fixed initial temperatures $T_{\rm i} 
\simeq 0.015$, 0.012, and 0.01. The data are calculated by $\Gamma_B = 1/T 
(dT/dH)_S$, and are shifted vertically by a value of 30 for clarify. For small 
$\theta$, e.g., $0.8^{\circ}$, $1.4^{\circ}$, and $2.8^{\circ}$, two critical fields 
$B_{c1}$ and $B_{c2}$ indicated by the red and blue arrows denote the low- 
and high-field phase transitions, respectively; while only a single phase transition 
$B_c$ indicated by a black arrow is observed for $\theta \geq 4^{\circ}$. The 
segment around each arrow gives the range of errorbar for the determined 
transition fields. 
}
\label{Fig:Gruneisen}
\end{figure*}
% ================================================== %

% ========= about Cm, S, Gamma_B, torque and k ========= %
\section{Finite-temperature characteristics of quantum spin states 
and transitions}

\subsection{Specific heat and isentropes}

We start with conventional thermodynamic 
quantities such as the specific heat $C_{\rm m}$ and magnetic entropy 
$S/\ln{2}$ in Fig.~\ref{Fig:Gruneisen}(a-d), where the contour plots 
can be used to map the temperature-field phase diagram with various 
angles $\theta$. 
As shown in Fig.~\ref{Fig:Gruneisen}(a), when the field is applied along 
the $\theta=0.8^{\circ}$ direction, the double-peaked $C_{\rm m}$ 
structure can be observed under a finite range of fields ($B/|K| \lesssim 
0.22$), with the high-$T$ and low-$T$ peaks correspond to two temperature 
scales $T_{\rm H}$ and $T_{\rm L}$: the short-range spin correlations 
establish at $T_{\rm H}$ and the long-range antiferromagnetic 
zigzag order is formed below $T_{\rm L}$, respectively. When the 
field $B/|K|$ is increased from 0 to 0.22, the low-$T$ $C_{\rm m}$ peak 
moves towards lower temperatures, indicating that the 
zigzag order gets gradually suppressed by the magnetic fields. 
On the other hand, as the field exceeds $B/|K|=0.22$, and below the 
polarization field, a low-$T$ peak emerges as indicated by $T_{\rm L}^{''}$, 
below which there exists a field-induced QSL phase (c.f., Appendix
\ref{Sec:KJGGP}).

The corresponding isentropes with $\theta = 0.8^{\circ}$ are shown in 
Fig.~\ref{Fig:Gruneisen}(c). The adiabatic $T$-$B$ curves exhibit distinct 
changes when entering (rapid increase of $T$) and leaving (a dip) the 
intermediate QSL regime. They clearly signal two QPTs from the zigzag 
order to the QSL phase then to the field-polarized phases, at {$B/|K| \simeq 
0.22$ and 0.62}, respectively. The transition fields determined with density 
matrix renormalization group (DMRG) calculations on the same geometry
\cite{Zhou2022arXiv} are denoted in the $T=0$ axis with solid dots,
where excellent agreements with the present finite-temperature results 
are seen.

The situation changes dramatically when the field angle increases to 
$\theta = 5^{\circ}$. As shown in Fig.~\ref{Fig:Gruneisen}(b,d), the results 
suggest that there is only one critical field between the zigzag ordered and 
field polarized phases, with no intermediate states any more. The behaviors 
of $C_{\rm m}$ and $S$ are quite similar to that of the in-plane-field case
\cite{Han2021}, except that the transition field is higher. Thus we find the 
intermediate QSL phase very sensitively depends on the angle $\theta$. 
To accurately determine the phase boundaries in the angle-field phase diagram, 
below we resort to the thermodynamic, experimentally accessible quantities 
and parameters.

\subsection{Gr\"uneisen parameter}

The magnetic Gr\"uneisen parameter $\Gamma_B$ 
has been employed 
to accurately determine the critical in-plane fields in $\alpha$-RuCl$_3$
\cite{Bachus2021}, which, however,  
poses challenges to many-body calculations. Here with the state-of-the-art XTRG 
method, we are able to compute this thermodynamic ratio and show the 
results in Fig.~\ref{Fig:Gruneisen}(e). The field-dependent $\Gamma_B 
= 1/T (dT/dH)_S$ are derived from the simulated isentropes starting from 
various initial temperatures (and a fixed field). A sign change structure 
in $\Gamma_B$ can be observed in Fig.~\ref{Fig:Gruneisen}(e) near the 
higher transition field $B_{c2}/|K| \simeq 0.62$ (indicated by the blue arrows), 
and it becomes more and more pronounced as 
temperature lowers, revealing a second-order phase transition from QSL 
to the polarized phase. On the other hand, in the relatively low-field 
regime with {$B_{c1}/|K| \simeq 0.22$}, a peak in $\Gamma_B$ is 
observed (indicated by a red arrow) that 
corresponds to a first-order QPT between 
ZZ and the QSL phases.

When the field is rotated within the ${ac}^*$-plane, the higher transition 
field shifts from $B/|K| \simeq 0.62$ to 0.06 as the angle $\theta$ 
changes from $0.8^{\circ}$ to ${20}^{\circ}$, which reflects that the 
polarization field is very sensitive to the angle $\theta$. The first-order 
QPT stays around $B_{c1}/|K| \simeq 0.23$ for small angles 
and merge to the second-order QPT at around $\theta \gtrsim 4^{\circ}$, 
where a tricritical point emerges. In Fig.~\ref{Fig:Illus}, we gather the 
transition fields estimated by $\Gamma_B$ and obtain the angle-field 
phase diagram. As also indicated in Fig.~\ref{Fig:Gruneisen}(e), the 
errorbars of the phase boundaries can be estimated as the difference 
in field strengths of the dips and peaks in $\Gamma_B$.

\subsection{Magnetic torque and magnetotropic susceptibility}
The torque magnetometry constitutes a sensitive technique to 
probe the magnetic anisotropies in quantum materials, and recently 
be used to study the intricate quantum spin states and transitions in 
$\alpha$-RuCl$_3$~\cite{Modic2018,Modic2021}. However, its 
numerical results are lacking, partly due to the challenges in 
its many-body simulations.

With thermal tensor networks, we can compute the magnetic torque
and its derivative, magnetotropic susceptibility, with a high accuracy.
As the free energy $F$ can be written as ${\rm d}F = -S{\rm d}T - 
P{\rm d}V - M{\rm d}B + \tau{\rm d}\theta$ where $\theta$ is the tilted 
angle of the magnetic field, the first derivative $\tau = \partial F/ \partial 
\theta$ represents the magnetic torque, which can be measured in 
$\alpha$-RuCl$_3$ experiments through $M\times B$~\cite{Modic2018}. 
Recently, resonant torsion magnetometry technique is also used to measure 
the magnetotropic susceptibility $k = \partial ^2F /\partial \theta^2$ (the 
second derivative of free energy)~\cite{Modic2021, Modic2022arXiv}. 
Following this line, below we perform XTRG calculations of the 
$K$-$J$-$\Gamma$-$\Gamma'$ model for $\alpha$-RuCl$_3$, 
investigate $\tau$ and $k$ at various temperatures and fields, 
and predict salient features of the two QPTs in the magnetotropic 
quantities to be checked in future high-field measurements.

% ================== Fig3: torque and k ================== %
\begin{figure}[h!]
\includegraphics[angle=0,width=0.9\linewidth]{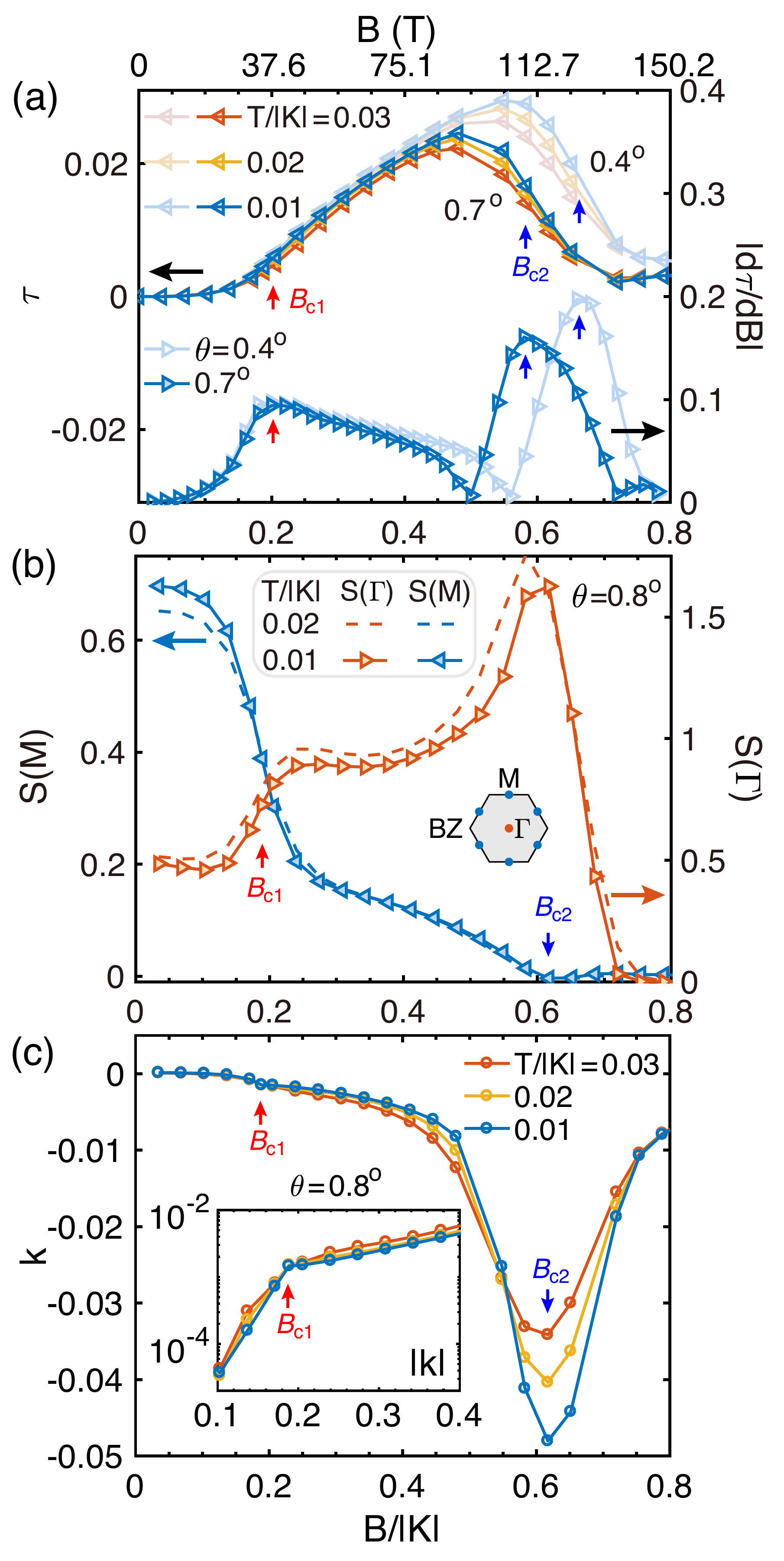}
\renewcommand{\figurename}{\textbf{Fig. }}
\caption{
(a) The calculated magnetic torque $\tau$ (the upside curves with left axis) 
and the absolute value of its derivative $|{\rm d}\tau/{\rm d}B|$ (the downside 
two with right axis) of $\alpha$-RuCl$_3$ model with fields applied along 
$\theta = 0.4^{\circ}$ and $0.7^{\circ}$ at $T\simeq0.03$, 0.02 and 0.01.
Two transition fields $B_{c1}$ and $B_{c2}$ are identified from the peak 
positions of $|{\rm d}\tau/{\rm d}B|$ indicated by the red and blue arrows, 
respectively.
(b) The static spin-structure factors $S($\textbf{k}$)$ (see the main text) 
for $\theta \simeq 0.8^{\circ}$ with $\textbf{k} =$ M and $\Gamma$ in the 
Brillouin zone (shown in the inset). The red arrow denotes a fast drop of 
$S({\rm M})$, indicating the suppression of the zigzag antiferromagnetic 
order at low temperatures, while the blue arrow corresponds to the field 
where both $S({\rm M})$ and $S(\Gamma)$ decrease towards zero.
(c) The calculated magnetotropic susceptibility $k$ for $\theta \simeq 
0.8^{\circ}$ at various low temperatures. The sharp dip corresponds
to the second-order phase transition denoted by the blue arrow, while 
a kink occurs at around {$B/|K| = 0.19$} signposted by the red arrow 
as zoomed in in the inset.
}
\label{Fig:Magnetotropic}
\end{figure}
% ================================================== %

In Fig.~\ref{Fig:Magnetotropic}(a), we show the magnetic torque 
$\tau(\theta/2) = (F_{\theta} - F_{0}) / \theta$ (where $F_0$ 
represents the free energy at zero-field) with $\theta=0.8^{\circ}$ 
and $1.4^{\circ}$, computed at low temperatures $T/|K|=0.03$, 
0.02 and 0.01. At low fields, $B < B_{c1}$, we find a relatively 
small value of $\tau$, which is understandable as the torques in 
two sublattices are expected to cancel each other in the 
antiferromagnetic ZZ phase, resulting in a nearly zero total net torque 
value. As fields further increase, the calculated $\tau$ gets enhanced 
rapidly as the ZZ order is suppressed in the intermediate QSL regime,
which eventually drops again to small values at high fields as the system 
enters to the polarized phase. This can be ascribed to the fact that the 
angle between induced moments and fields is almost zero. The transition 
fields can thus be determined from where the torque changes most
rapidly by computing the derivatives of $\tau$ with respect to the field $B$, 
i.e., $d\tau/d B$ shown in Fig.~\ref{Fig:Magnetotropic}(a). The red and blue 
arrows indicate the transition fields from ZZ to QSL and QSL to polarized 
phases, respectively.  
 
The behaviors of magnetic torque are also found consistent with the static 
spin-structure factor results
\begin{equation}
{S}(\textbf{k})=\sum_{j\in {N}, j \neq i_0} 
e^{ i \textbf{k} (\textbf{r}_j-\textbf{r}_{i_0})} 
(\langle S_{i_0} S_j\rangle - \langle 
S_{i_0}\rangle\langle S_{j}\rangle),
\end{equation}
where $i_0$ indicates a central reference site and the results are at relatively 
low temperature $T/|K| \simeq 0.02$ and 0.01. As shown in Fig.
\ref{Fig:Magnetotropic}(b), the zigzag spin correlations at small fields, e.g., 
$B < B_{c1}$ can be evidenced by the large $S({\rm M})$ value 
[with the $\rm M$ as well as $\Gamma$ point indicated in the inset 
of Fig.~\ref{Fig:Magnetotropic}(b)], which becomes suppressed in the 
intermediate QSL phase. The enhancement of $S(\Gamma)$ near $B_{c1}$ 
signals the buildup of uniform magnetization where the torque $\tau$ also 
increases rapidly in Fig.~\ref{Fig:Magnetotropic}(a). When the system enters 
the spin polarized phase at $B_{c2}$, the structure factor peaks at M and 
$\Gamma$ points both vanish as expected [Fig.~\ref{Fig:Magnetotropic}(b)].
 
The magnetotropic susceptibility $k$ can also be used to sensitively probe
the two quantum phase transitions. In Fig.~\ref{Fig:Magnetotropic}(c), we 
plot the results with $\theta = 0.8^{\circ}$ at $T/|K|=0.03$, 0.02, and 0.01. 
The parameter $k$, second-order derivative of the free energy with respect 
to the magnetic field orientation $\theta$, has intimate relation to the susceptibility 
$\chi$~\cite{Modic2022arXiv} and exhibits discontinuities at second-order 
phase transitions. In Fig.~\ref{Fig:Magnetotropic}(c), the sharp dip at around 
{$B\simeq B_{c2}$} denoted by the blue arrow corresponds to a second-order 
transition, while the low-field one, as emphasized in the inset, shows a kink at 
around $B \simeq B_{c1}$ which corresponds to a first-order phase transition. 
From the magnetotropic quantities $\tau$ and $k$, we determine the transition 
fields at $\theta=0.7^{\circ}$ and $0.8^{\circ}$ and show them also in Fig.~\ref{Fig:Illus}.
Besides, we have also computed the matrix product operator (MPO) entanglement
of the system, which provides accurate estimate of transition fields in accordance
with the results above (see Appendix \ref{Sec:SE}). With these finite-temperature 
simulations, we show that the high-field torque magnetometry measurements 
can be used to sensitively detect the two QPTs associated with the intermediate
QSL phase in future experimental studies.

% ============= Nature of high-field QSL ===================== %
\section{Gapless nature of the high-field QSL identified from thermodynamics}
\label{Sec:CmS}
As indicated by the dome-like feature in Fig.~\ref{Fig:Gruneisen}(a,c), 
there exist an intermediate QSL regime below the emergent 
low-temperature scale $T_{\rm L}^{''}$. To further reveal the nature 
of this intermediate phase, we push the calculations of  $C_{\rm m}$ 
and $S/\ln{2}$ to longer YC4 cylinders with $L$ up to 12.

\begin{figure}[t!]
\includegraphics[angle=0,width=0.88\linewidth]{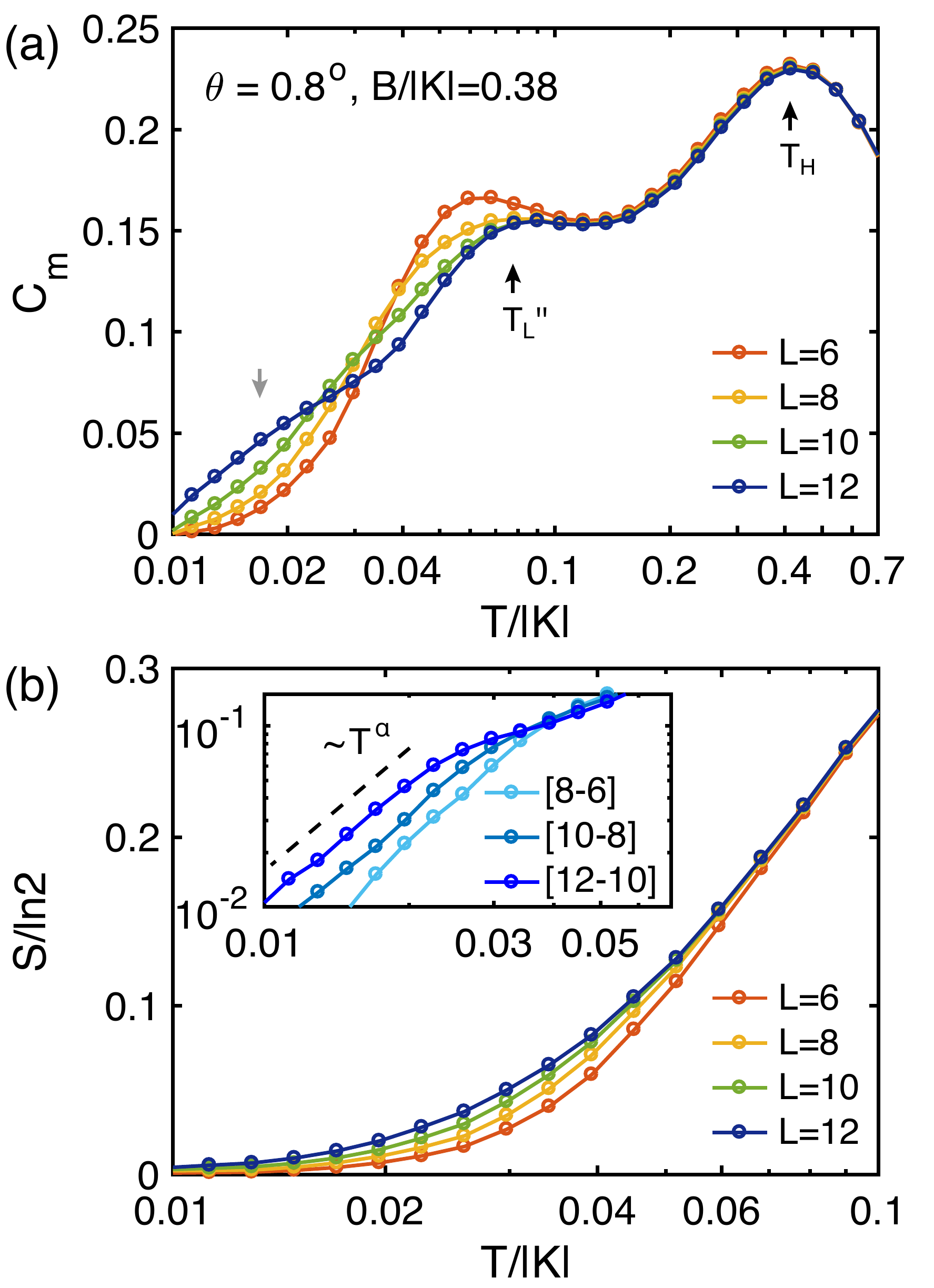}
\renewcommand{\figurename}{\textbf{Fig.}}
\caption{ (a) The computed specific heat $C_{\rm m}$
on various YC$4\times L\times 2$ geometries with different 
lengths $L$ ranging from 6 to 12 under a field $B/|K| = 0.38$ 
along $\theta=0.8^{\circ}$ away from the $c^*$-axis. The high- 
and low-temperature scale $T_{\rm H}$ and $T_{\rm L}^{''}$ are 
indicated. The grey arrow stress the enhancement of $C_{\rm m}$ 
at very low temperature $T/|K|<0.03$ as system length $L$ increases.
(b) The corresponding thermal entropy $S/\ln{2}$ results, with  
the subtracted data reflecting the bulk property at low temperature 
shown in the inset. The dashed line at $T/|K|<0.02$ represents a 
power-law fitting with $\alpha \simeq 1.5$, serving as a guide for the eye. 
}
\label{Fig:CmS}
\end{figure}

In Fig.~\ref{Fig:CmS}(a), we find the high- and low-temperature scales 
$T_{\rm H}$ and $T_{\rm L}^{''}$ change only slightly as we elongate the system 
from $L=6$ to $L=12$. The height of the peak at $T_{\rm L}^{''}$ gets lowered, 
while the $C_{\rm m}$ values for $T< T_{\rm L}^{''}$ are actually enhanced, 
which gives rise to a shoulder-like structure for the largest system size $L=12$ 
as indicated by the grey arrow below $T/|K| \simeq 0.03$. The corresponding 
entropy curves are shown in Fig.~\ref{Fig:CmS}(b), where we see that there 
are considerable amount of low-temperature entropies below $T/|K| \simeq 0.03$, 
indicating the strong spin fluctuations and large spin excitation density of states. 
In the inset of Fig.~\ref{Fig:CmS}(b), we subtract the results of two YC4 lattices 
with different (adjacent) lengths, e.g., the $[8-6]$ represent results obtained by 
subtracting YC$4\times 6 \times2$ data from the YC$4\times 8 \times2$. The 
obtained entropy results reflect the bulk property in the central columns and 
suffer less severe boundary effects, and a power-law 
behavior of entropy $S \sim T^{\alpha}$ can be clearly seen, which indicates 
that the high-field QSL has gapless low-energy excitations and there are 
considerable entropies released only below the temperature $T \simeq 
0.03 |K|$. Overall, the thermodynamic results here along the tilted angle 
point to the conclusion of a gapless QSL, consistent with previous DMRG 
results (restricted to out-of-plane fields)~\cite{Han2021}.

% =============== Fig4: Cm, S  ================= %

% =========== Conclusions and discussions ============ %
\section{Conclusions and discussions}
In the present work, we have calculated the experimentally relevant 
thermodynamic properties, i.e., magnetic specific heat, magnetocaloric 
effect characterized by the Gr\"uneisen parameters, magnetic torque, 
and the magnetotropic susceptibility of the primary candidate Kitaev magnet 
$\alpha$-RuCl$_3$ based on the realistic $K$-$J$-$\Gamma$-$\Gamma'$ 
model and through highly accurate XTRG method. Recently, a high-field 
magnetization measurement on $\alpha$-RuCl$_3$ up to 102 T have 
witnessed two phase transitions enclosing an intermediate phase
\cite{Zhou2022arXiv}, in agreement with the prediction based on 
the model calculations~\cite{Han2021}. 
Here we calculated further thermodynamic properties that provide a 
comprehensive angle-field phase diagram and useful guide for future
experimental studies. For $\theta<4^{\circ}$, we find two field-induced 
quantum phase transitions evidenced by various quantities. (i) The 
diverging Gr\"uneisen parameter $\Gamma_B$ shows a sign change 
behavior at high-field transition point $B_{c2}$,  suggesting a second-order 
phase transition. Exactly at the same field, the magnetotropic susceptibility 
$k$ features a sharp peak. 
(ii) The hump in $\Gamma_B$ at around $B_{c1}$ reflects a 
quantum phase transition possibly of first-order. There is also a peak 
in $|{\rm d}\tau/{\rm d}B|$ and a kink in $k$, which point to the 
same conclusion. On the other hand, for large $\theta \gtrsim
4^{\circ}$, only a single phase transition from antiferromagnetic 
to polarized phase is found, suggesting the absence of an 
intermediate QSL phase.

Moreover, it is noteworthy that besides the conventional candidate materials 
with Kitaev interactions, 
e.g., X$_2$IrO$_3$ (X = Na, Li, Cu)~\cite{Singh2010,Chaloupka2010,Singh2012,
Katukuri2014,Yamaji2014,Winter2016,Mehlawat2017,Abramchuk2017,Choi2019}, 
X$_3$LiIr$_2$O$_3$ (X = Ag, Cu, H) with Ir$^{4+}$ \cite{Todorova2011,
Roudebush2016,Kitagawa2018}, and XR$_3$ (X = Ru, Yb, Cr; R = Cl, I, Br) 
\cite{Winter2016, Winter2017NC, Wu2018, Cookmeyer2018, Kim2016, 
Suzuki2019, Ran2017, Wang2017, Ozel2019, Banerjee2016, HSKim2015,
Danrui2022, Imai2021, Hao2021, Xing2020, Sala2020, McGuire2015}, etc.,
some newly reported Kitaev family such as rare-earth chalcohalide REChX 
(RE = rare earth; Ch = O, S, Se, Te; X = F, Cl, Br, I)~\cite{Zhang2021CPL, Zhang2022PRR} and cobalt honeycomb oxides Na$_2$Co$_2$TeO$_6$
\cite{Lin2021NC, Yao2022PRL}, Na$_3$Co$_2$SbO$_6$~\cite{Liu2020PRL},
and BaCo$_2$(AsO$_4$)$_2$~\cite{Zhong2020SA}, etc., also offer a platform 
exhibiting highly anisotropic, bond-dependent exchange couplings. It would be 
worthwhile to explore their field-induced quantum spin states along the 
out-of-plane direction and generally tilted angles in the future, and the present 
study on angle-field phase diagram of the $K$-$J$-$\Gamma$-$\Gamma'$
model provides theoretical guide for experimental 
explorations in these intriguing quantum magnets.

% ============= acknowledgments ===================== %

\begin{acknowledgments}

{This work was supported by the National Natural Science Foundation of 
China (Grant Nos. 12222412, 11974036, 11834014, 12047503, and 12174386), 
Strategic Priority Research Program of CAS (Grant No. XDB28000000),
National Key R$\&$D Program of China (Grant No. 2018YFA0305800), 
CAS Project for Young Scientists in Basic Research (Grant No.~YSBR-057),
and China National Postdoctoral Program for Innovative Talents (Grant 
No. BX20220291). H.L. and W.L. are indebted to Xu-Guang Zhou, Shun-Yao Yu, 
Shou-Shu Gong, and Zheng-Xin Liu for stimulating discussions, and thank the 
HPC-ITP for the technical support and generous allocation of CPU time.}\\
\end{acknowledgments}

\appendix

\setcounter{equation}{0}
\setcounter{figure}{0}
\setcounter{table}{0}

\makeatletter

% Prefix a "S" to all equations, figures, tables and reset the counter
\renewcommand{\theequation}{A\arabic{equation}}
\renewcommand{\thefigure}{A\arabic{figure}}
\renewcommand{\theHfigure}{A\arabic{figure}}
\renewcommand{\bibnumfmt}[1]{[#1]}
\renewcommand{\citenumfont}[1]{#1}

% ============= Appendix.A ===================== %
\section{Realistic $\alpha$-RuCl$_3$ model and the high-field QSL}
\label{Sec:KJGGP}

{In the strongly correlated transition metal compounds, $\alpha$-RuCl$_3$ 
is believed to serve as a prototypical candidate material for the 
Kitaev model~\cite{Plumb2014,Sears2015,
Johnson2015,Banerjee2016,Banerjee2017,Ran2017,Sears2020}. As it undergoes 
a magnetic transition to antiferromagnetic order at a relatively low temperature, 
i.e., $T\simeq 7$~K~\cite{Sears2015,Johnson2015,Banerjee2017,Kubota2015},
people has taken efforts to find out the effective spin 
Hamiltonian of $\alpha$-RuCl$_3$, which includes not only the Kitaev term 
$K$, but also the Heisenberg interactions $J$, $J_2$, and $J_3$, and off-diagonal 
$\Gamma$ and $\Gamma'$ couplings~\cite{Winter2016,Winter2017NC,Wu2018,
Cookmeyer2018,Kim2016,Suzuki2019,Ran2017,Wang2017,Ozel2019,Banerjee2016,
HSKim2015,Janssen2017, Andrade2020}, which is important for gaining insights into the candidate Kitaev 
material. The Kitaev interaction in this compound 
has been widely accepted to be ferromagnetic~\cite{Winter2016,Winter2017NC,
Wu2018,Cookmeyer2018,Kim2016,Suzuki2019,Ran2017,Wang2017,Ozel2019,
Do2017,Motome2019,Sears2020}. However, the magnitudes of $K$ and even 
the signs of non-Kitaev terms were undetermined, and it is very challenging
to find a model that can 
accurately describe the realistic $\alpha$-RuCl$_3$~\cite{Laurell2020}.

We focus on the minimal $K$-$J$-$\Gamma$-$\Gamma'$ model
\cite{Gordon2019, Lee2020nc, Han2021, Janssen2017}, especially on its field-induced 
properties. In our previous work~\cite{Han2021}, we determine the parameters 
from fitting the thermodynamic properties, i.e., the double-peak feature 
of specific heat with two temperature scales at around 100~K and 7~K
\cite{Kubota2015,Do2017,Widmann2019}, and the anisotropic susceptibilities 
along $a$- and $c^*$-axis~\cite{Lampen-Kelley2018,Banerjee2017,Weber2016}. 
The determined parameter set is $K=-25$~meV, $\Gamma=0.3|K|$, $\Gamma'
=-0.02|K|$, $J=-0.1|K|$, with in-plane and out-of-plane Land\'e factor 
$g_{a}=2.5$ and $g_{c*}=2.3$, respectively. With this model, the 
low-temperature zigzag antiferromagnetic order~\cite{Banerjee2017} and its
magnetization curve can be well reproduced, which are found in quantitative 
agreement with  experiments~\cite{Johnson2015, Kubota2015}.
The transition fields that suppressed the zigzag order are also in accordance 
with experimental observations, along the in-plane direction~\cite{Kubota2015,
Johnson2015,Zheng2017,Banerjee2018} and with a tilted $\theta=35^{\circ}$ 
angle~\cite{Baek2017}. Besides, it was also found that the zigzag order gets 
suppressed at 35 T under out-of-plane fields ($\theta=0^{\circ}$), above which, 
and below a polarization field of 100~T level, a field-induced QSL phase 
emerges as evidenced by both density matrix renormalization group (DMRG) 
and variational Monte Carlo (VMC) method at ground state~\cite{Han2021}. 
Here in this work, we further extend the conclusion that the high-field QSL 
phase can extend to a finite range of $\theta$ angles.
}

% ============= Appendix.B ===================== %
\section{Exponential tensor renormalization group method}
\label{Sec:Benchmark}

The exponential tensor renormalization group (XTRG) method
\cite{Chen.b+:2017:SETTN,Chen2018} exploited in this work carries 
out the finite-temperature many-body simulations down to low 
temperature exponentially fast, which has been shown to be a 
highly efficient and very powerful tool in solving various 2D spin
lattice models~\cite{Chen2018,Chen2018b,Lih2019,Li2020b}, 
realistic quantum magnets~\cite{Li2020,Hu2020,Han2021,Gao2022}, 
and correlated fermion system~\cite{Chen2021,Lin2022}.
Below we synopsize the main idea of such method and provide some 
benchmark results on the realistic $K$-$J$-$\Gamma$-$\Gamma'$ model.

In XTRG, we start with the high-temperature density matrix $\rho({\tau_0})$ 
with the initial $\tau_0 \equiv |K|/T = 0.0025$ through a series expansion 
in thermal tensor networks~\cite{Chen.b+:2017:SETTN}, i.e.,
\begin{equation}
\rho({\tau_0}) = e^{-\tau_0 \hat{H}} \simeq 
\sum^{N_c}_{n=0}\frac{(-\tau_0)^n}{n!}\hat{H}^n,
\end{equation} 
where $N_c$ is the expansion order (often smaller than 10 in practice) 
and the $\rho({\tau_0})$ could converge to machine precision. Given the 
density matrix $\rho({\tau_0})$ represented in the form of a matrix product
operator (MPO), we keep squaring the MPO via tensor network contractions 
and thus cool down the system exponentially
as $\rho({\tau_n}) = \rho({\tau_{n-1}}) \cdot \rho({\tau_{n-1}})$
where $\tau_n = 2^n\tau_0$ ($n \geq 1$).
Based on this, various thermodynamic properties can be computed, including 
free energy $f$, internal energy $u$, magnetic thermal entropy {$S$}, and 
static spin-structure factors $S(\textbf{k})$, etc. We parallel perform the 
simulations with interleaved data points along the temperature axis,
and interpolate data between those sampling points.

 % ============ FigA1: MH curves  ============= %
\begin{figure}[h!]
\includegraphics[angle=0,width=0.85\linewidth]{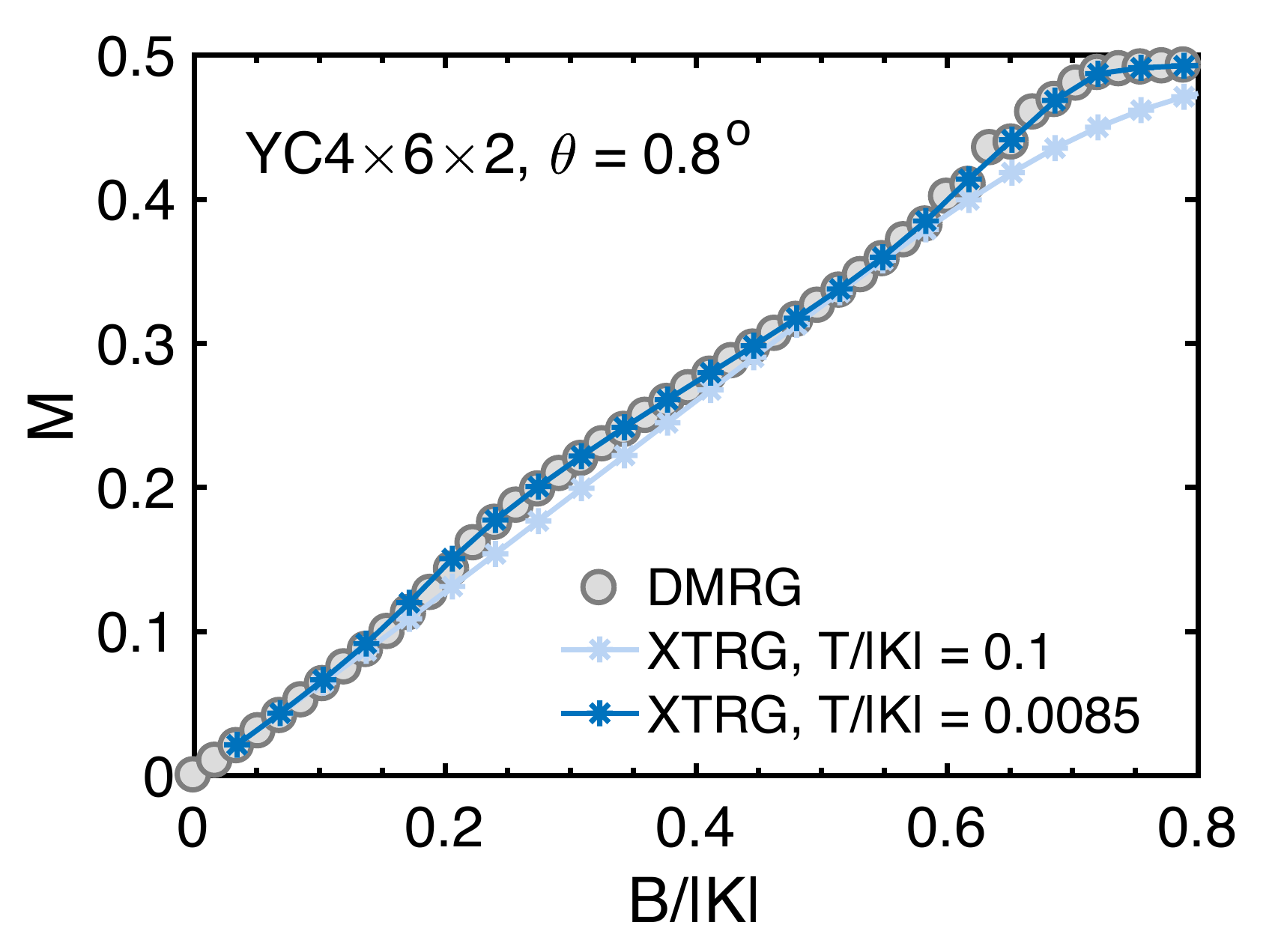}
\renewcommand{\figurename}{\textbf{Fig. }}
\caption{The low-temperature magnetization curves with field 
applied along $\theta=0.8^{\circ}$ of the $\alpha$-RuCl$_3$ 
model. When the temperature is sufficiently low, the results 
converge to the ground state curve computed with DMRG
\cite{Zhou2022arXiv} on the same YC$4\times6\times2$ geometry.
}
\label{Fig:MH}
\end{figure}

For the YC$4\times6\times2$ geometry considered in the main text,
we compare in Fig.~\ref{Fig:MH} the low-temperature magnetization 
curves ($\theta=0.8^{\circ}$ case) calculated by XTRG method with the DMRG results~\cite{Zhou2022arXiv}, where we find $T/|K|=0.0085$ data 
converges well with the DMRG data. This confirms that the XTRG
calculations can approach the low-temperature regime in the close
vicinity of the ground state.

% ============= Appendix.C ============== %
\section{Matrix product operator entanglement}
\label{Sec:SE}

 % ========= FigA2: MPO Entanglement  ========= %
\begin{figure}[h!]
\includegraphics[angle=0,width=0.85\linewidth]{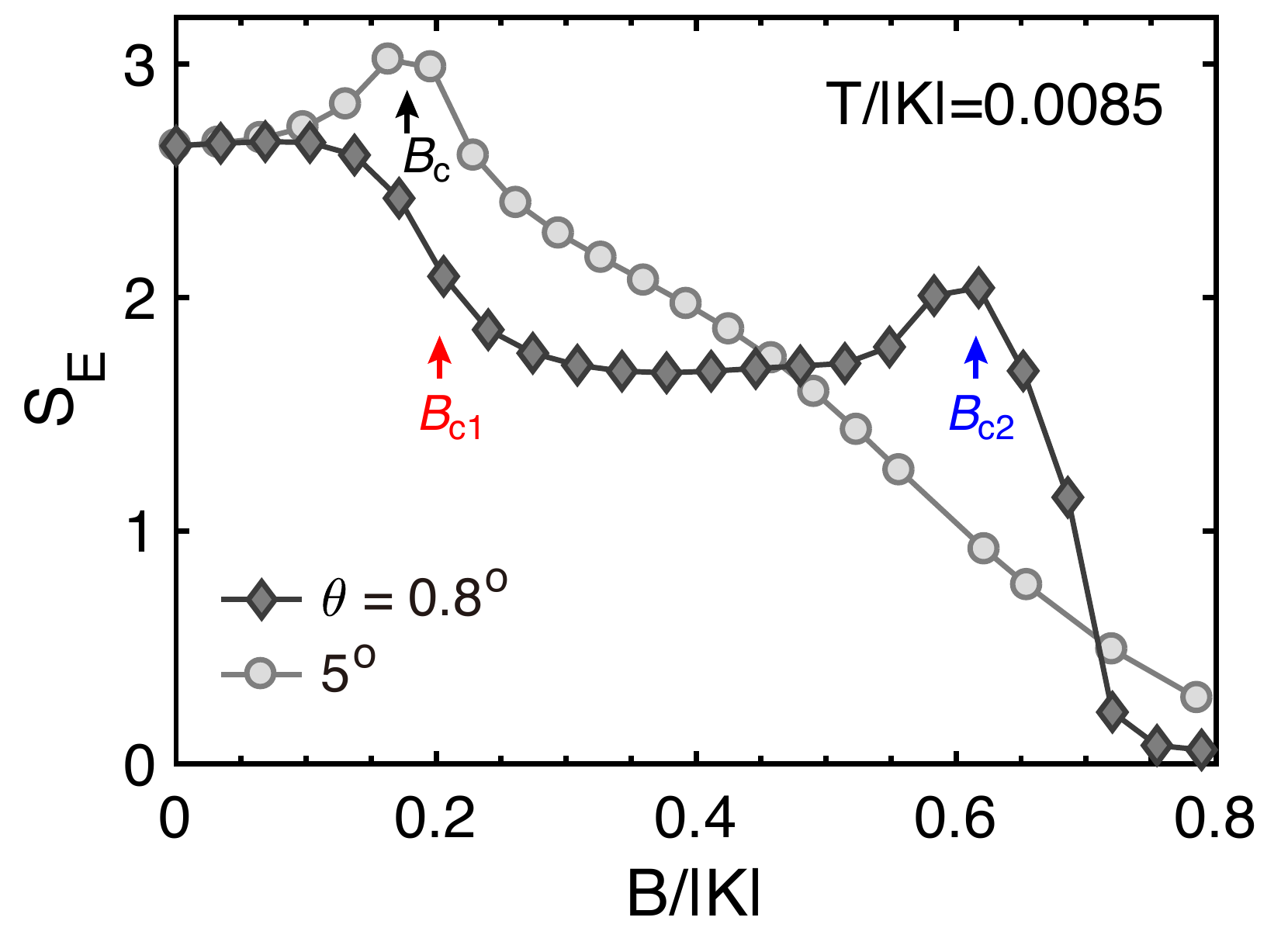}
\renewcommand{\figurename}{\textbf{Fig. }}
\caption{The bipartite MPO entanglement entropies $S_{\rm E}$ 
under various fields $B$ applied along $\theta=0.8^{\circ}$ and 
$\theta=5^{\circ}$, at a low temperature $T/|K| =0.0085$. A drop 
and a peak features can be seen at respectively the low and high 
transition fields $B_{c1}$ and $B_{c2}$ for $\theta=0.8^{\circ}$ 
(denoted by the red and blue arrows). For the $\theta=5^{\circ}$ 
case, there is only a single peak in the $S_{\rm E}$ curve located at $B_c$.
}
\label{Fig:Entanglement}
\end{figure}
% ========================================= %

The phase transitions can be detected sensitively by the entanglement 
entropy of the matrix product operator (MPO)~\cite{Chen2018}. 
Regarding the MPO as a supervector, we can take a Schmidt 
decomposition of the purified ``wavefunction'' and compute the 
entanglement entropy $S_{\rm E}$ between the two parts of the system.
Here we study its field-dependent behaviors for $\theta = 0.8^{\circ}$ 
and $5^{\circ}$, on the YC$4\times 6\times2$ lattice with the calculated 
results shown in Fig.~\ref{Fig:Entanglement}.

The MPO entanglement entropy $S_{\rm E}$ is expected to diverge at the 
second-order quantum critical point (QCP) in the low temperature limit. 
In finite-size calculations, it instead exhibits a peak near the QCP. In 
Fig.~\ref{Fig:Entanglement}, we show the low-temperature $S_{\rm E}$ 
vs. magnetic fields $B$, and find for $\theta=0.8^{\circ}$ case there exists 
a peak near $B_{c2}$. This is indicated by the blue arrow, with the 
determined field value consistent with that obtained from $\Gamma_B$ 
data in Fig.~2(e). In addition, it can be seen that $S_{\rm E}$ firstly shows 
an almost steady behavior at the low-field antiferromagnetic phase, then 
drops abruptly near $B_{c1}$ as indicated by the red arrow, reflecting that 
the low-field phase transition is likely of first order. For $\theta=5^{\circ}$ 
case, there is only a single peak for $S_{\rm E}$ vs. $B$ as shown in 
Fig.~\ref{Fig:Entanglement}, where a prominent peak at $B_c$ clearly 
signals the QCP between the zigzag and spin polarized phases.

% ======= Reference ======== %
\AtEndEnvironment{thebibliography}{
}
\bibliography{kitaevRef}

\begin{thebibliography}{92}%
\makeatletter
\providecommand \@ifxundefined [1]{%
 \@ifx{#1\undefined}
}%
\providecommand \@ifnum [1]{%
 \ifnum #1\expandafter \@firstoftwo
 \else \expandafter \@secondoftwo
 \fi
}%
\providecommand \@ifx [1]{%
 \ifx #1\expandafter \@firstoftwo
 \else \expandafter \@secondoftwo
 \fi
}%
\providecommand \natexlab [1]{#1}%
\providecommand \enquote  [1]{``#1''}%
\providecommand \bibnamefont  [1]{#1}%
\providecommand \bibfnamefont [1]{#1}%
\providecommand \citenamefont [1]{#1}%
\providecommand \href@noop [0]{\@secondoftwo}%
\providecommand \href [0]{\begingroup \@sanitize@url \@href}%
\providecommand \@href[1]{\@@startlink{#1}\@@href}%
\providecommand \@@href[1]{\endgroup#1\@@endlink}%
\providecommand \@sanitize@url [0]{\catcode `\\12\catcode `\$12\catcode
  `\&12\catcode `\#12\catcode `\^12\catcode `\_12\catcode `\%12\relax}%
\providecommand \@@startlink[1]{}%
\providecommand \@@endlink[0]{}%
\providecommand \url  [0]{\begingroup\@sanitize@url \@url }%
\providecommand \@url [1]{\endgroup\@href {#1}{\urlprefix }}%
\providecommand \urlprefix  [0]{URL }%
\providecommand \Eprint [0]{\href }%
\providecommand \doibase [0]{http://dx.doi.org/}%
\providecommand \selectlanguage [0]{\@gobble}%
\providecommand \bibinfo  [0]{\@secondoftwo}%
\providecommand \bibfield  [0]{\@secondoftwo}%
\providecommand \translation [1]{[#1]}%
\providecommand \BibitemOpen [0]{}%
\providecommand \bibitemStop [0]{}%
\providecommand \bibitemNoStop [0]{.\EOS\space}%
\providecommand \EOS [0]{\spacefactor3000\relax}%
\providecommand \BibitemShut  [1]{\csname bibitem#1\endcsname}%
\let\auto@bib@innerbib\@empty
%</preamble>
\bibitem [{\citenamefont {{Li}}\ \emph {et~al.}(2021)\citenamefont {{Li}},
  \citenamefont {{Zhang}}, \citenamefont {{Wang}}, \citenamefont {{Wu}},
  \citenamefont {{Gao}}, \citenamefont {{Qu}}, \citenamefont {{Liu}},
  \citenamefont {{Gong}},\ and\ \citenamefont {{Li}}}]{Han2021}%
  \BibitemOpen
  \bibfield  {author} {\bibinfo {author} {\bibfnamefont {Han}\ \bibnamefont
  {{Li}}}, \bibinfo {author} {\bibfnamefont {Hao-Kai}\ \bibnamefont {{Zhang}}},
  \bibinfo {author} {\bibfnamefont {Jiucai}\ \bibnamefont {{Wang}}}, \bibinfo
  {author} {\bibfnamefont {Han-Qing}\ \bibnamefont {{Wu}}}, \bibinfo {author}
  {\bibfnamefont {Yuan}\ \bibnamefont {{Gao}}}, \bibinfo {author}
  {\bibfnamefont {Dai-Wei}\ \bibnamefont {{Qu}}}, \bibinfo {author}
  {\bibfnamefont {Zheng-Xin}\ \bibnamefont {{Liu}}}, \bibinfo {author}
  {\bibfnamefont {Shou-Shu}\ \bibnamefont {{Gong}}}, \ and\ \bibinfo {author}
  {\bibfnamefont {Wei}\ \bibnamefont {{Li}}},\ }\bibfield  {title} {\enquote
  {\bibinfo {title} {{Identification of magnetic interactions and high-field
  quantum spin liquid in {\ensuremath{\alpha}}-RuCl$_{3}$}},}\ }\href {\doibase
  10.1038/s41467-021-24257-8} {\bibfield  {journal} {\bibinfo  {journal} {Nat.
  Commun.}\ }\textbf {\bibinfo {volume} {12}},\ \bibinfo {eid} {4007} (\bibinfo
  {year} {2021})}\BibitemShut {NoStop}%
\bibitem [{\citenamefont {Anderson}(1973)}]{Anderson1973}%
  \BibitemOpen
  \bibfield  {author} {\bibinfo {author} {\bibfnamefont {P.~W.}\ \bibnamefont
  {Anderson}},\ }\bibfield  {title} {\enquote {\bibinfo {title} {Resonating
  valence bonds: A new kind of insulator?}}\ }\href {\doibase
  https://doi.org/10.1016/0025-5408(73)90167-0} {\bibfield  {journal} {\bibinfo
   {journal} {Mater. Res. Bull.}\ }\textbf {\bibinfo {volume} {8}},\ \bibinfo
  {pages} {153 -- 160} (\bibinfo {year} {1973})}\BibitemShut {NoStop}%
\bibitem [{\citenamefont {{Balents}}(2010)}]{Balents2010}%
  \BibitemOpen
  \bibfield  {author} {\bibinfo {author} {\bibfnamefont {L.}~\bibnamefont
  {{Balents}}},\ }\bibfield  {title} {\enquote {\bibinfo {title} {{Spin liquids
  in frustrated magnets}},}\ }\href {\doibase 10.1038/nature08917} {\bibfield
  {journal} {\bibinfo  {journal} {\nat}\ }\textbf {\bibinfo {volume} {464}},\
  \bibinfo {pages} {199--208} (\bibinfo {year} {2010})}\BibitemShut {NoStop}%
\bibitem [{\citenamefont {Zhou}\ \emph {et~al.}(2017)\citenamefont {Zhou},
  \citenamefont {Kanoda},\ and\ \citenamefont {Ng}}]{Zhou2017}%
  \BibitemOpen
  \bibfield  {author} {\bibinfo {author} {\bibfnamefont {Y.}~\bibnamefont
  {Zhou}}, \bibinfo {author} {\bibfnamefont {K.}~\bibnamefont {Kanoda}}, \ and\
  \bibinfo {author} {\bibfnamefont {T.-K.}\ \bibnamefont {Ng}},\ }\bibfield
  {title} {\enquote {\bibinfo {title} {Quantum spin liquid states},}\ }\href
  {\doibase 10.1103/RevModPhys.89.025003} {\bibfield  {journal} {\bibinfo
  {journal} {Rev. Mod. Phys.}\ }\textbf {\bibinfo {volume} {89}},\ \bibinfo
  {pages} {025003} (\bibinfo {year} {2017})}\BibitemShut {NoStop}%
\bibitem [{\citenamefont {Wen}\ \emph {et~al.}(2019)\citenamefont {Wen},
  \citenamefont {Yu}, \citenamefont {Li}, \citenamefont {Yu},\ and\
  \citenamefont {Li}}]{JW2019QMats}%
  \BibitemOpen
  \bibfield  {author} {\bibinfo {author} {\bibfnamefont {Jinsheng}\
  \bibnamefont {Wen}}, \bibinfo {author} {\bibfnamefont {Shun-Li}\ \bibnamefont
  {Yu}}, \bibinfo {author} {\bibfnamefont {Shiyan}\ \bibnamefont {Li}},
  \bibinfo {author} {\bibfnamefont {Weiqiang}\ \bibnamefont {Yu}}, \ and\
  \bibinfo {author} {\bibfnamefont {Jian-Xin}\ \bibnamefont {Li}},\ }\bibfield
  {title} {\enquote {\bibinfo {title} {Experimental identification of quantum
  spin liquids},}\ }\href {\doibase 10.1038/s41535-019-0151-6} {\bibfield
  {journal} {\bibinfo  {journal} {npj Quantum Materials}\ }\textbf {\bibinfo
  {volume} {4}},\ \bibinfo {pages} {12} (\bibinfo {year} {2019})}\BibitemShut
  {NoStop}%
\bibitem [{\citenamefont {Broholm}\ \emph {et~al.}(2020)\citenamefont
  {Broholm}, \citenamefont {Cava}, \citenamefont {Kivelson}, \citenamefont
  {Nocera}, \citenamefont {Norman},\ and\ \citenamefont
  {Senthil}}]{Broholm2020Science}%
  \BibitemOpen
  \bibfield  {author} {\bibinfo {author} {\bibfnamefont {C.}~\bibnamefont
  {Broholm}}, \bibinfo {author} {\bibfnamefont {R.~J.}\ \bibnamefont {Cava}},
  \bibinfo {author} {\bibfnamefont {S.~A.}\ \bibnamefont {Kivelson}}, \bibinfo
  {author} {\bibfnamefont {D.~G.}\ \bibnamefont {Nocera}}, \bibinfo {author}
  {\bibfnamefont {M.~R.}\ \bibnamefont {Norman}}, \ and\ \bibinfo {author}
  {\bibfnamefont {T.}~\bibnamefont {Senthil}},\ }\bibfield  {title} {\enquote
  {\bibinfo {title} {Quantum spin liquids},}\ }\href {\doibase
  10.1126/science.aay0668} {\bibfield  {journal} {\bibinfo  {journal}
  {Science}\ }\textbf {\bibinfo {volume} {367}},\ \bibinfo {pages} {eaay0668}
  (\bibinfo {year} {2020})}\BibitemShut {NoStop}%
\bibitem [{\citenamefont {{Kitaev}}(2003)}]{Kitaev2003}%
  \BibitemOpen
  \bibfield  {author} {\bibinfo {author} {\bibfnamefont {A.}~\bibnamefont
  {{Kitaev}}},\ }\bibfield  {title} {\enquote {\bibinfo {title}
  {{Fault-tolerant quantum computation by anyons}},}\ }\href {\doibase
  10.1016/S0003-4916(02)00018-0} {\bibfield  {journal} {\bibinfo  {journal}
  {Ann. Phys.}\ }\textbf {\bibinfo {volume} {303}},\ \bibinfo {pages} {2--30}
  (\bibinfo {year} {2003})}\BibitemShut {NoStop}%
\bibitem [{\citenamefont {Kitaev}(2006)}]{Kitaev2006}%
  \BibitemOpen
  \bibfield  {author} {\bibinfo {author} {\bibfnamefont {A.}~\bibnamefont
  {Kitaev}},\ }\bibfield  {title} {\enquote {\bibinfo {title} {Anyons in an
  exactly solved model and beyond},}\ }\href {\doibase
  https://doi.org/10.1016/j.aop.2005.10.005} {\bibfield  {journal} {\bibinfo
  {journal} {Ann. Phys.}\ }\textbf {\bibinfo {volume} {321}},\ \bibinfo {pages}
  {2 -- 111} (\bibinfo {year} {2006})},\ \bibinfo {note} {january Special
  Issue}\BibitemShut {NoStop}%
\bibitem [{\citenamefont {Jackeli}\ and\ \citenamefont
  {Khaliullin}(2009)}]{Jackeli2009}%
  \BibitemOpen
  \bibfield  {author} {\bibinfo {author} {\bibfnamefont {G.}~\bibnamefont
  {Jackeli}}\ and\ \bibinfo {author} {\bibfnamefont {G.}~\bibnamefont
  {Khaliullin}},\ }\bibfield  {title} {\enquote {\bibinfo {title} {Mott
  insulators in the strong spin-orbit coupling limit: From {Heisenberg} to a
  quantum compass and {Kitaev} models},}\ }\href {\doibase
  10.1103/PhysRevLett.102.017205} {\bibfield  {journal} {\bibinfo  {journal}
  {Phys. Rev. Lett.}\ }\textbf {\bibinfo {volume} {102}},\ \bibinfo {pages}
  {017205} (\bibinfo {year} {2009})}\BibitemShut {NoStop}%
\bibitem [{\citenamefont {{Trebst}}()}]{Trebst2017arXiv}%
  \BibitemOpen
  \bibfield  {author} {\bibinfo {author} {\bibfnamefont {S.}~\bibnamefont
  {{Trebst}}},\ }\bibfield  {title} {\enquote {\bibinfo {title} {Kitaev
  materials},}\ }\href@noop {} {\ }\Eprint {http://arxiv.org/abs/1701.07056
  (2017)} {arXiv:1701.07056 (2017)} \BibitemShut {NoStop}%
\bibitem [{\citenamefont {Winter}\ \emph {et~al.}(2017)\citenamefont {Winter},
  \citenamefont {Tsirlin}, \citenamefont {Daghofer}, \citenamefont {van~den
  Brink}, \citenamefont {Singh}, \citenamefont {Gegenwart},\ and\ \citenamefont
  {Valent{\'{\i}}}}]{Winter2017}%
  \BibitemOpen
  \bibfield  {author} {\bibinfo {author} {\bibfnamefont {S.~M}\ \bibnamefont
  {Winter}}, \bibinfo {author} {\bibfnamefont {A.~A}\ \bibnamefont {Tsirlin}},
  \bibinfo {author} {\bibfnamefont {M.}~\bibnamefont {Daghofer}}, \bibinfo
  {author} {\bibfnamefont {J.}~\bibnamefont {van~den Brink}}, \bibinfo {author}
  {\bibfnamefont {Y.}~\bibnamefont {Singh}}, \bibinfo {author} {\bibfnamefont
  {P.}~\bibnamefont {Gegenwart}}, \ and\ \bibinfo {author} {\bibfnamefont
  {R.}~\bibnamefont {Valent{\'{\i}}}},\ }\bibfield  {title} {\enquote {\bibinfo
  {title} {Models and materials for generalized {Kitaev} magnetism},}\ }\href
  {\doibase 10.1088/1361-648x/aa8cf5} {\bibfield  {journal} {\bibinfo
  {journal} {J. Phys.: Condens. Matter}\ }\textbf {\bibinfo {volume} {29}},\
  \bibinfo {pages} {493002} (\bibinfo {year} {2017})}\BibitemShut {NoStop}%
\bibitem [{\citenamefont {Janssen}\ and\ \citenamefont
  {Vojta}(2019)}]{Janssen2019}%
  \BibitemOpen
  \bibfield  {author} {\bibinfo {author} {\bibfnamefont {L.}~\bibnamefont
  {Janssen}}\ and\ \bibinfo {author} {\bibfnamefont {M.}~\bibnamefont
  {Vojta}},\ }\bibfield  {title} {\enquote {\bibinfo {title}
  {{Heisenberg{\textendash}Kitaev} physics in magnetic fields},}\ }\href
  {\doibase 10.1088/1361-648x/ab283e} {\bibfield  {journal} {\bibinfo
  {journal} {J. Phys.: Condens. Matter}\ }\textbf {\bibinfo {volume} {31}},\
  \bibinfo {pages} {423002} (\bibinfo {year} {2019})}\BibitemShut {NoStop}%
\bibitem [{\citenamefont {{Motome}}\ \emph {et~al.}(2020)\citenamefont
  {{Motome}}, \citenamefont {{Sano}}, \citenamefont {{Jang}}, \citenamefont
  {{Sugita}},\ and\ \citenamefont {{Kato}}}]{Motome2020b}%
  \BibitemOpen
  \bibfield  {author} {\bibinfo {author} {\bibfnamefont {Yukitoshi}\
  \bibnamefont {{Motome}}}, \bibinfo {author} {\bibfnamefont {Ryoya}\
  \bibnamefont {{Sano}}}, \bibinfo {author} {\bibfnamefont {Seonghoon}\
  \bibnamefont {{Jang}}}, \bibinfo {author} {\bibfnamefont {Yusuke}\
  \bibnamefont {{Sugita}}}, \ and\ \bibinfo {author} {\bibfnamefont {Yasuyuki}\
  \bibnamefont {{Kato}}},\ }\bibfield  {title} {\enquote {\bibinfo {title}
  {{Materials design of Kitaev spin liquids beyond the Jackeli-Khaliullin
  mechanism}},}\ }\href {\doibase 10.1088/1361-648X/ab8525} {\bibfield
  {journal} {\bibinfo  {journal} {J. Phys.: Condens. Matter.}\ }\textbf
  {\bibinfo {volume} {32}},\ \bibinfo {eid} {404001} (\bibinfo {year}
  {2020})}\BibitemShut {NoStop}%
\bibitem [{\citenamefont {Sears}\ \emph {et~al.}(2015)\citenamefont {Sears},
  \citenamefont {Songvilay}, \citenamefont {Plumb}, \citenamefont {Clancy},
  \citenamefont {Qiu}, \citenamefont {Zhao}, \citenamefont {Parshall},\ and\
  \citenamefont {Kim}}]{Sears2015}%
  \BibitemOpen
  \bibfield  {author} {\bibinfo {author} {\bibfnamefont {J.~A.}\ \bibnamefont
  {Sears}}, \bibinfo {author} {\bibfnamefont {M.}~\bibnamefont {Songvilay}},
  \bibinfo {author} {\bibfnamefont {K.~W.}\ \bibnamefont {Plumb}}, \bibinfo
  {author} {\bibfnamefont {J.~P.}\ \bibnamefont {Clancy}}, \bibinfo {author}
  {\bibfnamefont {Y.}~\bibnamefont {Qiu}}, \bibinfo {author} {\bibfnamefont
  {Y.}~\bibnamefont {Zhao}}, \bibinfo {author} {\bibfnamefont {D.}~\bibnamefont
  {Parshall}}, \ and\ \bibinfo {author} {\bibfnamefont {Y.-J.}\ \bibnamefont
  {Kim}},\ }\bibfield  {title} {\enquote {\bibinfo {title} {Magnetic order in
  ${\alpha}$-{RuCl}$_{3}$: A honeycomb-lattice quantum magnet with strong
  spin-orbit coupling},}\ }\href {\doibase 10.1103/PhysRevB.91.144420}
  {\bibfield  {journal} {\bibinfo  {journal} {Phys. Rev. B}\ }\textbf {\bibinfo
  {volume} {91}},\ \bibinfo {pages} {144420} (\bibinfo {year}
  {2015})}\BibitemShut {NoStop}%
\bibitem [{\citenamefont {{Banerjee}}\ \emph {et~al.}(2017)\citenamefont
  {{Banerjee}}, \citenamefont {{Yan}}, \citenamefont {{Knolle}}, \citenamefont
  {{Bridges}}, \citenamefont {{Stone}}, \citenamefont {{Lumsden}},
  \citenamefont {{Mandrus}}, \citenamefont {{Tennant}}, \citenamefont
  {{Moessner}},\ and\ \citenamefont {{Nagler}}}]{Banerjee2017}%
  \BibitemOpen
  \bibfield  {author} {\bibinfo {author} {\bibfnamefont {A.}~\bibnamefont
  {{Banerjee}}}, \bibinfo {author} {\bibfnamefont {J.}~\bibnamefont {{Yan}}},
  \bibinfo {author} {\bibfnamefont {J.}~\bibnamefont {{Knolle}}}, \bibinfo
  {author} {\bibfnamefont {C.~A.}\ \bibnamefont {{Bridges}}}, \bibinfo {author}
  {\bibfnamefont {M.~B.}\ \bibnamefont {{Stone}}}, \bibinfo {author}
  {\bibfnamefont {M.~D.}\ \bibnamefont {{Lumsden}}}, \bibinfo {author}
  {\bibfnamefont {D.~G.}\ \bibnamefont {{Mandrus}}}, \bibinfo {author}
  {\bibfnamefont {D.~A.}\ \bibnamefont {{Tennant}}}, \bibinfo {author}
  {\bibfnamefont {R.}~\bibnamefont {{Moessner}}}, \ and\ \bibinfo {author}
  {\bibfnamefont {S.~E.}\ \bibnamefont {{Nagler}}},\ }\bibfield  {title}
  {\enquote {\bibinfo {title} {{Neutron scattering in the proximate quantum
  spin liquid {\ensuremath{\alpha}}-RuCl$_{3}$}},}\ }\href {\doibase
  10.1126/science.aah6015} {\bibfield  {journal} {\bibinfo  {journal}
  {Science}\ }\textbf {\bibinfo {volume} {356}},\ \bibinfo {pages} {1055--1059}
  (\bibinfo {year} {2017})}\BibitemShut {NoStop}%
\bibitem [{\citenamefont {Do}\ \emph {et~al.}(2017)\citenamefont {Do},
  \citenamefont {Park}, \citenamefont {Yoshitake}, \citenamefont {Nasu},
  \citenamefont {Motome}, \citenamefont {Kwon}, \citenamefont {Adroja},
  \citenamefont {Voneshen}, \citenamefont {Kim}, \citenamefont {Jang},
  \citenamefont {Park}, \citenamefont {Choi},\ and\ \citenamefont
  {Ji}}]{Do2017}%
  \BibitemOpen
  \bibfield  {author} {\bibinfo {author} {\bibfnamefont {S.-H.}\ \bibnamefont
  {Do}}, \bibinfo {author} {\bibfnamefont {S.-Y.}\ \bibnamefont {Park}},
  \bibinfo {author} {\bibfnamefont {J.}~\bibnamefont {Yoshitake}}, \bibinfo
  {author} {\bibfnamefont {J.}~\bibnamefont {Nasu}}, \bibinfo {author}
  {\bibfnamefont {Y.}~\bibnamefont {Motome}}, \bibinfo {author} {\bibfnamefont
  {Y.~S.}\ \bibnamefont {Kwon}}, \bibinfo {author} {\bibfnamefont {D.~T.}\
  \bibnamefont {Adroja}}, \bibinfo {author} {\bibfnamefont {D.~J.}\
  \bibnamefont {Voneshen}}, \bibinfo {author} {\bibfnamefont {K.}~\bibnamefont
  {Kim}}, \bibinfo {author} {\bibfnamefont {T.-H.}\ \bibnamefont {Jang}},
  \bibinfo {author} {\bibfnamefont {J.-H.}\ \bibnamefont {Park}}, \bibinfo
  {author} {\bibfnamefont {K.-Y.}\ \bibnamefont {Choi}}, \ and\ \bibinfo
  {author} {\bibfnamefont {S.}~\bibnamefont {Ji}},\ }\bibfield  {title}
  {\enquote {\bibinfo {title} {{Majorana fermions in the Kitaev quantum spin
  system $\alpha$-RuCl$_3$}},}\ }\href {\doibase 10.1038/nphys4264} {\bibfield
  {journal} {\bibinfo  {journal} {Nat. Phys.}\ }\textbf {\bibinfo {volume}
  {13}},\ \bibinfo {pages} {1079} (\bibinfo {year} {2017})}\BibitemShut
  {NoStop}%
\bibitem [{\citenamefont {{Banerjee}}\ \emph {et~al.}(2018)\citenamefont
  {{Banerjee}}, \citenamefont {{Lampen-Kelley}}, \citenamefont {{Knolle}},
  \citenamefont {{Balz}}, \citenamefont {{Aczel}}, \citenamefont {{Winn}},
  \citenamefont {{Liu}}, \citenamefont {{Pajerowski}}, \citenamefont {{Yan}},
  \citenamefont {{Bridges}}, \citenamefont {{Savici}}, \citenamefont
  {{Chakoumakos}}, \citenamefont {{Lumsden}}, \citenamefont {{Tennant}},
  \citenamefont {{Moessner}}, \citenamefont {{Mandrus}},\ and\ \citenamefont
  {{Nagler}}}]{Banerjee2018}%
  \BibitemOpen
  \bibfield  {author} {\bibinfo {author} {\bibfnamefont {A.}~\bibnamefont
  {{Banerjee}}}, \bibinfo {author} {\bibfnamefont {P.}~\bibnamefont
  {{Lampen-Kelley}}}, \bibinfo {author} {\bibfnamefont {J.}~\bibnamefont
  {{Knolle}}}, \bibinfo {author} {\bibfnamefont {C.}~\bibnamefont {{Balz}}},
  \bibinfo {author} {\bibfnamefont {A.}~\bibnamefont {{Aczel}}}, \bibinfo
  {author} {\bibfnamefont {B.}~\bibnamefont {{Winn}}}, \bibinfo {author}
  {\bibfnamefont {Y.}~\bibnamefont {{Liu}}}, \bibinfo {author} {\bibfnamefont
  {D.}~\bibnamefont {{Pajerowski}}}, \bibinfo {author} {\bibfnamefont
  {J.}~\bibnamefont {{Yan}}}, \bibinfo {author} {\bibfnamefont {C.~A.}\
  \bibnamefont {{Bridges}}}, \bibinfo {author} {\bibfnamefont {A.~T.}\
  \bibnamefont {{Savici}}}, \bibinfo {author} {\bibfnamefont {B.~C.}\
  \bibnamefont {{Chakoumakos}}}, \bibinfo {author} {\bibfnamefont {M.~D.}\
  \bibnamefont {{Lumsden}}}, \bibinfo {author} {\bibfnamefont {D.~A.}\
  \bibnamefont {{Tennant}}}, \bibinfo {author} {\bibfnamefont {R.}~\bibnamefont
  {{Moessner}}}, \bibinfo {author} {\bibfnamefont {D.~G.}\ \bibnamefont
  {{Mandrus}}}, \ and\ \bibinfo {author} {\bibfnamefont {S.~E.}\ \bibnamefont
  {{Nagler}}},\ }\bibfield  {title} {\enquote {\bibinfo {title} {{Excitations
  in the field-induced quantum spin liquid state of
  {\ensuremath{\alpha}}-{RuCl$_{3}$}}},}\ }\href {\doibase
  10.1038/s41535-018-0079-2} {\bibfield  {journal} {\bibinfo  {journal} {npj
  Quant. Mater.}\ }\textbf {\bibinfo {volume} {3}},\ \bibinfo {eid} {8}
  (\bibinfo {year} {2018})}\BibitemShut {NoStop}%
\bibitem [{\citenamefont {Balz}\ \emph {et~al.}(2021)\citenamefont {Balz},
  \citenamefont {Janssen}, \citenamefont {Lampen-Kelley}, \citenamefont
  {Banerjee}, \citenamefont {Liu}, \citenamefont {Yan}, \citenamefont
  {Mandrus}, \citenamefont {Vojta},\ and\ \citenamefont {Nagler}}]{Balz2021}%
  \BibitemOpen
  \bibfield  {author} {\bibinfo {author} {\bibfnamefont {C.}~\bibnamefont
  {Balz}}, \bibinfo {author} {\bibfnamefont {L.}~\bibnamefont {Janssen}},
  \bibinfo {author} {\bibfnamefont {P.}~\bibnamefont {Lampen-Kelley}}, \bibinfo
  {author} {\bibfnamefont {A.}~\bibnamefont {Banerjee}}, \bibinfo {author}
  {\bibfnamefont {Y.~H.}\ \bibnamefont {Liu}}, \bibinfo {author} {\bibfnamefont
  {J.-Q.}\ \bibnamefont {Yan}}, \bibinfo {author} {\bibfnamefont {D.~G.}\
  \bibnamefont {Mandrus}}, \bibinfo {author} {\bibfnamefont {M.}~\bibnamefont
  {Vojta}}, \ and\ \bibinfo {author} {\bibfnamefont {S.~E.}\ \bibnamefont
  {Nagler}},\ }\bibfield  {title} {\enquote {\bibinfo {title} {Field-induced
  intermediate ordered phase and anisotropic interlayer interactions in
  $\alpha$-{RuCl}$_3$},}\ }\href {\doibase 10.1103/PhysRevB.103.174417}
  {\bibfield  {journal} {\bibinfo  {journal} {Phys. Rev. B}\ }\textbf {\bibinfo
  {volume} {103}},\ \bibinfo {pages} {174417} (\bibinfo {year}
  {2021})}\BibitemShut {NoStop}%
\bibitem [{\citenamefont {Sears}\ \emph {et~al.}(2017)\citenamefont {Sears},
  \citenamefont {Zhao}, \citenamefont {Xu}, \citenamefont {Lynn},\ and\
  \citenamefont {Kim}}]{Sears2017}%
  \BibitemOpen
  \bibfield  {author} {\bibinfo {author} {\bibfnamefont {J.~A.}\ \bibnamefont
  {Sears}}, \bibinfo {author} {\bibfnamefont {Y.}~\bibnamefont {Zhao}},
  \bibinfo {author} {\bibfnamefont {Z.}~\bibnamefont {Xu}}, \bibinfo {author}
  {\bibfnamefont {J.~W.}\ \bibnamefont {Lynn}}, \ and\ \bibinfo {author}
  {\bibfnamefont {Y.-J.}\ \bibnamefont {Kim}},\ }\bibfield  {title} {\enquote
  {\bibinfo {title} {Phase diagram of {$\alpha$-RuCl$_3$} in an in-plane
  magnetic field},}\ }\href {\doibase 10.1103/PhysRevB.95.180411} {\bibfield
  {journal} {\bibinfo  {journal} {Phys. Rev. B}\ }\textbf {\bibinfo {volume}
  {95}},\ \bibinfo {pages} {180411(R)} (\bibinfo {year} {2017})}\BibitemShut
  {NoStop}%
\bibitem [{\citenamefont {Zheng}\ \emph {et~al.}(2017)\citenamefont {Zheng},
  \citenamefont {Ran}, \citenamefont {Li}, \citenamefont {Wang}, \citenamefont
  {Wang}, \citenamefont {Liu}, \citenamefont {Liu}, \citenamefont {Normand},
  \citenamefont {Wen},\ and\ \citenamefont {Yu}}]{Zheng2017}%
  \BibitemOpen
  \bibfield  {author} {\bibinfo {author} {\bibfnamefont {J.}~\bibnamefont
  {Zheng}}, \bibinfo {author} {\bibfnamefont {K.}~\bibnamefont {Ran}}, \bibinfo
  {author} {\bibfnamefont {T.}~\bibnamefont {Li}}, \bibinfo {author}
  {\bibfnamefont {J.}~\bibnamefont {Wang}}, \bibinfo {author} {\bibfnamefont
  {P.}~\bibnamefont {Wang}}, \bibinfo {author} {\bibfnamefont {B.}~\bibnamefont
  {Liu}}, \bibinfo {author} {\bibfnamefont {Z.-X.}\ \bibnamefont {Liu}},
  \bibinfo {author} {\bibfnamefont {B.}~\bibnamefont {Normand}}, \bibinfo
  {author} {\bibfnamefont {J.}~\bibnamefont {Wen}}, \ and\ \bibinfo {author}
  {\bibfnamefont {W.}~\bibnamefont {Yu}},\ }\bibfield  {title} {\enquote
  {\bibinfo {title} {Gapless spin excitations in the field-induced quantum spin
  liquid phase of ${\alpha}$-{RuCl}$_{3}$},}\ }\href {\doibase
  10.1103/PhysRevLett.119.227208} {\bibfield  {journal} {\bibinfo  {journal}
  {Phys. Rev. Lett.}\ }\textbf {\bibinfo {volume} {119}},\ \bibinfo {pages}
  {227208} (\bibinfo {year} {2017})}\BibitemShut {NoStop}%
\bibitem [{\citenamefont {Baek}\ \emph {et~al.}(2017)\citenamefont {Baek},
  \citenamefont {Do}, \citenamefont {Choi}, \citenamefont {Kwon}, \citenamefont
  {Wolter}, \citenamefont {Nishimoto}, \citenamefont {van~den Brink},\ and\
  \citenamefont {B\"uchner}}]{Baek2017}%
  \BibitemOpen
  \bibfield  {author} {\bibinfo {author} {\bibfnamefont {S.-H.}\ \bibnamefont
  {Baek}}, \bibinfo {author} {\bibfnamefont {S.-H.}\ \bibnamefont {Do}},
  \bibinfo {author} {\bibfnamefont {K.-Y.}\ \bibnamefont {Choi}}, \bibinfo
  {author} {\bibfnamefont {Y.~S.}\ \bibnamefont {Kwon}}, \bibinfo {author}
  {\bibfnamefont {A.~U.~B.}\ \bibnamefont {Wolter}}, \bibinfo {author}
  {\bibfnamefont {S.}~\bibnamefont {Nishimoto}}, \bibinfo {author}
  {\bibfnamefont {Jeroen}\ \bibnamefont {van~den Brink}}, \ and\ \bibinfo
  {author} {\bibfnamefont {B.}~\bibnamefont {B\"uchner}},\ }\bibfield  {title}
  {\enquote {\bibinfo {title} {Evidence for a field-induced quantum spin liquid
  in {$\alpha$-RuCl$_3$}},}\ }\href {\doibase 10.1103/PhysRevLett.119.037201}
  {\bibfield  {journal} {\bibinfo  {journal} {Phys. Rev. Lett.}\ }\textbf
  {\bibinfo {volume} {119}},\ \bibinfo {pages} {037201} (\bibinfo {year}
  {2017})}\BibitemShut {NoStop}%
\bibitem [{\citenamefont {{Jan{\v{s}}a}}\ \emph {et~al.}(2018)\citenamefont
  {{Jan{\v{s}}a}}, \citenamefont {{Zorko}}, \citenamefont {{Gomil{\v{s}}ek}},
  \citenamefont {{Pregelj}}, \citenamefont {{Kr{\"a}mer}}, \citenamefont
  {{Biner}}, \citenamefont {{Biffin}}, \citenamefont {{R{\"u}egg}},\ and\
  \citenamefont {{Klanj{\v{s}}ek}}}]{Jansa2018}%
  \BibitemOpen
  \bibfield  {author} {\bibinfo {author} {\bibfnamefont {N.}~\bibnamefont
  {{Jan{\v{s}}a}}}, \bibinfo {author} {\bibfnamefont {A.}~\bibnamefont
  {{Zorko}}}, \bibinfo {author} {\bibfnamefont {M.}~\bibnamefont
  {{Gomil{\v{s}}ek}}}, \bibinfo {author} {\bibfnamefont {M.}~\bibnamefont
  {{Pregelj}}}, \bibinfo {author} {\bibfnamefont {K.~W.}\ \bibnamefont
  {{Kr{\"a}mer}}}, \bibinfo {author} {\bibfnamefont {D.}~\bibnamefont
  {{Biner}}}, \bibinfo {author} {\bibfnamefont {A.}~\bibnamefont {{Biffin}}},
  \bibinfo {author} {\bibfnamefont {C.}~\bibnamefont {{R{\"u}egg}}}, \ and\
  \bibinfo {author} {\bibfnamefont {M.}~\bibnamefont {{Klanj{\v{s}}ek}}},\
  }\bibfield  {title} {\enquote {\bibinfo {title} {{Observation of two types of
  fractional excitation in the Kitaev honeycomb magnet}},}\ }\href {\doibase
  10.1038/s41567-018-0129-5} {\bibfield  {journal} {\bibinfo  {journal} {Nat.
  Phys.}\ }\textbf {\bibinfo {volume} {14}},\ \bibinfo {pages} {786--790}
  (\bibinfo {year} {2018})}\BibitemShut {NoStop}%
\bibitem [{\citenamefont {Wulferding}\ \emph {et~al.}(2020)\citenamefont
  {Wulferding}, \citenamefont {Choi}, \citenamefont {Do}, \citenamefont {Lee},
  \citenamefont {Lemmens}, \citenamefont {Faugeras}, \citenamefont {Gallais},\
  and\ \citenamefont {Choi}}]{Wulferding2020}%
  \BibitemOpen
  \bibfield  {author} {\bibinfo {author} {\bibfnamefont {D.}~\bibnamefont
  {Wulferding}}, \bibinfo {author} {\bibfnamefont {Y.}~\bibnamefont {Choi}},
  \bibinfo {author} {\bibfnamefont {S.-H.}\ \bibnamefont {Do}}, \bibinfo
  {author} {\bibfnamefont {C.~H.}\ \bibnamefont {Lee}}, \bibinfo {author}
  {\bibfnamefont {P.}~\bibnamefont {Lemmens}}, \bibinfo {author} {\bibfnamefont
  {C.}~\bibnamefont {Faugeras}}, \bibinfo {author} {\bibfnamefont
  {Y.}~\bibnamefont {Gallais}}, \ and\ \bibinfo {author} {\bibfnamefont
  {K.-Y.}\ \bibnamefont {Choi}},\ }\bibfield  {title} {\enquote {\bibinfo
  {title} {Magnon bound states versus anyonic {Majorana} excitations in the
  {Kitaev} honeycomb magnet $\alpha$-{RuCl}$_3$},}\ }\href {\doibase
  10.1038/s41467-020-15370-1} {\bibfield  {journal} {\bibinfo  {journal} {Nat.
  Commun.}\ }\textbf {\bibinfo {volume} {11}},\ \bibinfo {pages} {1603}
  (\bibinfo {year} {2020})}\BibitemShut {NoStop}%
\bibitem [{\citenamefont {Ponomaryov}\ \emph {et~al.}(2020)\citenamefont
  {Ponomaryov}, \citenamefont {Zviagina}, \citenamefont {Wosnitza},
  \citenamefont {Lampen-Kelley}, \citenamefont {Banerjee}, \citenamefont {Yan},
  \citenamefont {Bridges}, \citenamefont {Mandrus}, \citenamefont {Nagler},\
  and\ \citenamefont {Zvyagin}}]{Ponomaryov2020}%
  \BibitemOpen
  \bibfield  {author} {\bibinfo {author} {\bibfnamefont {A.~N.}\ \bibnamefont
  {Ponomaryov}}, \bibinfo {author} {\bibfnamefont {L.}~\bibnamefont
  {Zviagina}}, \bibinfo {author} {\bibfnamefont {J.}~\bibnamefont {Wosnitza}},
  \bibinfo {author} {\bibfnamefont {P.}~\bibnamefont {Lampen-Kelley}}, \bibinfo
  {author} {\bibfnamefont {A.}~\bibnamefont {Banerjee}}, \bibinfo {author}
  {\bibfnamefont {J.-Q.}\ \bibnamefont {Yan}}, \bibinfo {author} {\bibfnamefont
  {C.~A.}\ \bibnamefont {Bridges}}, \bibinfo {author} {\bibfnamefont {D.~G.}\
  \bibnamefont {Mandrus}}, \bibinfo {author} {\bibfnamefont {S.~E.}\
  \bibnamefont {Nagler}}, \ and\ \bibinfo {author} {\bibfnamefont {S.~A.}\
  \bibnamefont {Zvyagin}},\ }\bibfield  {title} {\enquote {\bibinfo {title}
  {Nature of magnetic excitations in the high-field phase of
  $\alpha$-{RuCl}$_3$},}\ }\href {\doibase 10.1103/PhysRevLett.125.037202}
  {\bibfield  {journal} {\bibinfo  {journal} {Phys. Rev. Lett.}\ }\textbf
  {\bibinfo {volume} {125}},\ \bibinfo {pages} {037202} (\bibinfo {year}
  {2020})}\BibitemShut {NoStop}%
\bibitem [{\citenamefont {Leahy}\ \emph {et~al.}(2017)\citenamefont {Leahy},
  \citenamefont {Pocs}, \citenamefont {Siegfried}, \citenamefont {Graf},
  \citenamefont {Do}, \citenamefont {Choi}, \citenamefont {Normand},\ and\
  \citenamefont {Lee}}]{Leahy2017}%
  \BibitemOpen
  \bibfield  {author} {\bibinfo {author} {\bibfnamefont {I.~A.}\ \bibnamefont
  {Leahy}}, \bibinfo {author} {\bibfnamefont {C.~A.}\ \bibnamefont {Pocs}},
  \bibinfo {author} {\bibfnamefont {P.~E.}\ \bibnamefont {Siegfried}}, \bibinfo
  {author} {\bibfnamefont {D.}~\bibnamefont {Graf}}, \bibinfo {author}
  {\bibfnamefont {S.-H.}\ \bibnamefont {Do}}, \bibinfo {author} {\bibfnamefont
  {K.-Y.}\ \bibnamefont {Choi}}, \bibinfo {author} {\bibfnamefont
  {B.}~\bibnamefont {Normand}}, \ and\ \bibinfo {author} {\bibfnamefont
  {M.}~\bibnamefont {Lee}},\ }\bibfield  {title} {\enquote {\bibinfo {title}
  {Anomalous thermal conductivity and magnetic torque response in the honeycomb
  magnet $\alpha$-{RuCl}$_{3}$},}\ }\href {\doibase
  10.1103/PhysRevLett.118.187203} {\bibfield  {journal} {\bibinfo  {journal}
  {Phys. Rev. Lett.}\ }\textbf {\bibinfo {volume} {118}},\ \bibinfo {pages}
  {187203} (\bibinfo {year} {2017})}\BibitemShut {NoStop}%
\bibitem [{\citenamefont {{Modic}}\ \emph {et~al.}(2021)\citenamefont
  {{Modic}}, \citenamefont {{McDonald}}, \citenamefont {{Ruff}}, \citenamefont
  {{Bachmann}}, \citenamefont {{Lai}}, \citenamefont {{Palmstrom}},
  \citenamefont {{Graf}}, \citenamefont {{Chan}}, \citenamefont {{Balakirev}},
  \citenamefont {{Betts}}, \citenamefont {{Boebinger}}, \citenamefont
  {{Schmidt}}, \citenamefont {{Lawler}}, \citenamefont {{Sokolov}},
  \citenamefont {{Moll}}, \citenamefont {{Ramshaw}},\ and\ \citenamefont
  {{Shekhter}}}]{Modic2021}%
  \BibitemOpen
  \bibfield  {author} {\bibinfo {author} {\bibfnamefont {K.~A.}\ \bibnamefont
  {{Modic}}}, \bibinfo {author} {\bibfnamefont {Ross~D.}\ \bibnamefont
  {{McDonald}}}, \bibinfo {author} {\bibfnamefont {J.~P.~C.}\ \bibnamefont
  {{Ruff}}}, \bibinfo {author} {\bibfnamefont {Maja~D.}\ \bibnamefont
  {{Bachmann}}}, \bibinfo {author} {\bibfnamefont {You}\ \bibnamefont {{Lai}}},
  \bibinfo {author} {\bibfnamefont {Johanna~C.}\ \bibnamefont {{Palmstrom}}},
  \bibinfo {author} {\bibfnamefont {David}\ \bibnamefont {{Graf}}}, \bibinfo
  {author} {\bibfnamefont {Mun~K.}\ \bibnamefont {{Chan}}}, \bibinfo {author}
  {\bibfnamefont {F.~F.}\ \bibnamefont {{Balakirev}}}, \bibinfo {author}
  {\bibfnamefont {J.~B.}\ \bibnamefont {{Betts}}}, \bibinfo {author}
  {\bibfnamefont {G.~S.}\ \bibnamefont {{Boebinger}}}, \bibinfo {author}
  {\bibfnamefont {Marcus}\ \bibnamefont {{Schmidt}}}, \bibinfo {author}
  {\bibfnamefont {Michael~J.}\ \bibnamefont {{Lawler}}}, \bibinfo {author}
  {\bibfnamefont {D.~A.}\ \bibnamefont {{Sokolov}}}, \bibinfo {author}
  {\bibfnamefont {Philip J.~W.}\ \bibnamefont {{Moll}}}, \bibinfo {author}
  {\bibfnamefont {B.~J.}\ \bibnamefont {{Ramshaw}}}, \ and\ \bibinfo {author}
  {\bibfnamefont {Arkady}\ \bibnamefont {{Shekhter}}},\ }\bibfield  {title}
  {\enquote {\bibinfo {title} {{Scale-invariant magnetic anisotropy in
  RuCl$_{3}$ at high magnetic fields}},}\ }\href {\doibase
  10.1038/s41567-020-1028-0} {\bibfield  {journal} {\bibinfo  {journal} {Nat.
  Phys.}\ }\textbf {\bibinfo {volume} {17}},\ \bibinfo {pages} {240--244}
  (\bibinfo {year} {2021})}\BibitemShut {NoStop}%
\bibitem [{\citenamefont {Kasahara}\ \emph
  {et~al.}(2018{\natexlab{a}})\citenamefont {Kasahara}, \citenamefont {Sugii},
  \citenamefont {Ohnishi}, \citenamefont {Shimozawa}, \citenamefont
  {Yamashita}, \citenamefont {Kurita}, \citenamefont {Tanaka}, \citenamefont
  {Nasu}, \citenamefont {Motome}, \citenamefont {Shibauchi},\ and\
  \citenamefont {Matsuda}}]{Kasahara2018Unusual}%
  \BibitemOpen
  \bibfield  {author} {\bibinfo {author} {\bibfnamefont {Y.}~\bibnamefont
  {Kasahara}}, \bibinfo {author} {\bibfnamefont {K.}~\bibnamefont {Sugii}},
  \bibinfo {author} {\bibfnamefont {T.}~\bibnamefont {Ohnishi}}, \bibinfo
  {author} {\bibfnamefont {M.}~\bibnamefont {Shimozawa}}, \bibinfo {author}
  {\bibfnamefont {M.}~\bibnamefont {Yamashita}}, \bibinfo {author}
  {\bibfnamefont {N.}~\bibnamefont {Kurita}}, \bibinfo {author} {\bibfnamefont
  {H.}~\bibnamefont {Tanaka}}, \bibinfo {author} {\bibfnamefont
  {J.}~\bibnamefont {Nasu}}, \bibinfo {author} {\bibfnamefont {Y.}~\bibnamefont
  {Motome}}, \bibinfo {author} {\bibfnamefont {T.}~\bibnamefont {Shibauchi}}, \
  and\ \bibinfo {author} {\bibfnamefont {Y.}~\bibnamefont {Matsuda}},\
  }\bibfield  {title} {\enquote {\bibinfo {title} {Unusual thermal {Hall}
  effect in a {Kitaev} spin liquid candidate ${\alpha}$-{RuCl}$_{3}$},}\ }\href
  {\doibase 10.1103/PhysRevLett.120.217205} {\bibfield  {journal} {\bibinfo
  {journal} {Phys. Rev. Lett.}\ }\textbf {\bibinfo {volume} {120}},\ \bibinfo
  {pages} {217205} (\bibinfo {year} {2018}{\natexlab{a}})}\BibitemShut
  {NoStop}%
\bibitem [{\citenamefont {Kasahara}\ \emph
  {et~al.}(2018{\natexlab{b}})\citenamefont {Kasahara}, \citenamefont
  {Ohnishi}, \citenamefont {Mizukami}, \citenamefont {Tanaka}, \citenamefont
  {Ma}, \citenamefont {Sugii}, \citenamefont {Kurita}, \citenamefont {Tanaka},
  \citenamefont {Nasu}, \citenamefont {Motome}, \citenamefont {Shibauchi},\
  and\ \citenamefont {Matsuda}}]{Kasahara2018}%
  \BibitemOpen
  \bibfield  {author} {\bibinfo {author} {\bibfnamefont {Y.}~\bibnamefont
  {Kasahara}}, \bibinfo {author} {\bibfnamefont {T.}~\bibnamefont {Ohnishi}},
  \bibinfo {author} {\bibfnamefont {Y.}~\bibnamefont {Mizukami}}, \bibinfo
  {author} {\bibfnamefont {O.}~\bibnamefont {Tanaka}}, \bibinfo {author}
  {\bibfnamefont {S.}~\bibnamefont {Ma}}, \bibinfo {author} {\bibfnamefont
  {K.}~\bibnamefont {Sugii}}, \bibinfo {author} {\bibfnamefont
  {N.}~\bibnamefont {Kurita}}, \bibinfo {author} {\bibfnamefont
  {H.}~\bibnamefont {Tanaka}}, \bibinfo {author} {\bibfnamefont
  {J.}~\bibnamefont {Nasu}}, \bibinfo {author} {\bibfnamefont {Y.}~\bibnamefont
  {Motome}}, \bibinfo {author} {\bibfnamefont {T.}~\bibnamefont {Shibauchi}}, \
  and\ \bibinfo {author} {\bibfnamefont {Y.}~\bibnamefont {Matsuda}},\
  }\bibfield  {title} {\enquote {\bibinfo {title} {Majorana quantization and
  half-integer thermal quantum {Hall} effect in a {Kitaev} spin liquid},}\
  }\href {\doibase 10.1038/s41586-018-0274-0} {\bibfield  {journal} {\bibinfo
  {journal} {Nature}\ }\textbf {\bibinfo {volume} {559}},\ \bibinfo {pages}
  {227--231} (\bibinfo {year} {2018}{\natexlab{b}})}\BibitemShut {NoStop}%
\bibitem [{\citenamefont {{Yokoi}}\ \emph {et~al.}(2021)\citenamefont
  {{Yokoi}}, \citenamefont {{Ma}}, \citenamefont {{Kasahara}}, \citenamefont
  {{Kasahara}}, \citenamefont {{Shibauchi}}, \citenamefont {{Kurita}},
  \citenamefont {{Tanaka}}, \citenamefont {{Nasu}}, \citenamefont {{Motome}},
  \citenamefont {{Hickey}}, \citenamefont {{Trebst}},\ and\ \citenamefont
  {{Matsuda}}}]{Yokoi2021Science}%
  \BibitemOpen
  \bibfield  {author} {\bibinfo {author} {\bibfnamefont {T.}~\bibnamefont
  {{Yokoi}}}, \bibinfo {author} {\bibfnamefont {S.}~\bibnamefont {{Ma}}},
  \bibinfo {author} {\bibfnamefont {Y.}~\bibnamefont {{Kasahara}}}, \bibinfo
  {author} {\bibfnamefont {S.}~\bibnamefont {{Kasahara}}}, \bibinfo {author}
  {\bibfnamefont {T.}~\bibnamefont {{Shibauchi}}}, \bibinfo {author}
  {\bibfnamefont {N.}~\bibnamefont {{Kurita}}}, \bibinfo {author}
  {\bibfnamefont {H.}~\bibnamefont {{Tanaka}}}, \bibinfo {author}
  {\bibfnamefont {J.}~\bibnamefont {{Nasu}}}, \bibinfo {author} {\bibfnamefont
  {Y.}~\bibnamefont {{Motome}}}, \bibinfo {author} {\bibfnamefont
  {C.}~\bibnamefont {{Hickey}}}, \bibinfo {author} {\bibfnamefont
  {S.}~\bibnamefont {{Trebst}}}, \ and\ \bibinfo {author} {\bibfnamefont
  {Y.}~\bibnamefont {{Matsuda}}},\ }\bibfield  {title} {\enquote {\bibinfo
  {title} {{Half-integer quantized anomalous thermal Hall effect in the Kitaev
  material candidate ${\alpha}$-{RuCl}$_{3}$}},}\ }\href {\doibase
  10.1126/science.aay5551} {\bibfield  {journal} {\bibinfo  {journal}
  {Science}\ }\textbf {\bibinfo {volume} {373}},\ \bibinfo {pages} {568--572}
  (\bibinfo {year} {2021})}\BibitemShut {NoStop}%
\bibitem [{\citenamefont {Yamashita}\ \emph {et~al.}(2020)\citenamefont
  {Yamashita}, \citenamefont {Gouchi}, \citenamefont {Uwatoko}, \citenamefont
  {Kurita},\ and\ \citenamefont {Tanaka}}]{Yamashita2020sample}%
  \BibitemOpen
  \bibfield  {author} {\bibinfo {author} {\bibfnamefont {M.}~\bibnamefont
  {Yamashita}}, \bibinfo {author} {\bibfnamefont {J.}~\bibnamefont {Gouchi}},
  \bibinfo {author} {\bibfnamefont {Y.}~\bibnamefont {Uwatoko}}, \bibinfo
  {author} {\bibfnamefont {N.}~\bibnamefont {Kurita}}, \ and\ \bibinfo {author}
  {\bibfnamefont {H.}~\bibnamefont {Tanaka}},\ }\bibfield  {title} {\enquote
  {\bibinfo {title} {Sample dependence of half-integer quantized thermal {Hall}
  effect in the {Kitaev} spin-liquid candidate $\alpha$-{RuCl}$_3$},}\ }\href
  {\doibase 10.1103/PhysRevB.102.220404} {\bibfield  {journal} {\bibinfo
  {journal} {Phys. Rev. B}\ }\textbf {\bibinfo {volume} {102}},\ \bibinfo
  {pages} {220404(R)} (\bibinfo {year} {2020})}\BibitemShut {NoStop}%
\bibitem [{\citenamefont {{Zhou}}\ \emph {et~al.}()\citenamefont {{Zhou}},
  \citenamefont {{Li}}, \citenamefont {{Matsuda}}, \citenamefont {{Matsuo}},
  \citenamefont {{Li}}, \citenamefont {{Kurita}}, \citenamefont {{Kindo}},\
  and\ \citenamefont {{Tanaka}}}]{Zhou2022arXiv}%
  \BibitemOpen
  \bibfield  {author} {\bibinfo {author} {\bibfnamefont {Xu-Guang}\
  \bibnamefont {{Zhou}}}, \bibinfo {author} {\bibfnamefont {Han}\ \bibnamefont
  {{Li}}}, \bibinfo {author} {\bibfnamefont {Yasuhiro~H.}\ \bibnamefont
  {{Matsuda}}}, \bibinfo {author} {\bibfnamefont {Akira}\ \bibnamefont
  {{Matsuo}}}, \bibinfo {author} {\bibfnamefont {Wei}\ \bibnamefont {{Li}}},
  \bibinfo {author} {\bibfnamefont {Nobuyuki}\ \bibnamefont {{Kurita}}},
  \bibinfo {author} {\bibfnamefont {Koichi}\ \bibnamefont {{Kindo}}}, \ and\
  \bibinfo {author} {\bibfnamefont {Hidekazu}\ \bibnamefont {{Tanaka}}},\
  }\bibfield  {title} {\enquote {\bibinfo {title} {{Intermediate Quantum Spin
  Liquid Phase in the Kitaev Material $\alpha$-RuCl$_3$ under High Magnetic
  Fields up to 100 T}},}\ }\href@noop {} {\ }\Eprint
  {http://arxiv.org/abs/2201.04597 (2022)} {arXiv:2201.04597 (2022)}
  \BibitemShut {NoStop}%
\bibitem [{\citenamefont {{Laurell}}\ and\ \citenamefont
  {{Okamoto}}(2020)}]{Laurell2020}%
  \BibitemOpen
  \bibfield  {author} {\bibinfo {author} {\bibfnamefont {P.}~\bibnamefont
  {{Laurell}}}\ and\ \bibinfo {author} {\bibfnamefont {S.}~\bibnamefont
  {{Okamoto}}},\ }\bibfield  {title} {\enquote {\bibinfo {title} {{Dynamical
  and thermal magnetic properties of the Kitaev spin liquid candidate
  {\ensuremath{\alpha}}-RuCl$_{3}$}},}\ }\href {\doibase
  10.1038/s41535-019-0203-y} {\bibfield  {journal} {\bibinfo  {journal} {npj
  Quant. Mater.}\ }\textbf {\bibinfo {volume} {5}},\ \bibinfo {eid} {2}
  (\bibinfo {year} {2020})}\BibitemShut {NoStop}%
\bibitem [{\citenamefont {Chen}\ \emph {et~al.}(2017)\citenamefont {Chen},
  \citenamefont {Liu}, \citenamefont {Chen},\ and\ \citenamefont
  {Li}}]{Chen.b+:2017:SETTN}%
  \BibitemOpen
  \bibfield  {author} {\bibinfo {author} {\bibfnamefont {B.-B.}\ \bibnamefont
  {Chen}}, \bibinfo {author} {\bibfnamefont {Y.-J.}\ \bibnamefont {Liu}},
  \bibinfo {author} {\bibfnamefont {Z.}~\bibnamefont {Chen}}, \ and\ \bibinfo
  {author} {\bibfnamefont {W.}~\bibnamefont {Li}},\ }\bibfield  {title}
  {\enquote {\bibinfo {title} {Series-expansion thermal tensor network approach
  for quantum lattice models},}\ }\href {\doibase 10.1103/PhysRevB.95.161104}
  {\bibfield  {journal} {\bibinfo  {journal} {Phys. Rev. B}\ }\textbf {\bibinfo
  {volume} {95}},\ \bibinfo {pages} {161104(R)} (\bibinfo {year}
  {2017})}\BibitemShut {NoStop}%
\bibitem [{\citenamefont {Chen}\ \emph {et~al.}(2018)\citenamefont {Chen},
  \citenamefont {Chen}, \citenamefont {Chen}, \citenamefont {Li},\ and\
  \citenamefont {Weichselbaum}}]{Chen2018}%
  \BibitemOpen
  \bibfield  {author} {\bibinfo {author} {\bibfnamefont {B.-B.}\ \bibnamefont
  {Chen}}, \bibinfo {author} {\bibfnamefont {L.}~\bibnamefont {Chen}}, \bibinfo
  {author} {\bibfnamefont {Z.}~\bibnamefont {Chen}}, \bibinfo {author}
  {\bibfnamefont {W.}~\bibnamefont {Li}}, \ and\ \bibinfo {author}
  {\bibfnamefont {A.}~\bibnamefont {Weichselbaum}},\ }\bibfield  {title}
  {\enquote {\bibinfo {title} {Exponential thermal tensor network approach for
  quantum lattice models},}\ }\href {\doibase 10.1103/PhysRevX.8.031082}
  {\bibfield  {journal} {\bibinfo  {journal} {Phys. Rev. X}\ }\textbf {\bibinfo
  {volume} {8}},\ \bibinfo {pages} {031082} (\bibinfo {year}
  {2018})}\BibitemShut {NoStop}%
\bibitem [{\citenamefont {Li}\ \emph {et~al.}(2019)\citenamefont {Li},
  \citenamefont {Chen}, \citenamefont {Chen}, \citenamefont {von Delft},
  \citenamefont {Weichselbaum},\ and\ \citenamefont {Li}}]{Lih2019}%
  \BibitemOpen
  \bibfield  {author} {\bibinfo {author} {\bibfnamefont {H.}~\bibnamefont
  {Li}}, \bibinfo {author} {\bibfnamefont {B.-B.}\ \bibnamefont {Chen}},
  \bibinfo {author} {\bibfnamefont {Z.}~\bibnamefont {Chen}}, \bibinfo {author}
  {\bibfnamefont {J.}~\bibnamefont {von Delft}}, \bibinfo {author}
  {\bibfnamefont {A.}~\bibnamefont {Weichselbaum}}, \ and\ \bibinfo {author}
  {\bibfnamefont {W.}~\bibnamefont {Li}},\ }\bibfield  {title} {\enquote
  {\bibinfo {title} {Thermal tensor renormalization group simulations of
  square-lattice quantum spin models},}\ }\href {\doibase
  10.1103/PhysRevB.100.045110} {\bibfield  {journal} {\bibinfo  {journal}
  {Phys. Rev. B}\ }\textbf {\bibinfo {volume} {100}},\ \bibinfo {pages}
  {045110} (\bibinfo {year} {2019})}\BibitemShut {NoStop}%
\bibitem [{\citenamefont {Kubota}\ \emph {et~al.}(2015)\citenamefont {Kubota},
  \citenamefont {Tanaka}, \citenamefont {Ono}, \citenamefont {Narumi},\ and\
  \citenamefont {Kindo}}]{Kubota2015}%
  \BibitemOpen
  \bibfield  {author} {\bibinfo {author} {\bibfnamefont {Y.}~\bibnamefont
  {Kubota}}, \bibinfo {author} {\bibfnamefont {H.}~\bibnamefont {Tanaka}},
  \bibinfo {author} {\bibfnamefont {T.}~\bibnamefont {Ono}}, \bibinfo {author}
  {\bibfnamefont {Y.}~\bibnamefont {Narumi}}, \ and\ \bibinfo {author}
  {\bibfnamefont {K.}~\bibnamefont {Kindo}},\ }\bibfield  {title} {\enquote
  {\bibinfo {title} {Successive magnetic phase transitions in
  {${\alpha}$-{RuCl}$_{3}$: XY-like} frustrated magnet on the honeycomb
  lattice},}\ }\href {\doibase 10.1103/PhysRevB.91.094422} {\bibfield
  {journal} {\bibinfo  {journal} {Phys. Rev. B}\ }\textbf {\bibinfo {volume}
  {91}},\ \bibinfo {pages} {094422} (\bibinfo {year} {2015})}\BibitemShut
  {NoStop}%
\bibitem [{\citenamefont {Widmann}\ \emph {et~al.}(2019)\citenamefont
  {Widmann}, \citenamefont {Tsurkan}, \citenamefont {Prishchenko},
  \citenamefont {Mazurenko}, \citenamefont {Tsirlin},\ and\ \citenamefont
  {Loidl}}]{Widmann2019}%
  \BibitemOpen
  \bibfield  {author} {\bibinfo {author} {\bibfnamefont {S.}~\bibnamefont
  {Widmann}}, \bibinfo {author} {\bibfnamefont {V.}~\bibnamefont {Tsurkan}},
  \bibinfo {author} {\bibfnamefont {D.~A.}\ \bibnamefont {Prishchenko}},
  \bibinfo {author} {\bibfnamefont {V.~G.}\ \bibnamefont {Mazurenko}}, \bibinfo
  {author} {\bibfnamefont {A.~A.}\ \bibnamefont {Tsirlin}}, \ and\ \bibinfo
  {author} {\bibfnamefont {A.}~\bibnamefont {Loidl}},\ }\bibfield  {title}
  {\enquote {\bibinfo {title} {Thermodynamic evidence of fractionalized
  excitations in ${\alpha}$-{RuCl}$_{3}$},}\ }\href {\doibase
  10.1103/PhysRevB.99.094415} {\bibfield  {journal} {\bibinfo  {journal} {Phys.
  Rev. B}\ }\textbf {\bibinfo {volume} {99}},\ \bibinfo {pages} {094415}
  (\bibinfo {year} {2019})}\BibitemShut {NoStop}%
\bibitem [{\citenamefont {Johnson}\ \emph {et~al.}(2015)\citenamefont
  {Johnson}, \citenamefont {Williams}, \citenamefont {Haghighirad},
  \citenamefont {Singleton}, \citenamefont {Zapf}, \citenamefont {Manuel},
  \citenamefont {Mazin}, \citenamefont {Li}, \citenamefont {Jeschke},
  \citenamefont {Valent\'{\i}},\ and\ \citenamefont {Coldea}}]{Johnson2015}%
  \BibitemOpen
  \bibfield  {author} {\bibinfo {author} {\bibfnamefont {R.~D.}\ \bibnamefont
  {Johnson}}, \bibinfo {author} {\bibfnamefont {S.~C.}\ \bibnamefont
  {Williams}}, \bibinfo {author} {\bibfnamefont {A.~A.}\ \bibnamefont
  {Haghighirad}}, \bibinfo {author} {\bibfnamefont {J.}~\bibnamefont
  {Singleton}}, \bibinfo {author} {\bibfnamefont {V.}~\bibnamefont {Zapf}},
  \bibinfo {author} {\bibfnamefont {P.}~\bibnamefont {Manuel}}, \bibinfo
  {author} {\bibfnamefont {I.~I.}\ \bibnamefont {Mazin}}, \bibinfo {author}
  {\bibfnamefont {Y.}~\bibnamefont {Li}}, \bibinfo {author} {\bibfnamefont
  {H.~O.}\ \bibnamefont {Jeschke}}, \bibinfo {author} {\bibfnamefont
  {R.}~\bibnamefont {Valent\'{\i}}}, \ and\ \bibinfo {author} {\bibfnamefont
  {R.}~\bibnamefont {Coldea}},\ }\bibfield  {title} {\enquote {\bibinfo {title}
  {Monoclinic crystal structure of $\alpha$-{RuCl}$_3$ and the zigzag
  antiferromagnetic ground state},}\ }\href {\doibase
  10.1103/PhysRevB.92.235119} {\bibfield  {journal} {\bibinfo  {journal} {Phys.
  Rev. B}\ }\textbf {\bibinfo {volume} {92}},\ \bibinfo {pages} {235119}
  (\bibinfo {year} {2015})}\BibitemShut {NoStop}%
\bibitem [{\citenamefont {{Weber}}\ \emph {et~al.}(2016)\citenamefont
  {{Weber}}, \citenamefont {{Schoop}}, \citenamefont {{Duppel}}, \citenamefont
  {{Lippmann}}, \citenamefont {{Nuss}},\ and\ \citenamefont
  {{Lotsch}}}]{Weber2016}%
  \BibitemOpen
  \bibfield  {author} {\bibinfo {author} {\bibfnamefont {D.}~\bibnamefont
  {{Weber}}}, \bibinfo {author} {\bibfnamefont {L.~M.}\ \bibnamefont
  {{Schoop}}}, \bibinfo {author} {\bibfnamefont {V.}~\bibnamefont {{Duppel}}},
  \bibinfo {author} {\bibfnamefont {J.~M.}\ \bibnamefont {{Lippmann}}},
  \bibinfo {author} {\bibfnamefont {J.}~\bibnamefont {{Nuss}}}, \ and\ \bibinfo
  {author} {\bibfnamefont {B.~V.}\ \bibnamefont {{Lotsch}}},\ }\bibfield
  {title} {\enquote {\bibinfo {title} {{Magnetic Properties of Restacked 2D
  Spin 1/2 honeycomb RuCl$_3$ Nanosheets}},}\ }\href {\doibase
  10.1021/acs.nanolett.6b00701} {\bibfield  {journal} {\bibinfo  {journal}
  {Nano Lett.}\ }\textbf {\bibinfo {volume} {16}},\ \bibinfo {pages}
  {3578--3584} (\bibinfo {year} {2016})}\BibitemShut {NoStop}%
\bibitem [{\citenamefont {Lampen-Kelley}\ \emph {et~al.}(2018)\citenamefont
  {Lampen-Kelley}, \citenamefont {Rachel}, \citenamefont {Reuther},
  \citenamefont {Yan}, \citenamefont {Banerjee}, \citenamefont {Bridges},
  \citenamefont {Cao}, \citenamefont {Nagler},\ and\ \citenamefont
  {Mandrus}}]{Lampen-Kelley2018}%
  \BibitemOpen
  \bibfield  {author} {\bibinfo {author} {\bibfnamefont {P.}~\bibnamefont
  {Lampen-Kelley}}, \bibinfo {author} {\bibfnamefont {S.}~\bibnamefont
  {Rachel}}, \bibinfo {author} {\bibfnamefont {J.}~\bibnamefont {Reuther}},
  \bibinfo {author} {\bibfnamefont {J.-Q.}\ \bibnamefont {Yan}}, \bibinfo
  {author} {\bibfnamefont {A.}~\bibnamefont {Banerjee}}, \bibinfo {author}
  {\bibfnamefont {C.~A.}\ \bibnamefont {Bridges}}, \bibinfo {author}
  {\bibfnamefont {H.~B.}\ \bibnamefont {Cao}}, \bibinfo {author} {\bibfnamefont
  {S.~E.}\ \bibnamefont {Nagler}}, \ and\ \bibinfo {author} {\bibfnamefont
  {D.}~\bibnamefont {Mandrus}},\ }\bibfield  {title} {\enquote {\bibinfo
  {title} {Anisotropic susceptibilities in the honeycomb {Kitaev} system
  $\alpha$-{RuCl}$_3$},}\ }\href {\doibase 10.1103/PhysRevB.98.100403}
  {\bibfield  {journal} {\bibinfo  {journal} {Phys. Rev. B}\ }\textbf {\bibinfo
  {volume} {98}},\ \bibinfo {pages} {100403(R)} (\bibinfo {year}
  {2018})}\BibitemShut {NoStop}%
\bibitem [{\citenamefont {{Sears}}\ \emph {et~al.}(2020)\citenamefont
  {{Sears}}, \citenamefont {{Chern}}, \citenamefont {{Kim}}, \citenamefont
  {{Bereciartua}}, \citenamefont {{Francoual}}, \citenamefont {{Kim}},\ and\
  \citenamefont {{Kim}}}]{Sears2020}%
  \BibitemOpen
  \bibfield  {author} {\bibinfo {author} {\bibfnamefont {Jennifer~A.}\
  \bibnamefont {{Sears}}}, \bibinfo {author} {\bibfnamefont {Li~Ern}\
  \bibnamefont {{Chern}}}, \bibinfo {author} {\bibfnamefont {Subin}\
  \bibnamefont {{Kim}}}, \bibinfo {author} {\bibfnamefont {Pablo~J.}\
  \bibnamefont {{Bereciartua}}}, \bibinfo {author} {\bibfnamefont {Sonia}\
  \bibnamefont {{Francoual}}}, \bibinfo {author} {\bibfnamefont {Yong~Baek}\
  \bibnamefont {{Kim}}}, \ and\ \bibinfo {author} {\bibfnamefont {Young-June}\
  \bibnamefont {{Kim}}},\ }\bibfield  {title} {\enquote {\bibinfo {title}
  {{Ferromagnetic Kitaev interaction and the origin of large magnetic
  anisotropy in {\ensuremath{\alpha}}-RuCl$_{3}$}},}\ }\href {\doibase
  10.1038/s41567-020-0874-0} {\bibfield  {journal} {\bibinfo  {journal} {Nat.
  Phys.}\ }\textbf {\bibinfo {volume} {16}},\ \bibinfo {pages} {837--840}
  (\bibinfo {year} {2020})}\BibitemShut {NoStop}%
\bibitem [{\citenamefont {Yu}\ \emph {et~al.}(2022)\citenamefont {Yu},
  \citenamefont {Li}, \citenamefont {Zhao}, \citenamefont {Li}, \citenamefont
  {Liu},\ and\ \citenamefont {Gong}}]{Yu2022}%
  \BibitemOpen
  \bibfield  {author} {\bibinfo {author} {\bibfnamefont {Shunyao}\ \bibnamefont
  {Yu}}, \bibinfo {author} {\bibfnamefont {Han}\ \bibnamefont {Li}}, \bibinfo
  {author} {\bibfnamefont {Qi-Rong}\ \bibnamefont {Zhao}}, \bibinfo {author}
  {\bibfnamefont {Wei}\ \bibnamefont {Li}}, \bibinfo {author} {\bibfnamefont
  {Zheng-Xin}\ \bibnamefont {Liu}}, \ and\ \bibinfo {author} {\bibfnamefont
  {Shou-Shu}\ \bibnamefont {Gong}},\ }\bibfield  {title} {\enquote {\bibinfo
  {title} {{Nematicity and Intermediate Quantum Spin Liquid Phase in the
  $K$-$J$-$\Gamma$-$\Gamma'$ model}, \textit{in preparation}},}\ }\href@noop {}
  {\  (\bibinfo {year} {2022})}\BibitemShut {NoStop}%
\bibitem [{\citenamefont {Bachus}\ \emph {et~al.}(2021)\citenamefont {Bachus},
  \citenamefont {Kaib}, \citenamefont {Jesche}, \citenamefont {Tsurkan},
  \citenamefont {Loidl}, \citenamefont {Winter}, \citenamefont {Tsirlin},
  \citenamefont {Valent\'{\i}},\ and\ \citenamefont {Gegenwart}}]{Bachus2021}%
  \BibitemOpen
  \bibfield  {author} {\bibinfo {author} {\bibfnamefont {S.}~\bibnamefont
  {Bachus}}, \bibinfo {author} {\bibfnamefont {D.~A.~S.}\ \bibnamefont {Kaib}},
  \bibinfo {author} {\bibfnamefont {A.}~\bibnamefont {Jesche}}, \bibinfo
  {author} {\bibfnamefont {V.}~\bibnamefont {Tsurkan}}, \bibinfo {author}
  {\bibfnamefont {A.}~\bibnamefont {Loidl}}, \bibinfo {author} {\bibfnamefont
  {S.~M.}\ \bibnamefont {Winter}}, \bibinfo {author} {\bibfnamefont {A.~A.}\
  \bibnamefont {Tsirlin}}, \bibinfo {author} {\bibfnamefont {R.}~\bibnamefont
  {Valent\'{\i}}}, \ and\ \bibinfo {author} {\bibfnamefont {P.}~\bibnamefont
  {Gegenwart}},\ }\bibfield  {title} {\enquote {\bibinfo {title}
  {Angle-dependent thermodynamics of $\alpha$-{RuCl}$_3$},}\ }\href {\doibase
  10.1103/PhysRevB.103.054440} {\bibfield  {journal} {\bibinfo  {journal}
  {Phys. Rev. B}\ }\textbf {\bibinfo {volume} {103}},\ \bibinfo {pages}
  {054440} (\bibinfo {year} {2021})}\BibitemShut {NoStop}%
\bibitem [{\citenamefont {Modic}\ \emph {et~al.}(2018)\citenamefont {Modic},
  \citenamefont {Bachmann}, \citenamefont {Ramshaw}, \citenamefont {Arnold},
  \citenamefont {Shirer}, \citenamefont {Estry}, \citenamefont {Betts},
  \citenamefont {Ghimire}, \citenamefont {Bauer}, \citenamefont {Schmidt},
  \citenamefont {Baenitz}, \citenamefont {Svanidze}, \citenamefont {McDonald},
  \citenamefont {Shekhter},\ and\ \citenamefont {Moll}}]{Modic2018}%
  \BibitemOpen
  \bibfield  {author} {\bibinfo {author} {\bibfnamefont {K.~A.}\ \bibnamefont
  {Modic}}, \bibinfo {author} {\bibfnamefont {Maja~D.}\ \bibnamefont
  {Bachmann}}, \bibinfo {author} {\bibfnamefont {B.~J.}\ \bibnamefont
  {Ramshaw}}, \bibinfo {author} {\bibfnamefont {F.}~\bibnamefont {Arnold}},
  \bibinfo {author} {\bibfnamefont {K.~R.}\ \bibnamefont {Shirer}}, \bibinfo
  {author} {\bibfnamefont {Amelia}\ \bibnamefont {Estry}}, \bibinfo {author}
  {\bibfnamefont {J.~B.}\ \bibnamefont {Betts}}, \bibinfo {author}
  {\bibfnamefont {Nirmal~J.}\ \bibnamefont {Ghimire}}, \bibinfo {author}
  {\bibfnamefont {E.~D.}\ \bibnamefont {Bauer}}, \bibinfo {author}
  {\bibfnamefont {Marcus}\ \bibnamefont {Schmidt}}, \bibinfo {author}
  {\bibfnamefont {Michael}\ \bibnamefont {Baenitz}}, \bibinfo {author}
  {\bibfnamefont {E.}~\bibnamefont {Svanidze}}, \bibinfo {author}
  {\bibfnamefont {Ross~D.}\ \bibnamefont {McDonald}}, \bibinfo {author}
  {\bibfnamefont {Arkady}\ \bibnamefont {Shekhter}}, \ and\ \bibinfo {author}
  {\bibfnamefont {Philip J.~W.}\ \bibnamefont {Moll}},\ }\bibfield  {title}
  {\enquote {\bibinfo {title} {Resonant torsion magnetometry in anisotropic
  quantum materials},}\ }\href {\doibase 10.1038/s41467-018-06412-w} {\bibfield
   {journal} {\bibinfo  {journal} {Nat. Commun.}\ }\textbf {\bibinfo {volume}
  {9}},\ \bibinfo {pages} {3975} (\bibinfo {year} {2018})}\BibitemShut
  {NoStop}%
\bibitem [{\citenamefont {{Shekhter}}\ \emph {et~al.}()\citenamefont
  {{Shekhter}}, \citenamefont {{McDonald}}, \citenamefont {{Ramshaw}},\ and\
  \citenamefont {{Modic}}}]{Modic2022arXiv}%
  \BibitemOpen
  \bibfield  {author} {\bibinfo {author} {\bibfnamefont {A.}~\bibnamefont
  {{Shekhter}}}, \bibinfo {author} {\bibfnamefont {R.~D.}\ \bibnamefont
  {{McDonald}}}, \bibinfo {author} {\bibfnamefont {B.~J.}\ \bibnamefont
  {{Ramshaw}}}, \ and\ \bibinfo {author} {\bibfnamefont {K.~A.}\ \bibnamefont
  {{Modic}}},\ }\bibfield  {title} {\enquote {\bibinfo {title} {{The
  magnetotropic susceptibility}},}\ }\href@noop {} {\ }\Eprint
  {http://arxiv.org/abs/2208.10038 (2022)} {arXiv:2208.10038 (2022)}
  \BibitemShut {NoStop}%
\bibitem [{\citenamefont {Singh}\ and\ \citenamefont
  {Gegenwart}(2010)}]{Singh2010}%
  \BibitemOpen
  \bibfield  {author} {\bibinfo {author} {\bibfnamefont {Y.}~\bibnamefont
  {Singh}}\ and\ \bibinfo {author} {\bibfnamefont {P.}~\bibnamefont
  {Gegenwart}},\ }\bibfield  {title} {\enquote {\bibinfo {title}
  {Antiferromagnetic {Mott} insulating state in single crystals of the
  honeycomb lattice material {${\text{Na}}_{2}{\text{IrO}}_{3}$}},}\ }\href
  {\doibase 10.1103/PhysRevB.82.064412} {\bibfield  {journal} {\bibinfo
  {journal} {Phys. Rev. B}\ }\textbf {\bibinfo {volume} {82}},\ \bibinfo
  {pages} {064412} (\bibinfo {year} {2010})}\BibitemShut {NoStop}%
\bibitem [{\citenamefont {Chaloupka}\ \emph {et~al.}(2010)\citenamefont
  {Chaloupka}, \citenamefont {Jackeli},\ and\ \citenamefont
  {Khaliullin}}]{Chaloupka2010}%
  \BibitemOpen
  \bibfield  {author} {\bibinfo {author} {\bibfnamefont {J.}~\bibnamefont
  {Chaloupka}}, \bibinfo {author} {\bibfnamefont {G.}~\bibnamefont {Jackeli}},
  \ and\ \bibinfo {author} {\bibfnamefont {G.}~\bibnamefont {Khaliullin}},\
  }\bibfield  {title} {\enquote {\bibinfo {title} {{Kitaev-Heisenberg} model on
  a honeycomb lattice: Possible exotic phases in iridium oxides
  {${A}_{2}{\mathrm{IrO}}_{3}$}},}\ }\href {\doibase
  10.1103/PhysRevLett.105.027204} {\bibfield  {journal} {\bibinfo  {journal}
  {Phys. Rev. Lett.}\ }\textbf {\bibinfo {volume} {105}},\ \bibinfo {pages}
  {027204} (\bibinfo {year} {2010})}\BibitemShut {NoStop}%
\bibitem [{\citenamefont {Singh}\ \emph {et~al.}(2012)\citenamefont {Singh},
  \citenamefont {Manni}, \citenamefont {Reuther}, \citenamefont {Berlijn},
  \citenamefont {Thomale}, \citenamefont {Ku}, \citenamefont {Trebst},\ and\
  \citenamefont {Gegenwart}}]{Singh2012}%
  \BibitemOpen
  \bibfield  {author} {\bibinfo {author} {\bibfnamefont {Y.}~\bibnamefont
  {Singh}}, \bibinfo {author} {\bibfnamefont {S.}~\bibnamefont {Manni}},
  \bibinfo {author} {\bibfnamefont {J.}~\bibnamefont {Reuther}}, \bibinfo
  {author} {\bibfnamefont {T.}~\bibnamefont {Berlijn}}, \bibinfo {author}
  {\bibfnamefont {R.}~\bibnamefont {Thomale}}, \bibinfo {author} {\bibfnamefont
  {W.}~\bibnamefont {Ku}}, \bibinfo {author} {\bibfnamefont {S.}~\bibnamefont
  {Trebst}}, \ and\ \bibinfo {author} {\bibfnamefont {P.}~\bibnamefont
  {Gegenwart}},\ }\bibfield  {title} {\enquote {\bibinfo {title} {Relevance of
  the {Heisenberg-Kitaev} model for the honeycomb lattice iridates
  {${A}_{2}{\mathrm{IrO}}_{3}$}},}\ }\href {\doibase
  10.1103/PhysRevLett.108.127203} {\bibfield  {journal} {\bibinfo  {journal}
  {Phys. Rev. Lett.}\ }\textbf {\bibinfo {volume} {108}},\ \bibinfo {pages}
  {127203} (\bibinfo {year} {2012})}\BibitemShut {NoStop}%
\bibitem [{\citenamefont {{Katukuri}}\ \emph {et~al.}(2014)\citenamefont
  {{Katukuri}}, \citenamefont {{Nishimoto}}, \citenamefont {{Yushankhai}},
  \citenamefont {{Stoyanova}}, \citenamefont {{Kandpal}}, \citenamefont
  {{Choi}}, \citenamefont {{Coldea}}, \citenamefont {{Rousochatzakis}},
  \citenamefont {{Hozoi}},\ and\ \citenamefont {{van den
  Brink}}}]{Katukuri2014}%
  \BibitemOpen
  \bibfield  {author} {\bibinfo {author} {\bibfnamefont {V.~M.}\ \bibnamefont
  {{Katukuri}}}, \bibinfo {author} {\bibfnamefont {S.}~\bibnamefont
  {{Nishimoto}}}, \bibinfo {author} {\bibfnamefont {V.}~\bibnamefont
  {{Yushankhai}}}, \bibinfo {author} {\bibfnamefont {A.}~\bibnamefont
  {{Stoyanova}}}, \bibinfo {author} {\bibfnamefont {H.}~\bibnamefont
  {{Kandpal}}}, \bibinfo {author} {\bibfnamefont {S.}~\bibnamefont {{Choi}}},
  \bibinfo {author} {\bibfnamefont {R.}~\bibnamefont {{Coldea}}}, \bibinfo
  {author} {\bibfnamefont {I.}~\bibnamefont {{Rousochatzakis}}}, \bibinfo
  {author} {\bibfnamefont {L.}~\bibnamefont {{Hozoi}}}, \ and\ \bibinfo
  {author} {\bibfnamefont {J.}~\bibnamefont {{van den Brink}}},\ }\bibfield
  {title} {\enquote {\bibinfo {title} {{Kitaev interactions between {$J = 1/2$}
  moments in honeycomb {Na$_{2}$IrO$_{3}$} are large and ferromagnetic:
  insights from $ab$ initio quantum chemistry calculations}},}\ }\href
  {\doibase 10.1088/1367-2630/16/1/013056} {\bibfield  {journal} {\bibinfo
  {journal} {New J. Phys.}\ }\textbf {\bibinfo {volume} {16}},\ \bibinfo {eid}
  {013056} (\bibinfo {year} {2014})}\BibitemShut {NoStop}%
\bibitem [{\citenamefont {Yamaji}\ \emph {et~al.}(2014)\citenamefont {Yamaji},
  \citenamefont {Nomura}, \citenamefont {Kurita}, \citenamefont {Arita},\ and\
  \citenamefont {Imada}}]{Yamaji2014}%
  \BibitemOpen
  \bibfield  {author} {\bibinfo {author} {\bibfnamefont {Y.}~\bibnamefont
  {Yamaji}}, \bibinfo {author} {\bibfnamefont {Y.}~\bibnamefont {Nomura}},
  \bibinfo {author} {\bibfnamefont {M.}~\bibnamefont {Kurita}}, \bibinfo
  {author} {\bibfnamefont {R.}~\bibnamefont {Arita}}, \ and\ \bibinfo {author}
  {\bibfnamefont {M.}~\bibnamefont {Imada}},\ }\bibfield  {title} {\enquote
  {\bibinfo {title} {First-principles study of the honeycomb-lattice iridates
  {${\mathrm{Na}}_{2}{\mathrm{IrO}}_{3}$} in the presence of strong spin-orbit
  interaction and electron correlations},}\ }\href {\doibase
  10.1103/PhysRevLett.113.107201} {\bibfield  {journal} {\bibinfo  {journal}
  {Phys. Rev. Lett.}\ }\textbf {\bibinfo {volume} {113}},\ \bibinfo {pages}
  {107201} (\bibinfo {year} {2014})}\BibitemShut {NoStop}%
\bibitem [{\citenamefont {Winter}\ \emph {et~al.}(2016)\citenamefont {Winter},
  \citenamefont {Li}, \citenamefont {Jeschke},\ and\ \citenamefont
  {Valent\'{\i}}}]{Winter2016}%
  \BibitemOpen
  \bibfield  {author} {\bibinfo {author} {\bibfnamefont {S.~M.}\ \bibnamefont
  {Winter}}, \bibinfo {author} {\bibfnamefont {Y.}~\bibnamefont {Li}}, \bibinfo
  {author} {\bibfnamefont {H.~O.}\ \bibnamefont {Jeschke}}, \ and\ \bibinfo
  {author} {\bibfnamefont {R.}~\bibnamefont {Valent\'{\i}}},\ }\bibfield
  {title} {\enquote {\bibinfo {title} {Challenges in design of {Kitaev}
  materials: Magnetic interactions from competing energy scales},}\ }\href
  {\doibase 10.1103/PhysRevB.93.214431} {\bibfield  {journal} {\bibinfo
  {journal} {Phys. Rev. B}\ }\textbf {\bibinfo {volume} {93}},\ \bibinfo
  {pages} {214431} (\bibinfo {year} {2016})}\BibitemShut {NoStop}%
\bibitem [{\citenamefont {Mehlawat}\ \emph {et~al.}(2017)\citenamefont
  {Mehlawat}, \citenamefont {Thamizhavel},\ and\ \citenamefont
  {Singh}}]{Mehlawat2017}%
  \BibitemOpen
  \bibfield  {author} {\bibinfo {author} {\bibfnamefont {K.}~\bibnamefont
  {Mehlawat}}, \bibinfo {author} {\bibfnamefont {A.}~\bibnamefont
  {Thamizhavel}}, \ and\ \bibinfo {author} {\bibfnamefont {Y.}~\bibnamefont
  {Singh}},\ }\bibfield  {title} {\enquote {\bibinfo {title} {Heat capacity
  evidence for proximity to the {Kitaev} quantum spin liquid in {${A}_{2}
  {\mathrm{IrO}}_{3}$ ($A=\mathrm{Na}$, Li)}},}\ }\href {\doibase
  10.1103/PhysRevB.95.144406} {\bibfield  {journal} {\bibinfo  {journal} {Phys.
  Rev. B}\ }\textbf {\bibinfo {volume} {95}},\ \bibinfo {pages} {144406}
  (\bibinfo {year} {2017})}\BibitemShut {NoStop}%
\bibitem [{\citenamefont {Abramchuk}\ \emph {et~al.}(2017)\citenamefont
  {Abramchuk}, \citenamefont {Ozsoy-Keskinbora}, \citenamefont {Krizan},
  \citenamefont {Metz}, \citenamefont {Bell},\ and\ \citenamefont
  {Tafti}}]{Abramchuk2017}%
  \BibitemOpen
  \bibfield  {author} {\bibinfo {author} {\bibfnamefont {Mykola}\ \bibnamefont
  {Abramchuk}}, \bibinfo {author} {\bibfnamefont {Cigdem}\ \bibnamefont
  {Ozsoy-Keskinbora}}, \bibinfo {author} {\bibfnamefont {Jason~W.}\
  \bibnamefont {Krizan}}, \bibinfo {author} {\bibfnamefont {Kenneth~R.}\
  \bibnamefont {Metz}}, \bibinfo {author} {\bibfnamefont {David~C.}\
  \bibnamefont {Bell}}, \ and\ \bibinfo {author} {\bibfnamefont {Fazel}\
  \bibnamefont {Tafti}},\ }\bibfield  {title} {\enquote {\bibinfo {title}
  {{Cu$_2$IrO$_3$}: A new magnetically frustrated honeycomb iridate},}\ }\href
  {\doibase 10.1021/jacs.7b06911} {\bibfield  {journal} {\bibinfo  {journal}
  {J. Am. Chem. Soc.}\ }\textbf {\bibinfo {volume} {139}},\ \bibinfo {pages}
  {15371--15376} (\bibinfo {year} {2017})}\BibitemShut {NoStop}%
\bibitem [{\citenamefont {Choi}\ \emph {et~al.}(2019)\citenamefont {Choi},
  \citenamefont {Lee}, \citenamefont {Lee}, \citenamefont {Yoon}, \citenamefont
  {Lee}, \citenamefont {Park}, \citenamefont {Ali}, \citenamefont {Singh},
  \citenamefont {Orain}, \citenamefont {Kim}, \citenamefont {Rhyee},
  \citenamefont {Chen}, \citenamefont {Chou},\ and\ \citenamefont
  {Choi}}]{Choi2019}%
  \BibitemOpen
  \bibfield  {author} {\bibinfo {author} {\bibfnamefont {Y.~S.}\ \bibnamefont
  {Choi}}, \bibinfo {author} {\bibfnamefont {C.~H.}\ \bibnamefont {Lee}},
  \bibinfo {author} {\bibfnamefont {S.}~\bibnamefont {Lee}}, \bibinfo {author}
  {\bibfnamefont {Sungwon}\ \bibnamefont {Yoon}}, \bibinfo {author}
  {\bibfnamefont {W.-J.}\ \bibnamefont {Lee}}, \bibinfo {author} {\bibfnamefont
  {J.}~\bibnamefont {Park}}, \bibinfo {author} {\bibfnamefont {Anzar}\
  \bibnamefont {Ali}}, \bibinfo {author} {\bibfnamefont {Yogesh}\ \bibnamefont
  {Singh}}, \bibinfo {author} {\bibfnamefont {Jean-Christophe}\ \bibnamefont
  {Orain}}, \bibinfo {author} {\bibfnamefont {Gareoung}\ \bibnamefont {Kim}},
  \bibinfo {author} {\bibfnamefont {Jong-Soo}\ \bibnamefont {Rhyee}}, \bibinfo
  {author} {\bibfnamefont {Wei-Tin}\ \bibnamefont {Chen}}, \bibinfo {author}
  {\bibfnamefont {Fangcheng}\ \bibnamefont {Chou}}, \ and\ \bibinfo {author}
  {\bibfnamefont {Kwang-Yong}\ \bibnamefont {Choi}},\ }\bibfield  {title}
  {\enquote {\bibinfo {title} {Exotic low-energy excitations emergent in the
  random kitaev magnet ${\mathrm{cu}}_{2}{\mathrm{iro}}_{3}$},}\ }\href
  {\doibase 10.1103/PhysRevLett.122.167202} {\bibfield  {journal} {\bibinfo
  {journal} {Phys. Rev. Lett.}\ }\textbf {\bibinfo {volume} {122}},\ \bibinfo
  {pages} {167202} (\bibinfo {year} {2019})}\BibitemShut {NoStop}%
\bibitem [{\citenamefont {Todorova}\ \emph {et~al.}(2011)\citenamefont
  {Todorova}, \citenamefont {Leineweber}, \citenamefont {Kienle}, \citenamefont
  {Duppel},\ and\ \citenamefont {Jansen}}]{Todorova2011}%
  \BibitemOpen
  \bibfield  {author} {\bibinfo {author} {\bibfnamefont {V.}~\bibnamefont
  {Todorova}}, \bibinfo {author} {\bibfnamefont {A.}~\bibnamefont
  {Leineweber}}, \bibinfo {author} {\bibfnamefont {L.}~\bibnamefont {Kienle}},
  \bibinfo {author} {\bibfnamefont {V.}~\bibnamefont {Duppel}}, \ and\ \bibinfo
  {author} {\bibfnamefont {M.}~\bibnamefont {Jansen}},\ }\bibfield  {title}
  {\enquote {\bibinfo {title} {On {AgRhO}$_2$, and the new quaternary
  delafossites {AgLi}$_{1/3}${M}$_{2/3}${O}$_2$, syntheses and analyses of real
  structures},}\ }\href {\doibase https://doi.org/10.1016/j.jssc.2011.03.014}
  {\bibfield  {journal} {\bibinfo  {journal} {J. Solid State Chem.}\ }\textbf
  {\bibinfo {volume} {184}},\ \bibinfo {pages} {1112--1119} (\bibinfo {year}
  {2011})}\BibitemShut {NoStop}%
\bibitem [{\citenamefont {Roudebush}\ \emph {et~al.}(2016)\citenamefont
  {Roudebush}, \citenamefont {Ross},\ and\ \citenamefont
  {Cava}}]{Roudebush2016}%
  \BibitemOpen
  \bibfield  {author} {\bibinfo {author} {\bibfnamefont {John~H.}\ \bibnamefont
  {Roudebush}}, \bibinfo {author} {\bibfnamefont {K.~A.}\ \bibnamefont {Ross}},
  \ and\ \bibinfo {author} {\bibfnamefont {R.~J.}\ \bibnamefont {Cava}},\
  }\bibfield  {title} {\enquote {\bibinfo {title} {Iridium containing honeycomb
  delafossites by topotactic cation exchange},}\ }\href {\doibase
  10.1039/C6DT00798H} {\bibfield  {journal} {\bibinfo  {journal} {Dalton
  Trans.}\ }\textbf {\bibinfo {volume} {45}},\ \bibinfo {pages} {8783--8789}
  (\bibinfo {year} {2016})}\BibitemShut {NoStop}%
\bibitem [{\citenamefont {{Kitagawa}}\ \emph {et~al.}(2018)\citenamefont
  {{Kitagawa}}, \citenamefont {{Takayama}}, \citenamefont {{Matsumoto}},
  \citenamefont {{Kato}}, \citenamefont {{Takano}}, \citenamefont
  {{Kishimoto}}, \citenamefont {{Bette}}, \citenamefont {{Dinnebier}},
  \citenamefont {{Jackeli}},\ and\ \citenamefont {{Takagi}}}]{Kitagawa2018}%
  \BibitemOpen
  \bibfield  {author} {\bibinfo {author} {\bibfnamefont {K.}~\bibnamefont
  {{Kitagawa}}}, \bibinfo {author} {\bibfnamefont {T.}~\bibnamefont
  {{Takayama}}}, \bibinfo {author} {\bibfnamefont {Y.}~\bibnamefont
  {{Matsumoto}}}, \bibinfo {author} {\bibfnamefont {A.}~\bibnamefont {{Kato}}},
  \bibinfo {author} {\bibfnamefont {R.}~\bibnamefont {{Takano}}}, \bibinfo
  {author} {\bibfnamefont {Y.}~\bibnamefont {{Kishimoto}}}, \bibinfo {author}
  {\bibfnamefont {S.}~\bibnamefont {{Bette}}}, \bibinfo {author} {\bibfnamefont
  {R.}~\bibnamefont {{Dinnebier}}}, \bibinfo {author} {\bibfnamefont
  {G.}~\bibnamefont {{Jackeli}}}, \ and\ \bibinfo {author} {\bibfnamefont
  {H.}~\bibnamefont {{Takagi}}},\ }\bibfield  {title} {\enquote {\bibinfo
  {title} {{A spin-orbital-entangled quantum liquid on a honeycomb lattice}},}\
  }\href {\doibase 10.1038/nature25482} {\bibfield  {journal} {\bibinfo
  {journal} {Nature}\ }\textbf {\bibinfo {volume} {554}},\ \bibinfo {pages}
  {341--345} (\bibinfo {year} {2018})}\BibitemShut {NoStop}%
\bibitem [{\citenamefont {{Winter}}\ \emph {et~al.}(2017)\citenamefont
  {{Winter}}, \citenamefont {{Riedl}}, \citenamefont {{Maksimov}},
  \citenamefont {{Chernyshev}}, \citenamefont {{Honecker}},\ and\ \citenamefont
  {{Valent{\'\i}}}}]{Winter2017NC}%
  \BibitemOpen
  \bibfield  {author} {\bibinfo {author} {\bibfnamefont {S.~M.}\ \bibnamefont
  {{Winter}}}, \bibinfo {author} {\bibfnamefont {K.}~\bibnamefont {{Riedl}}},
  \bibinfo {author} {\bibfnamefont {P.~A.}\ \bibnamefont {{Maksimov}}},
  \bibinfo {author} {\bibfnamefont {A.~L.}\ \bibnamefont {{Chernyshev}}},
  \bibinfo {author} {\bibfnamefont {A.}~\bibnamefont {{Honecker}}}, \ and\
  \bibinfo {author} {\bibfnamefont {R.}~\bibnamefont {{Valent{\'\i}}}},\
  }\bibfield  {title} {\enquote {\bibinfo {title} {{Breakdown of magnons in a
  strongly spin-orbital coupled magnet}},}\ }\href {\doibase
  10.1038/s41467-017-01177-0} {\bibfield  {journal} {\bibinfo  {journal} {Nat.
  Commun.}\ }\textbf {\bibinfo {volume} {8}},\ \bibinfo {eid} {1152} (\bibinfo
  {year} {2017})}\BibitemShut {NoStop}%
\bibitem [{\citenamefont {Wu}\ \emph {et~al.}(2018)\citenamefont {Wu},
  \citenamefont {Little}, \citenamefont {Aldape}, \citenamefont {Rees},
  \citenamefont {Thewalt}, \citenamefont {Lampen-Kelley}, \citenamefont
  {Banerjee}, \citenamefont {Bridges}, \citenamefont {Yan}, \citenamefont
  {Boone}, \citenamefont {Patankar}, \citenamefont {Goldhaber-Gordon},
  \citenamefont {Mandrus}, \citenamefont {Nagler}, \citenamefont {Altman},\
  and\ \citenamefont {Orenstein}}]{Wu2018}%
  \BibitemOpen
  \bibfield  {author} {\bibinfo {author} {\bibfnamefont {L.}~\bibnamefont
  {Wu}}, \bibinfo {author} {\bibfnamefont {A.}~\bibnamefont {Little}}, \bibinfo
  {author} {\bibfnamefont {E.~E.}\ \bibnamefont {Aldape}}, \bibinfo {author}
  {\bibfnamefont {D.}~\bibnamefont {Rees}}, \bibinfo {author} {\bibfnamefont
  {E.}~\bibnamefont {Thewalt}}, \bibinfo {author} {\bibfnamefont
  {P.}~\bibnamefont {Lampen-Kelley}}, \bibinfo {author} {\bibfnamefont
  {A.}~\bibnamefont {Banerjee}}, \bibinfo {author} {\bibfnamefont {C.~A.}\
  \bibnamefont {Bridges}}, \bibinfo {author} {\bibfnamefont {J.-Q.}\
  \bibnamefont {Yan}}, \bibinfo {author} {\bibfnamefont {D.}~\bibnamefont
  {Boone}}, \bibinfo {author} {\bibfnamefont {S.}~\bibnamefont {Patankar}},
  \bibinfo {author} {\bibfnamefont {D.}~\bibnamefont {Goldhaber-Gordon}},
  \bibinfo {author} {\bibfnamefont {D.}~\bibnamefont {Mandrus}}, \bibinfo
  {author} {\bibfnamefont {S.~E.}\ \bibnamefont {Nagler}}, \bibinfo {author}
  {\bibfnamefont {E.}~\bibnamefont {Altman}}, \ and\ \bibinfo {author}
  {\bibfnamefont {J.}~\bibnamefont {Orenstein}},\ }\bibfield  {title} {\enquote
  {\bibinfo {title} {Field evolution of magnons in
  $\ensuremath{\alpha}$-{RuCl}$_{3}$ by high-resolution polarized terahertz
  spectroscopy},}\ }\href {\doibase 10.1103/PhysRevB.98.094425} {\bibfield
  {journal} {\bibinfo  {journal} {Phys. Rev. B}\ }\textbf {\bibinfo {volume}
  {98}},\ \bibinfo {pages} {094425} (\bibinfo {year} {2018})}\BibitemShut
  {NoStop}%
\bibitem [{\citenamefont {Cookmeyer}\ and\ \citenamefont
  {Moore}(2018)}]{Cookmeyer2018}%
  \BibitemOpen
  \bibfield  {author} {\bibinfo {author} {\bibfnamefont {T.}~\bibnamefont
  {Cookmeyer}}\ and\ \bibinfo {author} {\bibfnamefont {J.~E.}\ \bibnamefont
  {Moore}},\ }\bibfield  {title} {\enquote {\bibinfo {title} {Spin-wave
  analysis of the low-temperature thermal {Hall} effect in the candidate
  {Kitaev} spin liquid $\ensuremath{\alpha}$-{RuCl}$_{3}$},}\ }\href {\doibase
  10.1103/PhysRevB.98.060412} {\bibfield  {journal} {\bibinfo  {journal} {Phys.
  Rev. B}\ }\textbf {\bibinfo {volume} {98}},\ \bibinfo {pages} {060412(R)}
  (\bibinfo {year} {2018})}\BibitemShut {NoStop}%
\bibitem [{\citenamefont {Kim}\ and\ \citenamefont {Kee}(2016)}]{Kim2016}%
  \BibitemOpen
  \bibfield  {author} {\bibinfo {author} {\bibfnamefont {H.-S.}\ \bibnamefont
  {Kim}}\ and\ \bibinfo {author} {\bibfnamefont {H.-Y.}\ \bibnamefont {Kee}},\
  }\bibfield  {title} {\enquote {\bibinfo {title} {Crystal structure and
  magnetism in $\ensuremath{\alpha}$-{RuCl}$_{3}$: An ab initio study},}\
  }\href {\doibase 10.1103/PhysRevB.93.155143} {\bibfield  {journal} {\bibinfo
  {journal} {Phys. Rev. B}\ }\textbf {\bibinfo {volume} {93}},\ \bibinfo
  {pages} {155143} (\bibinfo {year} {2016})}\BibitemShut {NoStop}%
\bibitem [{\citenamefont {Suzuki}\ and\ \citenamefont
  {Suga}(2019)}]{Suzuki2019}%
  \BibitemOpen
  \bibfield  {author} {\bibinfo {author} {\bibfnamefont {T.}~\bibnamefont
  {Suzuki}}\ and\ \bibinfo {author} {\bibfnamefont {S.-i.}\ \bibnamefont
  {Suga}},\ }\bibfield  {title} {\enquote {\bibinfo {title} {Erratum: Effective
  model with strong {Kitaev} interactions for
  $\ensuremath{\alpha}$-{RuCl}$_{3}$ {[Phys. Rev. B 97, 134424 (2018)]}},}\
  }\href {\doibase 10.1103/PhysRevB.99.249902} {\bibfield  {journal} {\bibinfo
  {journal} {Phys. Rev. B}\ }\textbf {\bibinfo {volume} {99}},\ \bibinfo
  {pages} {249902(E)} (\bibinfo {year} {2019})}\BibitemShut {NoStop}%
\bibitem [{\citenamefont {Ran}\ \emph {et~al.}(2017)\citenamefont {Ran},
  \citenamefont {Wang}, \citenamefont {Wang}, \citenamefont {Dong},
  \citenamefont {Ren}, \citenamefont {Bao}, \citenamefont {Li}, \citenamefont
  {Ma}, \citenamefont {Gan}, \citenamefont {Zhang}, \citenamefont {Park},
  \citenamefont {Deng}, \citenamefont {Danilkin}, \citenamefont {Yu},
  \citenamefont {Li},\ and\ \citenamefont {Wen}}]{Ran2017}%
  \BibitemOpen
  \bibfield  {author} {\bibinfo {author} {\bibfnamefont {K.}~\bibnamefont
  {Ran}}, \bibinfo {author} {\bibfnamefont {J.}~\bibnamefont {Wang}}, \bibinfo
  {author} {\bibfnamefont {W.}~\bibnamefont {Wang}}, \bibinfo {author}
  {\bibfnamefont {Z.-Y.}\ \bibnamefont {Dong}}, \bibinfo {author}
  {\bibfnamefont {X.}~\bibnamefont {Ren}}, \bibinfo {author} {\bibfnamefont
  {S.}~\bibnamefont {Bao}}, \bibinfo {author} {\bibfnamefont {S.}~\bibnamefont
  {Li}}, \bibinfo {author} {\bibfnamefont {Z.}~\bibnamefont {Ma}}, \bibinfo
  {author} {\bibfnamefont {Y.}~\bibnamefont {Gan}}, \bibinfo {author}
  {\bibfnamefont {Y.}~\bibnamefont {Zhang}}, \bibinfo {author} {\bibfnamefont
  {J.~T.}\ \bibnamefont {Park}}, \bibinfo {author} {\bibfnamefont
  {G.}~\bibnamefont {Deng}}, \bibinfo {author} {\bibfnamefont {S.}~\bibnamefont
  {Danilkin}}, \bibinfo {author} {\bibfnamefont {S.-L.}\ \bibnamefont {Yu}},
  \bibinfo {author} {\bibfnamefont {J.-X.}\ \bibnamefont {Li}}, \ and\ \bibinfo
  {author} {\bibfnamefont {J.}~\bibnamefont {Wen}},\ }\bibfield  {title}
  {\enquote {\bibinfo {title} {Spin-wave excitations evidencing the {Kitaev}
  interaction in single crystalline ${\alpha}$-{RuCl}$_{3}$},}\ }\href
  {\doibase 10.1103/PhysRevLett.118.107203} {\bibfield  {journal} {\bibinfo
  {journal} {Phys. Rev. Lett.}\ }\textbf {\bibinfo {volume} {118}},\ \bibinfo
  {pages} {107203} (\bibinfo {year} {2017})}\BibitemShut {NoStop}%
\bibitem [{\citenamefont {Wang}\ \emph {et~al.}(2017)\citenamefont {Wang},
  \citenamefont {Dong}, \citenamefont {Yu},\ and\ \citenamefont
  {Li}}]{Wang2017}%
  \BibitemOpen
  \bibfield  {author} {\bibinfo {author} {\bibfnamefont {W.}~\bibnamefont
  {Wang}}, \bibinfo {author} {\bibfnamefont {Z.-Y.}\ \bibnamefont {Dong}},
  \bibinfo {author} {\bibfnamefont {S.-L.}\ \bibnamefont {Yu}}, \ and\ \bibinfo
  {author} {\bibfnamefont {J.-X.}\ \bibnamefont {Li}},\ }\bibfield  {title}
  {\enquote {\bibinfo {title} {Theoretical investigation of magnetic dynamics
  in $\alpha$-{RuCl}$_3$},}\ }\href {\doibase 10.1103/PhysRevB.96.115103}
  {\bibfield  {journal} {\bibinfo  {journal} {Phys. Rev. B}\ }\textbf {\bibinfo
  {volume} {96}},\ \bibinfo {pages} {115103} (\bibinfo {year}
  {2017})}\BibitemShut {NoStop}%
\bibitem [{\citenamefont {Ozel}\ \emph {et~al.}(2019)\citenamefont {Ozel},
  \citenamefont {Belvin}, \citenamefont {Baldini}, \citenamefont {Kimchi},
  \citenamefont {Do}, \citenamefont {Choi},\ and\ \citenamefont
  {Gedik}}]{Ozel2019}%
  \BibitemOpen
  \bibfield  {author} {\bibinfo {author} {\bibfnamefont {I.~O.}\ \bibnamefont
  {Ozel}}, \bibinfo {author} {\bibfnamefont {C.~A.}\ \bibnamefont {Belvin}},
  \bibinfo {author} {\bibfnamefont {E.}~\bibnamefont {Baldini}}, \bibinfo
  {author} {\bibfnamefont {I.}~\bibnamefont {Kimchi}}, \bibinfo {author}
  {\bibfnamefont {S.-H.}\ \bibnamefont {Do}}, \bibinfo {author} {\bibfnamefont
  {K.-Y.}\ \bibnamefont {Choi}}, \ and\ \bibinfo {author} {\bibfnamefont
  {N.}~\bibnamefont {Gedik}},\ }\bibfield  {title} {\enquote {\bibinfo {title}
  {Magnetic field-dependent low-energy magnon dynamics in
  $\alpha$-{RuCl}$_3$},}\ }\href {\doibase 10.1103/PhysRevB.100.085108}
  {\bibfield  {journal} {\bibinfo  {journal} {Phys. Rev. B}\ }\textbf {\bibinfo
  {volume} {100}},\ \bibinfo {pages} {085108} (\bibinfo {year}
  {2019})}\BibitemShut {NoStop}%
\bibitem [{\citenamefont {{Banerjee}}\ \emph {et~al.}(2016)\citenamefont
  {{Banerjee}}, \citenamefont {{Bridges}}, \citenamefont {{Yan}}, \citenamefont
  {{Aczel}}, \citenamefont {{Li}}, \citenamefont {{Stone}}, \citenamefont
  {{Granroth}}, \citenamefont {{Lumsden}}, \citenamefont {{Yiu}}, \citenamefont
  {{Knolle}}, \citenamefont {{Bhattacharjee}}, \citenamefont {{Kovrizhin}},
  \citenamefont {{Moessner}}, \citenamefont {{Tennant}}, \citenamefont
  {{Mandrus}},\ and\ \citenamefont {{Nagler}}}]{Banerjee2016}%
  \BibitemOpen
  \bibfield  {author} {\bibinfo {author} {\bibfnamefont {A.}~\bibnamefont
  {{Banerjee}}}, \bibinfo {author} {\bibfnamefont {C.~A.}\ \bibnamefont
  {{Bridges}}}, \bibinfo {author} {\bibfnamefont {J.~Q.}\ \bibnamefont
  {{Yan}}}, \bibinfo {author} {\bibfnamefont {A.~A.}\ \bibnamefont {{Aczel}}},
  \bibinfo {author} {\bibfnamefont {L.}~\bibnamefont {{Li}}}, \bibinfo {author}
  {\bibfnamefont {M.~B.}\ \bibnamefont {{Stone}}}, \bibinfo {author}
  {\bibfnamefont {G.~E.}\ \bibnamefont {{Granroth}}}, \bibinfo {author}
  {\bibfnamefont {M.~D.}\ \bibnamefont {{Lumsden}}}, \bibinfo {author}
  {\bibfnamefont {Y.}~\bibnamefont {{Yiu}}}, \bibinfo {author} {\bibfnamefont
  {J.}~\bibnamefont {{Knolle}}}, \bibinfo {author} {\bibfnamefont
  {S.}~\bibnamefont {{Bhattacharjee}}}, \bibinfo {author} {\bibfnamefont
  {D.~L.}\ \bibnamefont {{Kovrizhin}}}, \bibinfo {author} {\bibfnamefont
  {R.}~\bibnamefont {{Moessner}}}, \bibinfo {author} {\bibfnamefont {D.~A.}\
  \bibnamefont {{Tennant}}}, \bibinfo {author} {\bibfnamefont {D.~G.}\
  \bibnamefont {{Mandrus}}}, \ and\ \bibinfo {author} {\bibfnamefont {S.~E.}\
  \bibnamefont {{Nagler}}},\ }\bibfield  {title} {\enquote {\bibinfo {title}
  {{Proximate Kitaev quantum spin liquid behaviour in a honeycomb magnet}},}\
  }\href {\doibase 10.1038/nmat4604} {\bibfield  {journal} {\bibinfo  {journal}
  {Nat. Mater.}\ }\textbf {\bibinfo {volume} {15}},\ \bibinfo {pages}
  {733--740} (\bibinfo {year} {2016})}\BibitemShut {NoStop}%
\bibitem [{\citenamefont {Kim}\ \emph {et~al.}(2015)\citenamefont {Kim},
  \citenamefont {V.}, \citenamefont {Catuneanu},\ and\ \citenamefont
  {Kee}}]{HSKim2015}%
  \BibitemOpen
  \bibfield  {author} {\bibinfo {author} {\bibfnamefont {Heung-Sik}\
  \bibnamefont {Kim}}, \bibinfo {author} {\bibfnamefont {Vijay~Shankar}\
  \bibnamefont {V.}}, \bibinfo {author} {\bibfnamefont {Andrei}\ \bibnamefont
  {Catuneanu}}, \ and\ \bibinfo {author} {\bibfnamefont {Hae-Young}\
  \bibnamefont {Kee}},\ }\bibfield  {title} {\enquote {\bibinfo {title}
  {{Kitaev magnetism in honeycomb RuCl$_{3}$ with intermediate spin-orbit
  coupling}},}\ }\href {\doibase 10.1103/PhysRevB.91.241110} {\bibfield
  {journal} {\bibinfo  {journal} {Phys. Rev. B}\ }\textbf {\bibinfo {volume}
  {91}},\ \bibinfo {pages} {241110(R)} (\bibinfo {year} {2015})}\BibitemShut
  {NoStop}%
\bibitem [{\citenamefont {Ni}\ \emph {et~al.}(2022)\citenamefont {Ni},
  \citenamefont {Gui}, \citenamefont {Powderly},\ and\ \citenamefont
  {Cava}}]{Danrui2022}%
  \BibitemOpen
  \bibfield  {author} {\bibinfo {author} {\bibfnamefont {Danrui}\ \bibnamefont
  {Ni}}, \bibinfo {author} {\bibfnamefont {Xin}\ \bibnamefont {Gui}}, \bibinfo
  {author} {\bibfnamefont {Kelly~M.}\ \bibnamefont {Powderly}}, \ and\ \bibinfo
  {author} {\bibfnamefont {Robert~J.}\ \bibnamefont {Cava}},\ }\bibfield
  {title} {\enquote {\bibinfo {title} {Honeycomb-structure {RuI}$_3$, a new
  quantum material related to $\alpha$-{RuCl}$_3$},}\ }\href {\doibase
  https://doi.org/10.1002/adma.202106831} {\bibfield  {journal} {\bibinfo
  {journal} {Adv. Mater.}\ }\textbf {\bibinfo {volume} {34}},\ \bibinfo {pages}
  {2106831} (\bibinfo {year} {2022})}\BibitemShut {NoStop}%
\bibitem [{\citenamefont {Imai}\ \emph {et~al.}(2022)\citenamefont {Imai},
  \citenamefont {Nawa}, \citenamefont {Shimizu}, \citenamefont {Yamada},
  \citenamefont {Fujihara}, \citenamefont {Aoyama}, \citenamefont {Takahashi},
  \citenamefont {Okuyama}, \citenamefont {Ohashi}, \citenamefont {Hagihala},
  \citenamefont {Torii}, \citenamefont {Morikawa}, \citenamefont {Terauchi},
  \citenamefont {Kawamata}, \citenamefont {Kato}, \citenamefont {Gotou},
  \citenamefont {Itoh}, \citenamefont {Sato},\ and\ \citenamefont
  {Ohgushi}}]{Imai2021}%
  \BibitemOpen
  \bibfield  {author} {\bibinfo {author} {\bibfnamefont {Yoshinori}\
  \bibnamefont {Imai}}, \bibinfo {author} {\bibfnamefont {Kazuhiro}\
  \bibnamefont {Nawa}}, \bibinfo {author} {\bibfnamefont {Yasuhiro}\
  \bibnamefont {Shimizu}}, \bibinfo {author} {\bibfnamefont {Wakana}\
  \bibnamefont {Yamada}}, \bibinfo {author} {\bibfnamefont {Hideyuki}\
  \bibnamefont {Fujihara}}, \bibinfo {author} {\bibfnamefont {Takuya}\
  \bibnamefont {Aoyama}}, \bibinfo {author} {\bibfnamefont {Ryotaro}\
  \bibnamefont {Takahashi}}, \bibinfo {author} {\bibfnamefont {Daisuke}\
  \bibnamefont {Okuyama}}, \bibinfo {author} {\bibfnamefont {Takamasa}\
  \bibnamefont {Ohashi}}, \bibinfo {author} {\bibfnamefont {Masato}\
  \bibnamefont {Hagihala}}, \bibinfo {author} {\bibfnamefont {Shuki}\
  \bibnamefont {Torii}}, \bibinfo {author} {\bibfnamefont {Daisuke}\
  \bibnamefont {Morikawa}}, \bibinfo {author} {\bibfnamefont {Masami}\
  \bibnamefont {Terauchi}}, \bibinfo {author} {\bibfnamefont {Takayuki}\
  \bibnamefont {Kawamata}}, \bibinfo {author} {\bibfnamefont {Masatsune}\
  \bibnamefont {Kato}}, \bibinfo {author} {\bibfnamefont {Hirotada}\
  \bibnamefont {Gotou}}, \bibinfo {author} {\bibfnamefont {Masayuki}\
  \bibnamefont {Itoh}}, \bibinfo {author} {\bibfnamefont {Taku~J.}\
  \bibnamefont {Sato}}, \ and\ \bibinfo {author} {\bibfnamefont {Kenya}\
  \bibnamefont {Ohgushi}},\ }\bibfield  {title} {\enquote {\bibinfo {title}
  {Zigzag magnetic order in the {Kitaev} spin-liquid candidate material
  {RuBr}$_{3}$ with a honeycomb lattice},}\ }\href {\doibase
  10.1103/PhysRevB.105.L041112} {\bibfield  {journal} {\bibinfo  {journal}
  {Phys. Rev. B}\ }\textbf {\bibinfo {volume} {105}},\ \bibinfo {pages}
  {L041112} (\bibinfo {year} {2022})}\BibitemShut {NoStop}%
\bibitem [{\citenamefont {{Hao}}\ \emph {et~al.}(2021)\citenamefont {{Hao}},
  \citenamefont {{Wo}}, \citenamefont {{Gu}}, \citenamefont {{Zhang}},
  \citenamefont {{Gu}}, \citenamefont {{Zheng}}, \citenamefont {{Zhao}},
  \citenamefont {{Xu}}, \citenamefont {{Lynn}}, \citenamefont {{Nakajima}},
  \citenamefont {{Murai}}, \citenamefont {{Wang}},\ and\ \citenamefont
  {{Zhao}}}]{Hao2021}%
  \BibitemOpen
  \bibfield  {author} {\bibinfo {author} {\bibfnamefont {Y.}~\bibnamefont
  {{Hao}}}, \bibinfo {author} {\bibfnamefont {H.}~\bibnamefont {{Wo}}},
  \bibinfo {author} {\bibfnamefont {Y.}~\bibnamefont {{Gu}}}, \bibinfo {author}
  {\bibfnamefont {X.}~\bibnamefont {{Zhang}}}, \bibinfo {author} {\bibfnamefont
  {Y.}~\bibnamefont {{Gu}}}, \bibinfo {author} {\bibfnamefont {S.}~\bibnamefont
  {{Zheng}}}, \bibinfo {author} {\bibfnamefont {Y.}~\bibnamefont {{Zhao}}},
  \bibinfo {author} {\bibfnamefont {G.}~\bibnamefont {{Xu}}}, \bibinfo {author}
  {\bibfnamefont {J.~W.}\ \bibnamefont {{Lynn}}}, \bibinfo {author}
  {\bibfnamefont {K.}~\bibnamefont {{Nakajima}}}, \bibinfo {author}
  {\bibfnamefont {N.}~\bibnamefont {{Murai}}}, \bibinfo {author} {\bibfnamefont
  {W.}~\bibnamefont {{Wang}}}, \ and\ \bibinfo {author} {\bibfnamefont
  {J.}~\bibnamefont {{Zhao}}},\ }\bibfield  {title} {\enquote {\bibinfo {title}
  {{Field-tuned magnetic structure and phase diagram of the honeycomb magnet
  YbCl$_{3}$}},}\ }\href {\doibase 10.1007/s11433-020-1626-3} {\bibfield
  {journal} {\bibinfo  {journal} {Sci. China Phys. Mech. Astron.}\ }\textbf
  {\bibinfo {volume} {64}},\ \bibinfo {eid} {237411} (\bibinfo {year}
  {2021})}\BibitemShut {NoStop}%
\bibitem [{\citenamefont {Xing}\ \emph {et~al.}(2020)\citenamefont {Xing},
  \citenamefont {Feng}, \citenamefont {Liu}, \citenamefont {Emmanouilidou},
  \citenamefont {Hu}, \citenamefont {Liu}, \citenamefont {Graf}, \citenamefont
  {Ramirez}, \citenamefont {Chen}, \citenamefont {Cao},\ and\ \citenamefont
  {Ni}}]{Xing2020}%
  \BibitemOpen
  \bibfield  {author} {\bibinfo {author} {\bibfnamefont {Jie}\ \bibnamefont
  {Xing}}, \bibinfo {author} {\bibfnamefont {Erxi}\ \bibnamefont {Feng}},
  \bibinfo {author} {\bibfnamefont {Yaohua}\ \bibnamefont {Liu}}, \bibinfo
  {author} {\bibfnamefont {Eve}\ \bibnamefont {Emmanouilidou}}, \bibinfo
  {author} {\bibfnamefont {Chaowei}\ \bibnamefont {Hu}}, \bibinfo {author}
  {\bibfnamefont {Jinyu}\ \bibnamefont {Liu}}, \bibinfo {author} {\bibfnamefont
  {David}\ \bibnamefont {Graf}}, \bibinfo {author} {\bibfnamefont {Arthur~P.}\
  \bibnamefont {Ramirez}}, \bibinfo {author} {\bibfnamefont {Gang}\
  \bibnamefont {Chen}}, \bibinfo {author} {\bibfnamefont {Huibo}\ \bibnamefont
  {Cao}}, \ and\ \bibinfo {author} {\bibfnamefont {Ni}~\bibnamefont {Ni}},\
  }\bibfield  {title} {\enquote {\bibinfo {title} {N\'eel-type
  antiferromagnetic order and magnetic field--temperature phase diagram in the
  spin-$\frac{1}{2}$ rare-earth honeycomb compound $\mathrm{YbCl}{}_{3}$},}\
  }\href {\doibase 10.1103/PhysRevB.102.014427} {\bibfield  {journal} {\bibinfo
   {journal} {Phys. Rev. B}\ }\textbf {\bibinfo {volume} {102}},\ \bibinfo
  {pages} {014427} (\bibinfo {year} {2020})}\BibitemShut {NoStop}%
\bibitem [{\citenamefont {{Sala}}\ \emph {et~al.}()\citenamefont {{Sala}},
  \citenamefont {{Stone}}, \citenamefont {{Rai}}, \citenamefont {{May}},
  \citenamefont {{Laurell}}, \citenamefont {{Garlea}}, \citenamefont {{Butch}},
  \citenamefont {{Lumsden}}, \citenamefont {{Ehlers}}, \citenamefont
  {{Pokharel}}, \citenamefont {{Mandrus}}, \citenamefont {{Parker}},
  \citenamefont {{Okamoto}}, \citenamefont {{Hal{\'a}sz}},\ and\ \citenamefont
  {{Christianson}}}]{Sala2020}%
  \BibitemOpen
  \bibfield  {author} {\bibinfo {author} {\bibfnamefont {G.}~\bibnamefont
  {{Sala}}}, \bibinfo {author} {\bibfnamefont {M.~B.}\ \bibnamefont {{Stone}}},
  \bibinfo {author} {\bibfnamefont {Binod~K.}\ \bibnamefont {{Rai}}}, \bibinfo
  {author} {\bibfnamefont {A.~F.}\ \bibnamefont {{May}}}, \bibinfo {author}
  {\bibfnamefont {Pontus}\ \bibnamefont {{Laurell}}}, \bibinfo {author}
  {\bibfnamefont {V.~O.}\ \bibnamefont {{Garlea}}}, \bibinfo {author}
  {\bibfnamefont {N.~P.}\ \bibnamefont {{Butch}}}, \bibinfo {author}
  {\bibfnamefont {M.~D.}\ \bibnamefont {{Lumsden}}}, \bibinfo {author}
  {\bibfnamefont {G.}~\bibnamefont {{Ehlers}}}, \bibinfo {author}
  {\bibfnamefont {G.}~\bibnamefont {{Pokharel}}}, \bibinfo {author}
  {\bibfnamefont {D.}~\bibnamefont {{Mandrus}}}, \bibinfo {author}
  {\bibfnamefont {D.~S.}\ \bibnamefont {{Parker}}}, \bibinfo {author}
  {\bibfnamefont {S.}~\bibnamefont {{Okamoto}}}, \bibinfo {author}
  {\bibfnamefont {G{\'a}bor~B.}\ \bibnamefont {{Hal{\'a}sz}}}, \ and\ \bibinfo
  {author} {\bibfnamefont {A.~D.}\ \bibnamefont {{Christianson}}},\ }\bibfield
  {title} {\enquote {\bibinfo {title} {{Van Hove singularity in the magnon
  spectrum of the antiferromagnetic quantum honeycomb lattice}},}\ }\href@noop
  {} {\ }\Eprint {http://arxiv.org/abs/2003.01754 (2020)} {arXiv:2003.01754
  (2020)} \BibitemShut {NoStop}%
\bibitem [{\citenamefont {McGuire}\ \emph {et~al.}(2015)\citenamefont
  {McGuire}, \citenamefont {Dixit}, \citenamefont {Cooper},\ and\ \citenamefont
  {Sales}}]{McGuire2015}%
  \BibitemOpen
  \bibfield  {author} {\bibinfo {author} {\bibfnamefont {Michael~A.}\
  \bibnamefont {McGuire}}, \bibinfo {author} {\bibfnamefont {Hemant}\
  \bibnamefont {Dixit}}, \bibinfo {author} {\bibfnamefont {Valentino~R.}\
  \bibnamefont {Cooper}}, \ and\ \bibinfo {author} {\bibfnamefont {Brian~C.}\
  \bibnamefont {Sales}},\ }\bibfield  {title} {\enquote {\bibinfo {title}
  {Coupling of crystal structure and magnetism in the layered, ferromagnetic
  insulator {CrI$_3$}},}\ }\href {\doibase 10.1021/cm504242t} {\bibfield
  {journal} {\bibinfo  {journal} {Chem. Mater.}\ }\textbf {\bibinfo {volume}
  {27}},\ \bibinfo {pages} {612--620} (\bibinfo {year} {2015})}\BibitemShut
  {NoStop}%
\bibitem [{\citenamefont {{Ji}}\ \emph {et~al.}(2021)\citenamefont {{Ji}},
  \citenamefont {{Sun}}, \citenamefont {{Cai}}, \citenamefont {{Wang}},
  \citenamefont {{Sun}}, \citenamefont {{Ren}}, \citenamefont {{Zhang}},
  \citenamefont {{Jin}},\ and\ \citenamefont {{Zhang}}}]{Zhang2021CPL}%
  \BibitemOpen
  \bibfield  {author} {\bibinfo {author} {\bibfnamefont {Jianting}\
  \bibnamefont {{Ji}}}, \bibinfo {author} {\bibfnamefont {Mengjie}\
  \bibnamefont {{Sun}}}, \bibinfo {author} {\bibfnamefont {Yanzhen}\
  \bibnamefont {{Cai}}}, \bibinfo {author} {\bibfnamefont {Yimeng}\
  \bibnamefont {{Wang}}}, \bibinfo {author} {\bibfnamefont {Yingqi}\
  \bibnamefont {{Sun}}}, \bibinfo {author} {\bibfnamefont {Wei}\ \bibnamefont
  {{Ren}}}, \bibinfo {author} {\bibfnamefont {Zheng}\ \bibnamefont {{Zhang}}},
  \bibinfo {author} {\bibfnamefont {Feng}\ \bibnamefont {{Jin}}}, \ and\
  \bibinfo {author} {\bibfnamefont {Qingming}\ \bibnamefont {{Zhang}}},\
  }\bibfield  {title} {\enquote {\bibinfo {title} {{Rare-Earth Chalcohalides: A
  Family of van der Waals Layered Kitaev Spin Liquid Candidates}},}\ }\href
  {\doibase 10.1088/0256-307X/38/4/047502} {\bibfield  {journal} {\bibinfo
  {journal} {Chinese Phys. Lett.}\ }\textbf {\bibinfo {volume} {38}},\ \bibinfo
  {eid} {047502} (\bibinfo {year} {2021})}\BibitemShut {NoStop}%
\bibitem [{\citenamefont {{Zhang}}\ \emph {et~al.}(2022)\citenamefont
  {{Zhang}}, \citenamefont {{Cai}}, \citenamefont {{Kang}}, \citenamefont
  {{Ouyang}}, \citenamefont {{Zhang}}, \citenamefont {{Zhang}}, \citenamefont
  {{Ji}}, \citenamefont {{Jin}},\ and\ \citenamefont {{Zhang}}}]{Zhang2022PRR}%
  \BibitemOpen
  \bibfield  {author} {\bibinfo {author} {\bibfnamefont {Zheng}\ \bibnamefont
  {{Zhang}}}, \bibinfo {author} {\bibfnamefont {Yanzhen}\ \bibnamefont
  {{Cai}}}, \bibinfo {author} {\bibfnamefont {Jing}\ \bibnamefont {{Kang}}},
  \bibinfo {author} {\bibfnamefont {Zhongwen}\ \bibnamefont {{Ouyang}}},
  \bibinfo {author} {\bibfnamefont {Zhitao}\ \bibnamefont {{Zhang}}}, \bibinfo
  {author} {\bibfnamefont {Anmin}\ \bibnamefont {{Zhang}}}, \bibinfo {author}
  {\bibfnamefont {Jianting}\ \bibnamefont {{Ji}}}, \bibinfo {author}
  {\bibfnamefont {Feng}\ \bibnamefont {{Jin}}}, \ and\ \bibinfo {author}
  {\bibfnamefont {Qingming}\ \bibnamefont {{Zhang}}},\ }\bibfield  {title}
  {\enquote {\bibinfo {title} {{Anisotropic exchange coupling and ground state
  phase diagram of Kitaev compound YbOCl}},}\ }\href {\doibase
  10.1103/PhysRevResearch.4.033006} {\bibfield  {journal} {\bibinfo  {journal}
  {Phys. Rev. Research}\ }\textbf {\bibinfo {volume} {4}},\ \bibinfo {eid}
  {033006} (\bibinfo {year} {2022})}\BibitemShut {NoStop}%
\bibitem [{\citenamefont {{Lin}}\ \emph {et~al.}(2021)\citenamefont {{Lin}},
  \citenamefont {{Jeong}}, \citenamefont {{Kim}}, \citenamefont {{Wang}},
  \citenamefont {{Huang}}, \citenamefont {{Masuda}}, \citenamefont {{Asai}},
  \citenamefont {{Itoh}}, \citenamefont {{G{\"u}nther}}, \citenamefont
  {{Russina}}, \citenamefont {{Lu}}, \citenamefont {{Sheng}}, \citenamefont
  {{Wang}}, \citenamefont {{Wang}}, \citenamefont {{Wang}}, \citenamefont
  {{Ren}}, \citenamefont {{Xi}}, \citenamefont {{Tong}}, \citenamefont
  {{Ling}}, \citenamefont {{Liu}}, \citenamefont {{Wu}}, \citenamefont {{Mei}},
  \citenamefont {{Qu}}, \citenamefont {{Zhou}}, \citenamefont {{Wang}},
  \citenamefont {{Park}}, \citenamefont {{Wan}},\ and\ \citenamefont
  {{Ma}}}]{Lin2021NC}%
  \BibitemOpen
  \bibfield  {author} {\bibinfo {author} {\bibfnamefont {Gaoting}\ \bibnamefont
  {{Lin}}}, \bibinfo {author} {\bibfnamefont {Jaehong}\ \bibnamefont
  {{Jeong}}}, \bibinfo {author} {\bibfnamefont {Chaebin}\ \bibnamefont
  {{Kim}}}, \bibinfo {author} {\bibfnamefont {Yao}\ \bibnamefont {{Wang}}},
  \bibinfo {author} {\bibfnamefont {Qing}\ \bibnamefont {{Huang}}}, \bibinfo
  {author} {\bibfnamefont {Takatsugu}\ \bibnamefont {{Masuda}}}, \bibinfo
  {author} {\bibfnamefont {Shinichiro}\ \bibnamefont {{Asai}}}, \bibinfo
  {author} {\bibfnamefont {Shinichi}\ \bibnamefont {{Itoh}}}, \bibinfo {author}
  {\bibfnamefont {Gerrit}\ \bibnamefont {{G{\"u}nther}}}, \bibinfo {author}
  {\bibfnamefont {Margarita}\ \bibnamefont {{Russina}}}, \bibinfo {author}
  {\bibfnamefont {Zhilun}\ \bibnamefont {{Lu}}}, \bibinfo {author}
  {\bibfnamefont {Jieming}\ \bibnamefont {{Sheng}}}, \bibinfo {author}
  {\bibfnamefont {Le}~\bibnamefont {{Wang}}}, \bibinfo {author} {\bibfnamefont
  {Jiucai}\ \bibnamefont {{Wang}}}, \bibinfo {author} {\bibfnamefont {Guohua}\
  \bibnamefont {{Wang}}}, \bibinfo {author} {\bibfnamefont {Qingyong}\
  \bibnamefont {{Ren}}}, \bibinfo {author} {\bibfnamefont {Chuanying}\
  \bibnamefont {{Xi}}}, \bibinfo {author} {\bibfnamefont {Wei}\ \bibnamefont
  {{Tong}}}, \bibinfo {author} {\bibfnamefont {Langsheng}\ \bibnamefont
  {{Ling}}}, \bibinfo {author} {\bibfnamefont {Zhengxin}\ \bibnamefont
  {{Liu}}}, \bibinfo {author} {\bibfnamefont {Liusuo}\ \bibnamefont {{Wu}}},
  \bibinfo {author} {\bibfnamefont {Jiawei}\ \bibnamefont {{Mei}}}, \bibinfo
  {author} {\bibfnamefont {Zhe}\ \bibnamefont {{Qu}}}, \bibinfo {author}
  {\bibfnamefont {Haidong}\ \bibnamefont {{Zhou}}}, \bibinfo {author}
  {\bibfnamefont {Xiaoqun}\ \bibnamefont {{Wang}}}, \bibinfo {author}
  {\bibfnamefont {Je-Geun}\ \bibnamefont {{Park}}}, \bibinfo {author}
  {\bibfnamefont {Yuan}\ \bibnamefont {{Wan}}}, \ and\ \bibinfo {author}
  {\bibfnamefont {Jie}\ \bibnamefont {{Ma}}},\ }\bibfield  {title} {\enquote
  {\bibinfo {title} {{Field-induced quantum spin disordered state in spin-1/2
  honeycomb magnet Na$_{2}$Co$_{2}$TeO$_{6}$}},}\ }\href {\doibase
  10.1038/s41467-021-25567-7} {\bibfield  {journal} {\bibinfo  {journal} {Nat.
  Commun.}\ }\textbf {\bibinfo {volume} {12}},\ \bibinfo {eid} {5559} (\bibinfo
  {year} {2021})}\BibitemShut {NoStop}%
\bibitem [{\citenamefont {Yao}\ \emph {et~al.}(2022)\citenamefont {Yao},
  \citenamefont {Iida}, \citenamefont {Kamazawa},\ and\ \citenamefont
  {Li}}]{Yao2022PRL}%
  \BibitemOpen
  \bibfield  {author} {\bibinfo {author} {\bibfnamefont {Weiliang}\
  \bibnamefont {Yao}}, \bibinfo {author} {\bibfnamefont {Kazuki}\ \bibnamefont
  {Iida}}, \bibinfo {author} {\bibfnamefont {Kazuya}\ \bibnamefont {Kamazawa}},
  \ and\ \bibinfo {author} {\bibfnamefont {Yuan}\ \bibnamefont {Li}},\
  }\bibfield  {title} {\enquote {\bibinfo {title} {Excitations in the ordered
  and paramagnetic states of honeycomb magnet {Na$_{2}$Co$_{2}$TeO$_{6}$}},}\
  }\href {\doibase 10.1103/PhysRevLett.129.147202} {\bibfield  {journal}
  {\bibinfo  {journal} {Phys. Rev. Lett.}\ }\textbf {\bibinfo {volume} {129}},\
  \bibinfo {pages} {147202} (\bibinfo {year} {2022})}\BibitemShut {NoStop}%
\bibitem [{\citenamefont {Liu}\ \emph {et~al.}(2020)\citenamefont {Liu},
  \citenamefont {Chaloupka},\ and\ \citenamefont {Khaliullin}}]{Liu2020PRL}%
  \BibitemOpen
  \bibfield  {author} {\bibinfo {author} {\bibfnamefont {Huimei}\ \bibnamefont
  {Liu}}, \bibinfo {author} {\bibfnamefont {Ji\ifmmode
  \check{r}\else~\v{r}\fi{}\'{\i}}\ \bibnamefont {Chaloupka}}, \ and\ \bibinfo
  {author} {\bibfnamefont {Giniyat}\ \bibnamefont {Khaliullin}},\ }\bibfield
  {title} {\enquote {\bibinfo {title} {Kitaev spin liquid in $3d$ transition
  metal compounds},}\ }\href {\doibase 10.1103/PhysRevLett.125.047201}
  {\bibfield  {journal} {\bibinfo  {journal} {Phys. Rev. Lett.}\ }\textbf
  {\bibinfo {volume} {125}},\ \bibinfo {pages} {047201} (\bibinfo {year}
  {2020})}\BibitemShut {NoStop}%
\bibitem [{\citenamefont {{Zhong}}\ \emph {et~al.}(2020)\citenamefont
  {{Zhong}}, \citenamefont {{Gao}}, \citenamefont {{Ong}},\ and\ \citenamefont
  {{Cava}}}]{Zhong2020SA}%
  \BibitemOpen
  \bibfield  {author} {\bibinfo {author} {\bibfnamefont {Ruidan}\ \bibnamefont
  {{Zhong}}}, \bibinfo {author} {\bibfnamefont {Tong}\ \bibnamefont {{Gao}}},
  \bibinfo {author} {\bibfnamefont {Nai~Phuan}\ \bibnamefont {{Ong}}}, \ and\
  \bibinfo {author} {\bibfnamefont {Robert~J.}\ \bibnamefont {{Cava}}},\
  }\bibfield  {title} {\enquote {\bibinfo {title} {{Weak-field induced
  nonmagnetic state in a {Co}-based honeycomb}},}\ }\href {\doibase
  10.1126/sciadv.aay6953} {\bibfield  {journal} {\bibinfo  {journal} {Sci.
  Adv.}\ }\textbf {\bibinfo {volume} {6}},\ \bibinfo {pages} {eaay6953}
  (\bibinfo {year} {2020})}\BibitemShut {NoStop}%
\bibitem [{\citenamefont {Plumb}\ \emph {et~al.}(2014)\citenamefont {Plumb},
  \citenamefont {Clancy}, \citenamefont {Sandilands}, \citenamefont {Shankar},
  \citenamefont {Hu}, \citenamefont {Burch}, \citenamefont {Kee},\ and\
  \citenamefont {Kim}}]{Plumb2014}%
  \BibitemOpen
  \bibfield  {author} {\bibinfo {author} {\bibfnamefont {K.~W.}\ \bibnamefont
  {Plumb}}, \bibinfo {author} {\bibfnamefont {J.~P.}\ \bibnamefont {Clancy}},
  \bibinfo {author} {\bibfnamefont {L.~J.}\ \bibnamefont {Sandilands}},
  \bibinfo {author} {\bibfnamefont {V.~V.}\ \bibnamefont {Shankar}}, \bibinfo
  {author} {\bibfnamefont {Y.~F.}\ \bibnamefont {Hu}}, \bibinfo {author}
  {\bibfnamefont {K.~S.}\ \bibnamefont {Burch}}, \bibinfo {author}
  {\bibfnamefont {H.-Y.}\ \bibnamefont {Kee}}, \ and\ \bibinfo {author}
  {\bibfnamefont {Y.-J.}\ \bibnamefont {Kim}},\ }\bibfield  {title} {\enquote
  {\bibinfo {title} {${\alpha}$-{RuCl}$_{3}$: A spin-orbit assisted {Mott}
  insulator on a honeycomb lattice},}\ }\href {\doibase
  10.1103/PhysRevB.90.041112} {\bibfield  {journal} {\bibinfo  {journal} {Phys.
  Rev. B}\ }\textbf {\bibinfo {volume} {90}},\ \bibinfo {pages} {041112(R)}
  (\bibinfo {year} {2014})}\BibitemShut {NoStop}%
\bibitem [{\citenamefont {Janssen}\ \emph {et~al.}(2017)\citenamefont
  {Janssen}, \citenamefont {Andrade},\ and\ \citenamefont
  {Vojta}}]{Janssen2017}%
  \BibitemOpen
  \bibfield  {author} {\bibinfo {author} {\bibfnamefont {Lukas}\ \bibnamefont
  {Janssen}}, \bibinfo {author} {\bibfnamefont {Eric~C.}\ \bibnamefont
  {Andrade}}, \ and\ \bibinfo {author} {\bibfnamefont {Matthias}\ \bibnamefont
  {Vojta}},\ }\bibfield  {title} {\enquote {\bibinfo {title} {Magnetization
  processes of zigzag states on the honeycomb lattice: Identifying spin models
  for {${\alpha}$-{RuCl}$_{3}$ and ${\mathrm{Na}}_{2}{\mathrm{IrO}}_{3}$}},}\
  }\href {\doibase 10.1103/PhysRevB.96.064430} {\bibfield  {journal} {\bibinfo
  {journal} {Phys. Rev. B}\ }\textbf {\bibinfo {volume} {96}},\ \bibinfo
  {pages} {064430} (\bibinfo {year} {2017})}\BibitemShut {NoStop}%
\bibitem [{\citenamefont {Andrade}\ \emph {et~al.}(2020)\citenamefont
  {Andrade}, \citenamefont {Janssen},\ and\ \citenamefont
  {Vojta}}]{Andrade2020}%
  \BibitemOpen
  \bibfield  {author} {\bibinfo {author} {\bibfnamefont {Eric~C.}\ \bibnamefont
  {Andrade}}, \bibinfo {author} {\bibfnamefont {Lukas}\ \bibnamefont
  {Janssen}}, \ and\ \bibinfo {author} {\bibfnamefont {Matthias}\ \bibnamefont
  {Vojta}},\ }\bibfield  {title} {\enquote {\bibinfo {title} {Susceptibility
  anisotropy and its disorder evolution in models for {Kitaev} materials},}\
  }\href {\doibase 10.1103/PhysRevB.102.115160} {\bibfield  {journal} {\bibinfo
   {journal} {Phys. Rev. B}\ }\textbf {\bibinfo {volume} {102}},\ \bibinfo
  {pages} {115160} (\bibinfo {year} {2020})}\BibitemShut {NoStop}%
\bibitem [{\citenamefont {Yoshitake}\ \emph {et~al.}(2020)\citenamefont
  {Yoshitake}, \citenamefont {Nasu}, \citenamefont {Kato},\ and\ \citenamefont
  {Motome}}]{Motome2019}%
  \BibitemOpen
  \bibfield  {author} {\bibinfo {author} {\bibfnamefont {J.}~\bibnamefont
  {Yoshitake}}, \bibinfo {author} {\bibfnamefont {J.}~\bibnamefont {Nasu}},
  \bibinfo {author} {\bibfnamefont {Y.}~\bibnamefont {Kato}}, \ and\ \bibinfo
  {author} {\bibfnamefont {Y.}~\bibnamefont {Motome}},\ }\bibfield  {title}
  {\enquote {\bibinfo {title} {Majorana-magnon crossover by a magnetic field in
  the {Kitaev} model: Continuous-time quantum {Monte Carlo} study},}\ }\href
  {\doibase 10.1103/PhysRevB.101.100408} {\bibfield  {journal} {\bibinfo
  {journal} {Phys. Rev. B}\ }\textbf {\bibinfo {volume} {101}},\ \bibinfo
  {pages} {100408(R)} (\bibinfo {year} {2020})}\BibitemShut {NoStop}%
\bibitem [{\citenamefont {Gordon}\ \emph {et~al.}(2019)\citenamefont {Gordon},
  \citenamefont {Catuneanu}, \citenamefont {S{\o}rensen},\ and\ \citenamefont
  {Kee}}]{Gordon2019}%
  \BibitemOpen
  \bibfield  {author} {\bibinfo {author} {\bibfnamefont {Jacob~S.}\
  \bibnamefont {Gordon}}, \bibinfo {author} {\bibfnamefont {Andrei}\
  \bibnamefont {Catuneanu}}, \bibinfo {author} {\bibfnamefont {Erik~S.}\
  \bibnamefont {S{\o}rensen}}, \ and\ \bibinfo {author} {\bibfnamefont
  {Hae-Young}\ \bibnamefont {Kee}},\ }\bibfield  {title} {\enquote {\bibinfo
  {title} {Theory of the field-revealed {Kitaev} spin liquid},}\ }\href
  {\doibase 10.1038/s41467-019-10405-8} {\bibfield  {journal} {\bibinfo
  {journal} {Nat. Commun.}\ }\textbf {\bibinfo {volume} {10}},\ \bibinfo
  {pages} {2470} (\bibinfo {year} {2019})}\BibitemShut {NoStop}%
\bibitem [{\citenamefont {Lee}\ \emph {et~al.}(2020)\citenamefont {Lee},
  \citenamefont {Kaneko}, \citenamefont {Chern}, \citenamefont {Okubo},
  \citenamefont {Yamaji}, \citenamefont {Kawashima},\ and\ \citenamefont
  {Kim}}]{Lee2020nc}%
  \BibitemOpen
  \bibfield  {author} {\bibinfo {author} {\bibfnamefont {Hyun-Yong}\
  \bibnamefont {Lee}}, \bibinfo {author} {\bibfnamefont {Ryui}\ \bibnamefont
  {Kaneko}}, \bibinfo {author} {\bibfnamefont {Li~Ern}\ \bibnamefont {Chern}},
  \bibinfo {author} {\bibfnamefont {Tsuyoshi}\ \bibnamefont {Okubo}}, \bibinfo
  {author} {\bibfnamefont {Youhei}\ \bibnamefont {Yamaji}}, \bibinfo {author}
  {\bibfnamefont {Naoki}\ \bibnamefont {Kawashima}}, \ and\ \bibinfo {author}
  {\bibfnamefont {Yong~Baek}\ \bibnamefont {Kim}},\ }\bibfield  {title}
  {\enquote {\bibinfo {title} {Magnetic field induced quantum phases in a
  tensor network study of {Kitaev} magnets},}\ }\href {\doibase
  10.1038/s41467-020-15320-x} {\bibfield  {journal} {\bibinfo  {journal} {Nat.
  Commun.}\ }\textbf {\bibinfo {volume} {11}},\ \bibinfo {pages} {1639}
  (\bibinfo {year} {2020})}\BibitemShut {NoStop}%
\bibitem [{\citenamefont {Chen}\ \emph {et~al.}(2019)\citenamefont {Chen},
  \citenamefont {Qu}, \citenamefont {Li}, \citenamefont {Chen}, \citenamefont
  {Gong}, \citenamefont {von Delft}, \citenamefont {Weichselbaum},\ and\
  \citenamefont {Li}}]{Chen2018b}%
  \BibitemOpen
  \bibfield  {author} {\bibinfo {author} {\bibfnamefont {L.}~\bibnamefont
  {Chen}}, \bibinfo {author} {\bibfnamefont {D.-W.}\ \bibnamefont {Qu}},
  \bibinfo {author} {\bibfnamefont {H.}~\bibnamefont {Li}}, \bibinfo {author}
  {\bibfnamefont {B.-B.}\ \bibnamefont {Chen}}, \bibinfo {author}
  {\bibfnamefont {S.-S.}\ \bibnamefont {Gong}}, \bibinfo {author}
  {\bibfnamefont {J.}~\bibnamefont {von Delft}}, \bibinfo {author}
  {\bibfnamefont {A.}~\bibnamefont {Weichselbaum}}, \ and\ \bibinfo {author}
  {\bibfnamefont {W.}~\bibnamefont {Li}},\ }\bibfield  {title} {\enquote
  {\bibinfo {title} {Two temperature scales in the triangular lattice
  {Heisenberg} antiferromagnet},}\ }\href {\doibase 10.1103/PhysRevB.99.140404}
  {\bibfield  {journal} {\bibinfo  {journal} {Phys. Rev. B}\ }\textbf {\bibinfo
  {volume} {99}},\ \bibinfo {pages} {140404(R)} (\bibinfo {year}
  {2019})}\BibitemShut {NoStop}%
\bibitem [{\citenamefont {{Li}}\ \emph {et~al.}(2020)\citenamefont {{Li}},
  \citenamefont {{Qu}}, \citenamefont {{Zhang}}, \citenamefont {{Jia}},
  \citenamefont {{Gong}}, \citenamefont {{Qi}},\ and\ \citenamefont
  {{Li}}}]{Li2020b}%
  \BibitemOpen
  \bibfield  {author} {\bibinfo {author} {\bibfnamefont {Han}\ \bibnamefont
  {{Li}}}, \bibinfo {author} {\bibfnamefont {Dai-Wei}\ \bibnamefont {{Qu}}},
  \bibinfo {author} {\bibfnamefont {Hao-Kai}\ \bibnamefont {{Zhang}}}, \bibinfo
  {author} {\bibfnamefont {Yi-Zhen}\ \bibnamefont {{Jia}}}, \bibinfo {author}
  {\bibfnamefont {Shou-Shu}\ \bibnamefont {{Gong}}}, \bibinfo {author}
  {\bibfnamefont {Yang}\ \bibnamefont {{Qi}}}, \ and\ \bibinfo {author}
  {\bibfnamefont {Wei}\ \bibnamefont {{Li}}},\ }\bibfield  {title} {\enquote
  {\bibinfo {title} {{Universal thermodynamics in the Kitaev fractional
  liquid}},}\ }\href {\doibase 10.1103/PhysRevResearch.2.043015} {\bibfield
  {journal} {\bibinfo  {journal} {Phys. Rev. Research}\ }\textbf {\bibinfo
  {volume} {2}},\ \bibinfo {eid} {043015} (\bibinfo {year} {2020})}\BibitemShut
  {NoStop}%
\bibitem [{\citenamefont {Li}\ \emph {et~al.}(2020)\citenamefont {Li},
  \citenamefont {Liao}, \citenamefont {Chen}, \citenamefont {Zeng},
  \citenamefont {Sheng}, \citenamefont {Qi}, \citenamefont {Meng},\ and\
  \citenamefont {Li}}]{Li2020}%
  \BibitemOpen
  \bibfield  {author} {\bibinfo {author} {\bibfnamefont {H.}~\bibnamefont
  {Li}}, \bibinfo {author} {\bibfnamefont {Y.-D.}\ \bibnamefont {Liao}},
  \bibinfo {author} {\bibfnamefont {B.-B.}\ \bibnamefont {Chen}}, \bibinfo
  {author} {\bibfnamefont {X.-T.}\ \bibnamefont {Zeng}}, \bibinfo {author}
  {\bibfnamefont {X.-L.}\ \bibnamefont {Sheng}}, \bibinfo {author}
  {\bibfnamefont {Y.}~\bibnamefont {Qi}}, \bibinfo {author} {\bibfnamefont
  {Z.~Y.}\ \bibnamefont {Meng}}, \ and\ \bibinfo {author} {\bibfnamefont
  {W.}~\bibnamefont {Li}},\ }\bibfield  {title} {\enquote {\bibinfo {title}
  {{Kosterlitz-Thouless} melting of magnetic order in the triangular quantum
  {Ising} material {TmMgGaO$_4$}},}\ }\href {\doibase
  10.1038/s41467-020-14907-8} {\bibfield  {journal} {\bibinfo  {journal} {Nat.
  Commun.}\ }\textbf {\bibinfo {volume} {11}},\ \bibinfo {pages} {1111}
  (\bibinfo {year} {2020})}\BibitemShut {NoStop}%
\bibitem [{\citenamefont {{Hu}}\ \emph {et~al.}(2020)\citenamefont {{Hu}},
  \citenamefont {{Ma}}, \citenamefont {{Liao}}, \citenamefont {{Li}},
  \citenamefont {{Ma}}, \citenamefont {{Cui}}, \citenamefont {{Shangguan}},
  \citenamefont {{Huang}}, \citenamefont {{Qi}}, \citenamefont {{Li}},
  \citenamefont {{Meng}}, \citenamefont {{Wen}},\ and\ \citenamefont
  {{Yu}}}]{Hu2020}%
  \BibitemOpen
  \bibfield  {author} {\bibinfo {author} {\bibfnamefont {Ze}~\bibnamefont
  {{Hu}}}, \bibinfo {author} {\bibfnamefont {Zhen}\ \bibnamefont {{Ma}}},
  \bibinfo {author} {\bibfnamefont {Yuan-Da}\ \bibnamefont {{Liao}}}, \bibinfo
  {author} {\bibfnamefont {Han}\ \bibnamefont {{Li}}}, \bibinfo {author}
  {\bibfnamefont {Chunsheng}\ \bibnamefont {{Ma}}}, \bibinfo {author}
  {\bibfnamefont {Yi}~\bibnamefont {{Cui}}}, \bibinfo {author} {\bibfnamefont
  {Yanyan}\ \bibnamefont {{Shangguan}}}, \bibinfo {author} {\bibfnamefont
  {Zhentao}\ \bibnamefont {{Huang}}}, \bibinfo {author} {\bibfnamefont {Yang}\
  \bibnamefont {{Qi}}}, \bibinfo {author} {\bibfnamefont {Wei}\ \bibnamefont
  {{Li}}}, \bibinfo {author} {\bibfnamefont {Zi~Yang}\ \bibnamefont {{Meng}}},
  \bibinfo {author} {\bibfnamefont {Jinsheng}\ \bibnamefont {{Wen}}}, \ and\
  \bibinfo {author} {\bibfnamefont {Weiqiang}\ \bibnamefont {{Yu}}},\
  }\bibfield  {title} {\enquote {\bibinfo {title} {{Evidence of the
  Berezinskii-Kosterlitz-Thouless phase in a frustrated magnet}},}\ }\href
  {\doibase 10.1038/s41467-020-19380-x} {\bibfield  {journal} {\bibinfo
  {journal} {Nat. Commun.}\ }\textbf {\bibinfo {volume} {11}},\ \bibinfo {eid}
  {5631} (\bibinfo {year} {2020})}\BibitemShut {NoStop}%
\bibitem [{\citenamefont {{Gao}}\ \emph {et~al.}(2022)\citenamefont {{Gao}},
  \citenamefont {{Fan}}, \citenamefont {{Li}}, \citenamefont {{Yang}},
  \citenamefont {{Zeng}}, \citenamefont {{Sheng}}, \citenamefont {{Zhong}},
  \citenamefont {{Qi}}, \citenamefont {{Wan}},\ and\ \citenamefont
  {{Li}}}]{Gao2022}%
  \BibitemOpen
  \bibfield  {author} {\bibinfo {author} {\bibfnamefont {Yuan}\ \bibnamefont
  {{Gao}}}, \bibinfo {author} {\bibfnamefont {Yu-Chen}\ \bibnamefont {{Fan}}},
  \bibinfo {author} {\bibfnamefont {Han}\ \bibnamefont {{Li}}}, \bibinfo
  {author} {\bibfnamefont {Fan}\ \bibnamefont {{Yang}}}, \bibinfo {author}
  {\bibfnamefont {Xu-Tao}\ \bibnamefont {{Zeng}}}, \bibinfo {author}
  {\bibfnamefont {Xian-Lei}\ \bibnamefont {{Sheng}}}, \bibinfo {author}
  {\bibfnamefont {Ruidan}\ \bibnamefont {{Zhong}}}, \bibinfo {author}
  {\bibfnamefont {Yang}\ \bibnamefont {{Qi}}}, \bibinfo {author} {\bibfnamefont
  {Yuan}\ \bibnamefont {{Wan}}}, \ and\ \bibinfo {author} {\bibfnamefont {Wei}\
  \bibnamefont {{Li}}},\ }\bibfield  {title} {\enquote {\bibinfo {title} {{Spin
  supersolidity in nearly ideal easy-axis triangular quantum antiferromagnet
  Na$_{2}$BaCo(PO$_{4}$)$_{2}$}},}\ }\href {\doibase
  10.1038/s41535-022-00500-3} {\bibfield  {journal} {\bibinfo  {journal} {npj
  Quant. Mater.}\ }\textbf {\bibinfo {volume} {7}},\ \bibinfo {eid} {89}
  (\bibinfo {year} {2022})}\BibitemShut {NoStop}%
\bibitem [{\citenamefont {{Chen}}\ \emph {et~al.}(2021)\citenamefont {{Chen}},
  \citenamefont {{Chen}}, \citenamefont {{Chen}}, \citenamefont {{Cui}},
  \citenamefont {{Zhai}}, \citenamefont {{Weichselbaum}}, \citenamefont {{von
  Delft}}, \citenamefont {{Meng}},\ and\ \citenamefont {{Li}}}]{Chen2021}%
  \BibitemOpen
  \bibfield  {author} {\bibinfo {author} {\bibfnamefont {Bin-Bin}\ \bibnamefont
  {{Chen}}}, \bibinfo {author} {\bibfnamefont {Chuang}\ \bibnamefont {{Chen}}},
  \bibinfo {author} {\bibfnamefont {Ziyu}\ \bibnamefont {{Chen}}}, \bibinfo
  {author} {\bibfnamefont {Jian}\ \bibnamefont {{Cui}}}, \bibinfo {author}
  {\bibfnamefont {Yueyang}\ \bibnamefont {{Zhai}}}, \bibinfo {author}
  {\bibfnamefont {Andreas}\ \bibnamefont {{Weichselbaum}}}, \bibinfo {author}
  {\bibfnamefont {Jan}\ \bibnamefont {{von Delft}}}, \bibinfo {author}
  {\bibfnamefont {Zi~Yang}\ \bibnamefont {{Meng}}}, \ and\ \bibinfo {author}
  {\bibfnamefont {Wei}\ \bibnamefont {{Li}}},\ }\bibfield  {title} {\enquote
  {\bibinfo {title} {{Quantum many-body simulations of the two-dimensional
  Fermi-Hubbard model in ultracold optical lattices}},}\ }\href {\doibase
  10.1103/PhysRevB.103.L041107} {\bibfield  {journal} {\bibinfo  {journal}
  {Phys. Rev. B}\ }\textbf {\bibinfo {volume} {103}},\ \bibinfo {eid} {L041107}
  (\bibinfo {year} {2021})}\BibitemShut {NoStop}%
\bibitem [{\citenamefont {{Lin}}\ \emph {et~al.}(2022)\citenamefont {{Lin}},
  \citenamefont {{Chen}}, \citenamefont {{Li}}, \citenamefont {{Meng}},\ and\
  \citenamefont {{Shi}}}]{Lin2022}%
  \BibitemOpen
  \bibfield  {author} {\bibinfo {author} {\bibfnamefont {Xiyue}\ \bibnamefont
  {{Lin}}}, \bibinfo {author} {\bibfnamefont {Bin-Bin}\ \bibnamefont {{Chen}}},
  \bibinfo {author} {\bibfnamefont {Wei}\ \bibnamefont {{Li}}}, \bibinfo
  {author} {\bibfnamefont {Zi~Yang}\ \bibnamefont {{Meng}}}, \ and\ \bibinfo
  {author} {\bibfnamefont {Tao}\ \bibnamefont {{Shi}}},\ }\bibfield  {title}
  {\enquote {\bibinfo {title} {{Exciton Proliferation and Fate of the
  Topological Mott Insulator in a Twisted Bilayer Graphene Lattice Model}},}\
  }\href {\doibase 10.1103/PhysRevLett.128.157201} {\bibfield  {journal}
  {\bibinfo  {journal} {Phys. Rev. Lett.}\ }\textbf {\bibinfo {volume} {128}},\
  \bibinfo {eid} {157201} (\bibinfo {year} {2022})}\BibitemShut {NoStop}%
\end{thebibliography}%

\end{document}